\def\sqr#1#2{{\vcenter{\vbox{\hrule height.#2pt\hbox{\vrule width.#2pt
height#1pt \kern#1pt \vrule width.#2pt}\hrule height.#2pt}}}}
\def\d{\partial}

\def\w{\mathchoice\sqr45\sqr45\sqr{2.1}3\sqr{1.5}3\,}
\def\=d{\,{\buildrel\rm def\over =}\,}
\def\psq{{\overline{\psi}}}

\documentclass{article}
\usepackage{amssymb}\usepackage{amsmath}
\newcommand{\CC}{\mathbb C}
\newcommand{\RR}{\mathbb R}
\newcommand{\ZZ}{\mathbb Z}
\newcommand{\NN}{\mathbb N}
\newcommand{\supp}{{\rm supp\>}}
\begin{document}
\title{The Master Ward Identity}
\author{M. D\"utsch\thanks{Work supported by the Deutsche
Forschungsgemeinschaft.} and F.-M. Boas\thanks{The second author was
involved in the treatment of BRST-symmetry.}\\[1mm]
Institut f\"ur Theoretische Physik\\
Universit\"at G\"ottingen\\
Bunsenstrasse 9\\
D-37073 G\"ottingen, Germany\\
{\tt duetsch@theorie.physik.uni-goettingen.de}\\
{\tt boas.franz-marc@bcg.com}}
\date{}

\maketitle
\begin{abstract}
In the framework of perturbative quantum field theory (QFT)
we propose a new, universal (re)normalization
condition (called 'master Ward identity') which expresses
the symmetries of the underlying classical theory. It implies
for example the field equations, energy-momentum, charge-
and ghost-number conservation, renormalized equal-time
commutation relations and BRST-symmetry.

It seems that the master Ward identity can nearly always be satisfied,
the only exceptions we know are the usual anomalies. We prove
the compatibility of the master Ward identity with the other
(re)normalization conditions of causal perturbation theory, and for
pure massive theories we show that the 'central solution' of Epstein
and Glaser fulfills the master Ward identity, if the UV-scaling
behavior of its individual terms is not relatively lowered.

Application of the master Ward identity to the
BRST-current of non-Abelian gauge theories generates an identity
(called 'master BRST-identity') which contains the information
which is needed for a local construction of the
algebra of observables, i.e. the elimination
of the unphysical fields and the construction of physical states in
the presence of an adiabatically switched off interaction.

{\bf PACS.} 11.10.Cd Field theory: axiomatic approach,
11.10.Gh Field theory: Renormalization,
11.15.Bt Gauge field theories: General properties of perturbation theory

\end{abstract}

\tableofcontents

\section{Introduction}

A perturbative interacting quantum field theory is usually constructed in
terms of time ordered products ('$T$-products') $T(W_1,...,W_n)(x_1,...,x_n)$
of Wick polynomials $W_1(x_1),...,W_n(x_n)$ of free fields.
The $T$-products are ill-defined for coinciding points because they
are (operator-valued) distributions. In the
framework of the inductive construction
of Bogoliubov \cite{BS} and Epstein/Glaser \cite{EG}
('causal perturbation theory') this can be formulated
as follows: the $T$-products of $n$-factors are known by
induction as operator-valued distributions up to the
total diagonal $D_n\=d\{(x_1,...,x_n)\>|\>x_1=...=x_n\}$. The problem
of renormalization is located in the extension of the $T$-products to $D_n$,
for every $n$. This extension is always possible, but it is non-unique.
The freedom is restricted by normalization conditions. They require
that symmetries which are present outside $D_n$ are maintained in the
extension and that, for each term in the $T$-product, the order of
the singularity at $D_n$ is not increased by the extension. (The
latter implies that an interaction with mass dimension $\leq 4$ yields a
renormalizable theory (by power counting)). Epstein/Glaser \cite{EG}
(see also \cite{BF}) give a general formula (\ref{extension}) for
the extension to $D_n$ which satisfies the last
requirement, but the other normalization conditions
are not taken into account. So, {\it the main problem of
perturbative renormalization is to prove that there is an extension
which fulfills all normalization conditions}. In the framework
of algebraic renormalization the corresponding problem is treated
by means of the 'quantum action principle' \cite{L}, \cite{Lam},
\cite{PS}, which states that
the variation of Green's functions (under a change of coordinates,
a variation of the fields or a variation of a parameter)
is equal to the insertion of a (local or space-time integrated)
composite field operator. Recently a local algebraic
operator formulation of certain cases of the quantum
action principle has been given by using
causal perturbation theory, and the connection to our normalization
conditions has been clarified \cite{DF1}.

The {\bf master Ward identity} (we will use the abbreviation 'MWI')
is a universal normalization condition supplementing the obvious
ones. It is an explicit expression for
\begin{equation}
  \d^\nu_x T(W,W_1,...,W_n)(x,x_1,...,x_n)- T(\d^\nu W,W_1,...,W_n)
(x,x_1,...,x_n).\label{[d,T]}
\end{equation}
Generally this difference cannot vanish for the following
reason\footnote{In particular this argument,
the name 'master Ward identity' and
the application of the MWI to the computation of a
rigorous substitute for the equal-time commutator of interacting
fields (\ref{etc}) are due to Klaus Fredenhagen.}:
the Wick polynomials are built
up from free fields, whereas the $T$-products are the building stones
of the perturbative interacting fields \cite{BS}. However, the field
equations of free and interacting fields are different.

Computing the difference (\ref{[d,T]}) by means of the Feynman rules
and the normalization condition {\bf (N3)} (see sect. 2.2),
it can be expressed solely by terms which contain the difference
$\d^\nu_x\langle\Omega,T(\phi,\chi)(x,x_l)\Omega\rangle-
\langle\Omega,T(\d^\nu\phi,\chi)
(x,x_l)\Omega\rangle$ of Feynman propagators,
where $\Omega$ is the Fock vacuum
and $\phi,\>\chi$ are free fields. The MWI requires that this
structure is preserved in the process of renormalization
(sect. 2). For tree diagrams this is automatically satisfied,
but for loop diagrams it is a hard task to show that there exists a
normalization which fulfills the MWI and the other normalization
conditions (sect. 3). Unfortunately there are a few examples where this is
impossible. However, the only obstructions we know are the usual,
well-known anomalies of perturbative QFT (sect. 5).

The {\it master Ward identity expresses all symmetries which can be
traced back to the field equations in classical field
theory}\footnote{In \cite{DF2} we extensively work out the
MWI in classical field theory. There the MWI can be formulated
non-perturbatively: it is a consequence of the field equations and the
fact that classical fields may be multiplied point-wise. Hence, the
classical MWI holds always true. (This, together with the fact that in the
perturbative expansion of classical fields solely tree diagrams appear,
agrees with the triviality of the quantum MWI for tree diagrams.) The
classical formulation of the MWI shows that there is a close
connection to the Schwinger-Dyson equations.}.
In particular we will demonstrate that the MWI implies

- the field equations for the interacting fields (sect. 4.1),

- conservation of the energy-momentum tensor (sect. 5.2),

- charge conservation in the presence of spinor fields (sect.4.2),

- ghost number conservation in the presence of fermionic ghost fields
(sect. 4.2), and

- the {\bf master  BRST-identity} (sect. 4.4-5), which contains the
full information of BRST-symmetry \cite{BRS} for massless and massive
gauge fields.

The field equations, conservation of the energy-momentum tensor,
charge and ghost number conservation have already been proved by using
other methods of renormalization (see e.g. \cite{Z} for the field
field equation and \cite{Lo} for the energy-momentum tensor, both are
based on BPHZ-renormalization) or/and in the framework of causal
perturbation theory \cite{St3},\cite{DKS1}, \cite{P:emt}, \cite{DF}.
Using the normal products of Zimmermann \cite{Z}, Lowenstein has proved
that it is allowed to take a partial derivative out of a Green's function,
if the degree of BPHZ-subtraction is lowered by one, see appendix B of
\cite{Lo}. However, we are not aware of a formulation of the
MWI in its full generality in any method of renormalization.

Also the master BRST-identity is new to our knowledge. It is the
answer to the obvious question: what results for
\begin{equation}
  [Q_0,T(W_1,...,W_n)(x_1,...,x_n)]_\mp\label{MBRST}
\end{equation}
if the MWI is satisfied?
Thereby, $W_1,...,W_n$ are arbitrary Wick monomials, $Q_0$ is the
generator of the BRST-transformation of the free fields and
$[\cdot ,\cdot]_\mp$ means the $\ZZ_2$-graded (with
respect to the ghost number) commutator.
There have been other approaches to formulate BRST-symmetry in the
framework of causal perturbation theory. In particular the
'perturbative gauge invariance' of \cite{DHKS}, \cite{DS}
and \cite{S-wiley}, which was further developed
by \cite{stora}, \cite{G} and \cite{DSchroer}, suffices
for a consistent construction of the $S$-matrix in the adiabatic limit,
provided this limit exists. However, this assumption holds certainly
not true in massless non-Abelian gauge theories, it seems that the
confinement is out of the reach of perturbation theory. In massive
non-Abelian gauge theories the instability of physical particles
($W$- and $Z$-bosons, muon and tau etc.) is an obstacle for an
$S$-matrix description with adiabatic limit. Our way out
is to {\it construct the observables locally}
(i.e. with the interaction adiabatically switched off, sect. 4.5),
as we have done it for QED \cite{DF}. For our operator
formalism the BRST-charge operator of Kugo and Ojima \cite{KO}
seems to be the adequate tool to define the BRST-transformation.
But in contrast to this reference we do not
perform the adiabatic limit and, hence, avoid the infrared divergences.
The mentioned perturbative gauge invariance \cite{DHKS}, \cite{DS}
does not suffice for our local construction of observables in
non-Abelian (massless or massive) gauge theories.
But we show that ghost number conservation and the master BRST-identity
contain all information which is needed for this construction.
In particular we will see that the master BRST-identity implies the
perturbative gauge invariance of \cite{DHKS}, \cite{DS} (and even
the generalization proposed in \cite{D1}, which is called
'generalized (free perturbative operator) gauge invariance' in
\cite{DSchroer}).

In spite of all these important implications of the MWI, it is
difficult to give a direct physical interpretation of this identity
(in its full generality) or to formulate the symmetry which is expressed
by it. We give two partial answers:\\
- In classical field theory the MWI can be understood as the most
general identity which can be obtained from the field equations and
the fact that classical fields may be multiplied point-wise and
factorize: $(AB)_{\cal L}(x)=A_{\cal L}(x)B_{\cal L}(x)$ (where
$A,B$ are field polynomials and $A_{\cal L},\> B_{\cal L}$ are
the corresponding fields to the interaction ${\cal L}$), see \cite{DF2}.
But quantum fields may not be multiplied point-wise (because they are
distributions) and, hence, the quantum MWI contains much more
information than only the field equations.\\
- A particular case of the MWI is a formula for
$\d_{x_0}T(A_{\cal L}(x)B_{\cal L}(y))-
T(\d_{x_0}A_{\cal L}(x)B_{\cal L}(y))$, where $A_{\cal L},B_{\cal L}$
are interacting quantum fields to the interaction ${\cal L}$
(i.e. non-local formal power series in free fields)
and $T(...)$ means time ordering in $x$ and $y$. This difference can be
interpreted as a {\it rigorous substitute for
the equal-time commutator of $A_{\cal L}$
and $B_{\cal L}$}:\footnote{We recall the well-known fact
that interacting fields to a sharp time do not exist, i.e.
$\int d^3x\,f(\vec{x})\int dx^0\,\delta (x_0-t)A_{\cal L}(x),\>t\in\RR,
\>f\in {\cal D}(\RR^3)$, is mathematically ill-defined.} defining
heuristically $\tilde T(A_{\cal L}(x)B_{\cal L}(y))\=d\Theta (x^0-y^0)
A_{\cal L}(x)B_{\cal L}(y)+\Theta (y^0-x^0)B_{\cal L}(y)A_{\cal L}(x)$
we in fact obtain
\begin{equation}
\d_{x_0}\tilde{T}(A_{\cal L}(x)B_{\cal L}(y))-
\tilde{T}(\d_{x_0}A_{\cal L}(x)B_{\cal L}(y))
=\delta (x_0-y_0)[A_{\cal L}(x),B_{\cal L}(y)].\label{etc}
\end{equation}
However, $\tilde T(...)$ is problematic: $\Theta (x^0-y^0)A(x)B(y)$
exists if $A$ and $B$ are free fields, but this does not hold
for $A$ and $B$ being Wick polynomials, and for interacting fields
the situation is even worse. In addition $\tilde T(...)$ is non-covariant:
for a Lorentz covariant $T$-product (denoted by $T(...)$ in the
following) and a free scalar field $\phi$
we must have the relation $\d_\mu T(\phi(x)\d_\nu\phi(y))-
T(\d_\mu\phi(x)\d_\nu\phi(y))=Cg_{\mu\nu}\delta (x-y)$ (where $C$
is an undetermined constant), which is obviously not satisfied by
$\tilde T(...)$. But fortunately there is the possibility
that the non-covariant terms (i.e. the terms coming
from $\tilde T(...)-T(...)$) cancel out with other unwanted terms.
This indeed happens for the (interacting) quark currents
$j^\mu_{a{\cal L}}$ in QCD: the (heuristic) equal-time
commutator $[j^0_{a{\cal L}}(t,\vec{x}),j^k_{a{\cal L}}(t,\vec{y})],
k=1,2,3$ has an 'anomalous' term, the Schwinger term:
\begin{equation}
[j^0_{a{\cal L}}(t,\vec{x}),j^k_{a{\cal L}}(t,\vec{y})]
=i f_{abc}j^k_{c{\cal L}}(t,\vec{x})\delta^{(3)}(\vec{x}-\vec{y})+
\sum_{l=1}^3 S_{ab}^{kl}\d^l\delta^{(3)}(\vec{x}-\vec{y}),\label{schwinger}
\end{equation}
where $S_{ab}^{kl}\in\CC$ is constant.
In formula 11-89 of \cite{IZ} it is postulated that the non-covariant terms
of $\tilde T(...)$ are compensated by the Schwinger terms:
\begin{equation}
\d_\mu^xT(j^\mu_{a{\cal L}}(x)j^\nu_{b{\cal L}}(y))-
T(\d_\mu j^\mu_{a{\cal L}}(x)j^\nu_{b{\cal L}}(y))=if_{abc}
j^\nu_{c{\cal L}}(x)\delta^{(4)}(x-y).\label{IZ}
\end{equation}
We will show that this identiy is in fact a consequence of the MWI
(sect. 4.3).

We return to the crucial question whether the MWI
can be satisfied in agreement with all other normalization
conditions. The compatibility with the other  normalization
conditions can be proved generally (sect. 3.2).
If all fields are massive there is a distinguished normalization of the
$T$-products, the so-called central solution \cite{EG}. We prove that
the central solutions fulfil the MWI (and all other normalization
conditions) if the UV-scaling behaviour of its individual terms
is not relatively lowered
(sect. 3.3). This assumption holds mostly true.
However, e.g. for the axial and pseudo-scalar triangle-diagram
(\ref{axan}) it is violated, and this makes possible
the appearance of the axial anomaly.\\

\section{Formulation of the master Ward identity}

\subsection{The symbolical algebra with internal and external derivative}

Let $\{\phi^{(k)}\>|\> k=1,...,M\}$ be the free quantum fields in
terms of which the model is defined. We assume that this set is closed
with respect to taking the adjoint operator. In the larger set
$\Phi\=d\{\d^a\phi^{(k)}\>|\> k=1,...,M,\> a\in\NN_0^4\}$ ($\d^a$
is a partial derivative of arbitrary order) we {\it neglect
the free field equations} and write $\Phi$ as a sequence
$(\phi_l)_{l\in\NN}$. To each $\phi_l$ we associate a
symbol $\varphi_l\equiv {\rm sym}(\phi_l),\> l\in\NN$.
Let ${\cal P}$ be the unital, Abelian
$*$-algebra\footnote{In the case of Fermi fields the symbols
anticommute and we call them 'fermionic symbols'.}
generated by these symbols. Thereby the symbols corresponding to a free
quantum field and to its partial derivatives are linearly independent. The
$*$-operation in ${\cal P}$ corresponds to taking the adjoint of the free
field operators: $\varphi_l^*\equiv {\rm sym}(\phi_l)^*\=d
{\rm sym}(\phi_l^+)$. We define an {\it internal derivative}
$\d^\mu:{\cal P}\rightarrow {\cal P}$ by $\d^\mu\varphi_l
\equiv \d^\mu {\rm sym}(\phi_l)\=d {\rm sym}(\d^\mu\phi_l)$ and
the requirements that
$\d^\mu$ is linear and a derivation. Now we divide ${\cal P}$ by the ideal
$J$ which is generated by the free field equations (with respect to
the internal derivative) and denote the resulting unital, Abelian
$*$-algebra by ${\cal P}_0$,
\begin{equation}
{\cal P}_0\=d\frac{\mathcal{P}}{J}.
\end{equation}
Let $\pi$ be the projection $\pi:\mathcal{P}\rightarrow\mathcal{P}_0:
A\rightarrow A+J$. Internal derivatives in $\mathcal{P}_0$
are defined by\footnote{Note that this definition is independent from
  the choice of the representative $A$.}
$\d^\mu\pi(A)\=d\pi(\d^\mu A)$, and in this sense
the free field equations are valid in ${\cal P}_0$.

In addition we introduce an {\it external derivative}\footnote{This
external derivative has nothing to do with the exterior derivative of
differential geometry.}
$\tilde\d^\mu$ on ${\cal P}_0$ which generates
new symbols $\tilde\d^a A$ ($A\in {\cal P}_0$, $a\in \NN_0^4$,
i.e. $\tilde\d^a$ means a higher external derivative of order
$|a|=a_0+a_1+a_2+a_3$) and is required to be linear and a derivation.
In particular we set $\tilde\d^a {\bf 1}\=d 0,\>\forall a\not= 0$.
The Abelian, unital $*$-algebra (anticommuting in the case of Fermi fields)
which is generated by these new symbols is denoted by $\tilde {\cal P}_0$:
\begin{equation}
\tilde {\cal P}_0\=d \bigvee\{\tilde\d^a A\,|\,A\in {\cal P}_0,
a\in \NN_0^4\}.
\end{equation}
Next we extend the external and internal derivatives and the
$*$-operation to maps $\tilde {\cal P}_0\rightarrow\tilde {\cal P}_0$.
For the former two we set
\begin{equation}
  \tilde\d^b\tilde\d^a A\=d
\tilde\d^{(b+a)}A,\quad\quad
\d^b\tilde\d^aA\=d\tilde\d^a\d^bA,\quad\forall A\in {\cal P}_0,
\end{equation}
and require that $\tilde\d^\mu$ and $\d^\mu$ are linear and
derivations. The $*$-operation is extended by $(\tilde\d^a A)^*\=d
\tilde\d^a (A^*)\quad (A\in {\cal P}_0)$ and by requiring the usual
algebraic relations: anti-linearity, $(BC)^*=C^*B^*$ and $B^{**}=B,\>
\forall B,C\in \tilde{\cal P}_0$.

Finally we introduce the space\footnote{A fermionic $V\in\tilde {\cal P}_0$
is paired (to $V\otimes f$) with a Grassmann-valued test function $f$,
see e.g. appendix D of \cite{S}.}
\begin{equation}
{\cal D}(\RR^4,\tilde {\cal P}_0)\cong \tilde {\cal P}_0
\otimes {\cal D}(\RR^4).\label{space}
\end{equation}
The internal and external derivatives are defined on this space as the
operators $\d^\mu\otimes {\bf 1}$ and $\tilde\d^\mu\otimes {\bf 1}$.\\
\\
{\it Remark}: There exists a surjective algebra $*$-homomorphism
$\tilde\sigma:\tilde\mathcal{P}_0\rightarrow\mathcal{P}$. This becomes
clear from the formalism developed in \cite{DF2}. Namely,
we prove in appendix A of \cite{DF2} that there exists a map
$\sigma:\mathcal{P}_0\rightarrow \mathcal{P}$ (i.e. 'from free fields
to fields') with the properties:\\
(i) $\pi\circ \sigma=\bf{1}$.\\
(ii) $\sigma$ is an algebra $*$-homomorphism, i.e. $\sigma$ is linear,
$\sigma (AB)=\sigma (A)\sigma (B)$ and $\sigma(A^*)=\sigma(A)^*$.\\
(iii) The Lorentz transformation commutes with $\sigma\pi$.\\
(iv) $\sigma\pi (\mathcal{P}_1)\subset
\mathcal{P}_1$, where $\mathcal{P}_1$ is the sub vector space of
$\mathcal{P}$ with basis $(\varphi_l)_l$, i.e. the 'one-factor symbols'.\\
(v) $\sigma$ does not increase the mass dimension of the fields,
i.e. $\sigma\pi (B)$ is a sum of terms with mass dimension
$\leq\mathrm{dim}\>(B)$.
In particular we find $\sigma\pi(\varphi)=\varphi$, if
$\varphi\in\mathcal{P}_1$ corresponds to a free field without
any derivative.\\
(vi) $\bigvee\{\d^a\sigma (A)|
A\in \mathcal{P}_0,a\in\NN_0^d\}=\mathcal{P}$.\\
We now extend $\sigma$ to a map $\tilde\sigma:\tilde\mathcal{P}_0
\rightarrow\mathcal{P}$ by setting
\begin{equation}
  \tilde\sigma(\tilde\d^a A)\=d\d^a\sigma (A),\quad A\in\mathcal{P}_0,
\label{tilde-sigma}
\end{equation}
and requiring that $\tilde\sigma$ is an algebra $*$-homomorphism.
The property (vi) means that $\tilde\sigma$ is surjective. However,
$\tilde\sigma$ is not injective. This follows from the following
simple example: let
$\varphi\in\mathcal{P}_1$ correspond to a free Klein Gordon field
(without any derivative). Usually it
holds $\sigma\pi(\d^\nu\varphi)=\d^\nu\varphi$.
Then, $\tilde\sigma(\tilde\d^\nu\pi\varphi)=
\d^\nu\varphi=\tilde\sigma(\d^\nu\pi\varphi)$.
This example does not appear if one introduces an additional field
$\varphi^\mu$ which replaces $\d^\mu\varphi$ (see sects. 2.3 and
3.2 of \cite{DF2}), but $\tilde\sigma$ is not injective in that case, too.

\subsection{Inductive construction of time ordered products,
basic normalization conditions (N0)-(N3)}

The time-ordered product $T_n$ (also called '$T$-product') is a
{\bf linear}, symmetrical\footnote{To distinguish the symmetry of $T_n$
from other symmetries we sometimes call it 'permutation symmetry'.} map from
${\cal D}(\RR^4,\tilde {\cal P}_0)^{\otimes n}$ into the (unbounded) operators
on the Fock space of the free quantum fields\footnote{In \cite{DF2}
and \cite{DF3}
the arguments of $T_n$ are elements of ${\cal D}(\RR^4,{\cal P})^
{\otimes n}$. The map $\tilde\sigma$ (\ref{tilde-sigma}) connects
the two formalisms.}. In particular linearity implies
$T_n((\d_\nu V^\nu)g\otimes ...)=0$ if $\d_\nu V^\nu$ vanishes due to
the free field equations.
All $T$-products $T_n(f_1\otimes...\otimes f_n),\>
f_j\in {\cal D}(\RR^4,\tilde {\cal P}_0),\>
n\in\NN,$ have the same domain ${\cal D}$ which is a dense subspace of
the Fock space and which is invariant under all $T$-products \cite{EG}.
Physicists use mostly 'unsmeared $T$-products', which are defined by
\begin{equation}
\int dx_1...dx_n\,T_n(V_1,...,V_n)(x_1,...,x_n)g_1(x_1)...g_n(x_n)\=d
T_n(V_1g_1\otimes...\otimes V_ng_n),\label{T(x)}
\end{equation}
where $g_1,...,g_n\in {\cal D}(\RR^4),\>V_1,...,V_n\in\tilde{\cal P}_0$.
More precisely $(V_1,...,V_n)\longrightarrow T_n(V_1,...,V_n)$ is
a linear and symmetrical map from $(\tilde {\cal P}_0)^{\otimes n}$
into the operator-valued distributions.

Let $\tilde {\cal P}_0\ni V=\prod_k \tilde\d^{a^{(k)}}{\rm sym}
(\phi_{j_k})$ (where $\phi_{j_k}\in\Phi,\>\forall k$).
Then we define $T_1$ by
\begin{equation}
T_1(Vg)\=d\int dx\,:\prod_k \d^{a^{(k)}}\phi_{j_k}:(x)g(x),\quad
T_1({\bf 1}g)\=d\int dx\,g(x),
\quad g\in {\cal D}(\RR^4),\label{T_1}
\end{equation}
and by linearity, where the double dots mean normal ordering of the free
field operators. We point out that $T_1$ is not injective, because
$T_1((\tilde\d^a V)Wg)=T_1((\d^a V)Wg),\>V,W\in\tilde{\cal P}_0$. However,
$T_1$ is injective if it is restricted to ${\cal D}(\RR^4,{\cal P}_0)$.

The $T$-products are required to satisfy causal factorization\footnote{This
is the reason for the name 'time ordered product'.}
\begin{gather}
{\bf (Causality)}\quad T_n(f_1\otimes...\otimes f_n)=
T_k(f_1\otimes...\otimes f_k)T_{n-k}(f_{k+1}\otimes...\otimes f_n)
\notag\\
\mathrm{if}\quad \bigl(\supp f_1\cup...\cup\supp f_k\bigr)\cap
\bigl((\supp f_{k+1}\cup...\cup\supp f_n)+\bar V_{-}\bigr)=
\emptyset,\label{caus}
\end{gather}
where $\bar V_{-}$ is the closed backward light cone in Minkowski space.
Causality enables us to construct inductively the $T$-products of
higher orders $n\geq 2$: if the time ordered products of less than
$n$ factors are everywhere defined, the time ordered product of $n$ factors is
uniquely determined up to the total diagonal $D_n\=d\{(x_1,...,x_n)\>|\>
x_1=...=x_n\}$. Thus renormalization
amounts to an extension, for every $n$, of time ordered products to $D_n$.
This extension is always possible, but it is non-unique. It can
be done such that the following {\bf normalization conditions} hold. (Note
that these conditions are automatically fulfilled on ${\cal D}(\RR^{4n}
\setminus D_n)$ due do the inductive procedure and causal factorization.)

$\bullet$ {\bf Poincare covariance}: Let $U$ be a unitary positive energy
representation of the Poincare group ${\cal P}_+^\uparrow$ in Fock space.
$U$ induces an automorphic action $\alpha$ of ${\cal P}_+^\uparrow$ on
${\cal D}(\RR^4,{\cal P}_0)$ by the definition
\begin{equation}
T_1(\alpha_L(f))\=d {\rm Ad}\,U(L)(T_1(f)),\quad\quad\forall
f\in {\cal D}(\RR^4,{\cal P}_0),\>L\in {\cal P}_+^\uparrow,\label{alpha}
\end{equation}
because $T_1$ is injective on this subspace. We extend $\alpha_L$ to
${\cal D}(\RR^4,\tilde{\cal P}_0)$ by the prescription that
$(\tilde\d^m\otimes {\bf 1}) f$ transforms in the same way as
$(\d^m\otimes {\bf 1}) f,\quad m\in\NN_0$ (where $\d^m$ ($\tilde\d^m$ resp.)
denotes the $m$-th power of the gradient $\d$ ($\tilde\d$ resp.)).
More precisely let $f=\sum_i
V_i\otimes g_i,\>V_i\in {\cal P}_0,\>g_i\in {\cal D}(\RR^4)$. From
(\ref{alpha}) we know the transformation of $(\d^m\otimes {\bf 1}) f$,
which can be written in the form
\begin{equation}
\alpha_{(\Lambda ,a)}(\d^m\otimes {\bf 1}) f=
\sum_{i,j}(\d^m V)_i\otimes D(\Lambda)_{ji}
g_{(\Lambda ,a)\,j},\quad L\equiv (\Lambda ,a)\in {\cal P}_+^\uparrow,
\quad m\in\NN_0,
\label{alpha1}
\end{equation}
where $g_{(\Lambda ,a)}(x)\=d g(\Lambda^{-1}(x-a))$. Then we define
\begin{equation}
\alpha_{(\Lambda ,a)} (\tilde\d^m\otimes {\bf 1})f\=d\sum_{i,j}
(\tilde\d^m V)_i\otimes D(\Lambda)_{ji}g_{(\Lambda ,a)\,j},
\quad (\Lambda ,a)\in {\cal P}_+^\uparrow,\quad m\in\NN_0.\label{alpha2}
\end{equation}
One easily verifies $\alpha_{L_1L_2}=\alpha_{L_1}\alpha_{L_2}$ and
that equation (\ref{alpha}) holds true for the extended $\alpha_L$, i.e.
for all $f\in {\cal D}(\RR^4,\tilde{\cal P}_0)$. The normalization condition
expressing the Poincare covariance of the time ordered products reads
\begin{equation}
{\bf (N1)}\quad {\rm Ad}\,U(L)(T(f_1\otimes...\otimes f_n))=
T(\alpha_L(f_1)\otimes...\otimes \alpha_L(f_n)),
\quad L\in {\cal P}_+^\uparrow.\nonumber
\end{equation}
For pure massive theories the so-called 'central solution/extension'
(see \cite{EG} and sect. 3.3) is Poincare covariant. For theories
with massless fields the existence of a Poincare covariant extension
has been proved (in \cite{St2} and in the second paper of \cite{DHKS})
by tracing it back to a cohomological problem;
an explicit solution has been given in \cite{BPP}.

$\bullet$ {\bf Unitarity}: To explain what we mean by 'unitarity' we introduce
the $S$-matrix (as a formal power series) which is the generating functional
of the $T$-products
\begin{equation}
S(f)={\bf 1}+\sum_{n=1}^\infty\frac{i^n}{n!}
T_n\bigl( f\otimes ...\otimes f\bigr),\quad
f\in {\cal D}(\RR^4,\tilde {\cal P}_0).\label{S}
\end{equation}
Since the zeroth order term does not vanish, it
has a unique inverse in the sense of formal power series
\begin{equation}
S(f)^{-1}={\bf 1}+\sum_{n=1}^\infty\frac{(-i)^n}{n!}
\bar T_n\bigl( f\otimes ...\otimes f\bigr),\label{S:inverse}
\end{equation}
where the 'anti-chronological products' $\bar T(...)$ can be
expressed in terms of the time ordered products
\begin{equation}
\bar T_n\bigl( f_1\otimes ...\otimes f_n\bigr)\=d
\sum_{P\in {\cal P}(\{1,...,n\})}(-1)^{|P|+n}\prod_{p\in P}
T_{|p|}\bigl( \otimes_{j\in p}f_j)\ .\label{T:bar}
\end{equation}
(Here ${\cal P}(\{1,...,n\})$ is the set of all ordered partitions of
$\{1,...,n\}$, $|P|$ is the number of subsets in $P$
and $|p|$ is the number of elements of $p$). The
reason for the word 'anti-chronological' is that the
$\bar T$-products satisfy anti-causal factorization, which means
(\ref{caus}) with reversed order of the factors on the r.h.s..
Unitarity of the $S$-matrix is expressed by
\begin{equation}
S(f)^+=S(f^*)^{-1}
\end{equation}
($^+$ means the adjoint on ${\cal D}$, $\>(\phi,B\psi)=(B^+
\phi,\psi),\quad\phi,\psi\in {\cal D}$.)
Hence, for the $T$-products we require the normalization condition
\begin{equation}
{\bf (N2)}\quad\quad\quad T_n\bigl( f_1\otimes ...\otimes f_n\bigr)^+=
\bar T_n\bigl( f_1^*\otimes ...\otimes f_n^*\bigr),\nonumber
\end{equation}
which can easily be satisfied by symmetrizing an arbitrary normalized
$T$-product (see \cite{EG}).

$\bullet$ {\bf Relation to $T$-products of sub-polynomials}:
Let ${\cal G}\subset\mathcal{P}_0$ be a linearly independent set of
generators of $\mathcal{P}_0$,
i.e. ${\cal G}$ is a (vector space) basis of $\pi\mathcal{P}_1$
(see the Remark in sect. 2.1 for the definition of $\mathcal{P}_1$).
Then $\tilde{\cal G}\=d\{\tilde\d^a\varphi\,|\,\varphi\in {\cal G}, a\in
\NN_0^4\}$ is a set of linearly independent generators of $\tilde{\cal P}_0$.
We define the commutator 'function' $\Delta_{\varphi,\chi}$ by
\begin{equation}
i\int dx\,dy\,h(x)g(y)\Delta_{\varphi,\chi}(x-y)
\=d [T_1(\varphi h),T_1(\chi g)],\quad \varphi,\chi\in
\tilde{\cal G}.\label{def:Delta}
\end{equation}
Every $V\in\mathcal{P}_0$ can uniquely be written as a polynomial in
the generators ${\cal G}$. By partial differentiation in this sense we
obtain a 'sub-polynomial' $\frac{\d V}{\d\varphi},\>\forall\varphi\in
{\cal G}$.\footnote{If $\varphi\in {\cal G}$ is a fermionic symbol,
  the derivative $\frac{\d}{\d\varphi}$ is a {\it graded} derivation.}
For $f(x)=\sum_iV_if_i(x),\>f_i\in {\cal D}(\RR^4),\>V_i\in
{\cal P}_0$, we set
\begin{equation}
\frac{\d f}{\d\varphi}\=d\sum_i\frac{\d V_i}{\d\varphi}f_i(x).
\label{der2}
\end{equation}
For $\psi\in \tilde {\cal G}$ we analogously define $\frac{\d}{\d\psi}$
to be a linear derivation $\tilde{\cal P}_0\rightarrow \tilde{\cal P}_0$
(${\cal D}(\RR^4,\tilde {\cal P}_0)\rightarrow {\cal D}(\RR^4,
\tilde {\cal P}_0)$ resp.) with
\begin{equation}
\frac{\d(\tilde\d^a \chi)}{\d(\tilde\d^b\varphi)}\=d \delta_{a,b}
\frac{\d \chi}{\d \varphi}=\delta_{a,b}\delta_{\chi,\varphi}
,\quad \chi,\varphi\in {\cal G}.\label{der3}
\end{equation}
Generally we call $W\in\tilde {\cal P}_0$ a subpolynomial of
$V\in\tilde {\cal P}_0$ iff it is of the form
$W=\frac{\d^k V}{\d\varphi_{i_1}...\d\varphi_{i_k}}$
for some $k\in\NN_0,\>\varphi_{i_1},...,\varphi_{i_k}\in \tilde {\cal G}$.
The derivation property of the commutator $[\cdot ,T_1(\chi g)]$
implies
\begin{equation}
[T_1(f),T_1(\chi g)]=
i\sum_{\psi\in \tilde{\cal G}}T_1\Bigl(\frac{\d f}{\d\psi}
\Delta_{\psi,\chi}\star g\Bigr),\quad\forall f\in {\cal D}(\RR^4,
\tilde {\cal P}_0),\>\chi\in\tilde {\cal G},\>g\in {\cal D}(\RR^4),
\label{der4}
\end{equation}
where $\star$ means convolution.

We now generalize the normalization condition {\bf (N3)} of
\cite{DF} to the present framework: we require
\begin{eqnarray}
({\bf N3})\quad &&[T_n(f_1\otimes ...\otimes f_n),T_1(\chi g)]=
\nonumber\\
&&i\sum_{l=1}^n\sum_{\psi\in\tilde{\cal G}}
T_n(f_1\otimes ...\otimes\frac{\d f_l}
{\d\psi}\Delta_{\psi,\chi}\star g\otimes ...\otimes f_n)
\label{N3}
\end{eqnarray}
where $f_1,...,f_n\in {\cal D}(\RR^4,\tilde {\cal P}_0),\chi\in
\tilde{\cal G}$.
The r.h.s. is well-defined because $\Delta_{\psi,\chi}\star g$
is a smooth function.

We point out that the defining properties of the $T$-products given
so far (linearity, symmetry, causality, {\bf (N1)}, {\bf (N2)}
and {\bf (N3)}) are purely algebraic conditions, they are independent
from the choice of a state. In the realization of the
$T$-products as operators in Fock space, {\bf (N3)} becomes equivalent
to the translation of the 'causal Wick expansion' of Epstein and
Glaser\footnote{Epstein/Glaser do not use this name, but it appears
  e.g. in \cite{BF}.} into our formalism, see \cite{EG} sect. 4.

$\bullet$ {\bf Scaling degree}: {\bf (N3)} gives the relation
to time ordered products of sub-polynomials.
Once these are known (in an inductive procedure), only the C-number part
of the $T$-product (which is equal to the Fock vacuum expectation value
of the $T$-product) has to be fixed. Due to translation invariance
this scalar distribution depends on the
relative coordinates only. Hence, the extension of the
(operator valued) $T$-product to $D_n$ is reduced to the extension of a
C-number distribution $t_0\in {\cal D}'(\RR^{4(n-1)}\setminus \{0\})$ to
$t\in {\cal D}'(\RR^{4(n-1)})$. (We call $t$ an extension of $t_0$ if
$t(f)=t_0(f),\>\forall f\in {\cal D}(\RR^{4(n-1)}\setminus \{0\})$).
The singularity of $t_0(y)$ and $t(y)$
at $y=0$ is classified in terms of Steinmann's scaling degree \cite{SD,BF}
\begin{equation}
{\rm sd}(t)\=d {\rm inf}\{\delta\in \RR\>,\>\lim_{\lambda\downarrow 0}
\lambda^\delta t(\lambda x)=0\}.\label{4.3a}
\end{equation}
Note
\begin{equation}
  {\rm sd}(\d^at)\leq {\rm sd}(t)+|a|\quad\quad\mathrm{and}
\quad\quad {\rm sd}(\d^a\delta^{(m)})=m+|a|,\label{sd}
\end{equation}
where $\delta^{(m)}$ denotes the $m$-dimensional $\delta$-distribution.
By definition ${\rm sd}(t_0)\leq {\rm sd}(t)$, and the
possible extensions are restricted by requiring
\begin{equation}
({\bf N0})\quad\quad\quad\quad\quad\quad
{\rm sd}(t_0)={\rm sd}(t).\label{4.3b}
\end{equation}
Then the extension is unique for ${\rm sd}(t_0)<4(n-1)$, and in the
general case
there remains the freedom to add derivatives of the $\delta$-distribution
up to order $({\rm sd}(t_0)-4(n-1))$. In formula:
\begin{equation}
t(y)+\sum_{|a|\leq {\rm sd}(t_0)-4(n-1)}C_a\d^a\delta (y)\label{4.3c}
\end{equation}
is the general solution, where $t$ is a special extension \cite{BF,P,EG},
and the constants $C_a$ are restricted by {\bf (N1)}, {\bf (N2)},
permutation symmetries and the normalization conditions
$\tilde {\rm\bf (N)}$ (normalization of time ordered products of symbols
with external derivative) and {\bf (N)} (MWI) below. For an
interaction $\mathcal{L}$ with UV-dimension
${\rm dim}({\cal L})\leq 4$ the requirement (\ref{4.3b}) implies
renormalizability by power counting, i.e. the number of indeterminate
constants $C_a$ in $(T_n((g\mathcal{L})^{\otimes n}))_n\>(g\in
\mathcal{D}(\RR^4))$ does not increase by going over to higher
perturbative orders $n$.

In the seminal paper \cite{EG} Epstein and Glaser prove that there exists
an extension to $D_n$ which fulfills the normalization
conditions\footnote{Here we neglect that Epstein and Glaser work with
  a different formalism, in particular they do not have the external
  derivative $\tilde\d$.}
{\bf (N0)-(N3)}, but they say only few about further symmetries which
should be maintained in the extension, e.g. the field equations or
gauge invariance. The MWI is a universal normalization
condition which summarizes the request for most of this 'further symmetries'.

\subsection{Normalization of time-ordered products of symbols
with external derivative}

The aim of this subsection is to fix the normalization of time-ordered
products of symbols with external derivative(s) in terms of
time-ordered products without external derivative. This fixation is a
necessary ingredient of the formulation
of the MWI, because $T$-products of symbols with
external derivatives unavoidably appear in the MWI. Heuristically the
external derivative is a derivative which acts {\it after} having done the
time-ordered contractions of the corresponding symbols (free fields resp.),
e.g.
\begin{gather}
T_{n+1}((\tilde\partial^\nu V)g\otimes W_1f_1\otimes...\otimes
W_nf_n)=\notag\\
\int dx\,dx_1...dx_n\,g(x)f_1(x_1)...f_n(x_n)\d^\nu_x
T(V,W_1,...,W_n)(x,x_1,...,x_n)\equiv\notag\\
-T_{n+1}(V\d^\nu g\otimes W_1f_1\otimes...\otimes W_nf_n),
\label{heur:tilde-d}
\end{gather}
where $V,W_1,...,W_n\in \tilde {\cal P}_0$. However, there are other
time ordered products involving factors with external derivatives
such as $(\tilde\partial^\nu V)W$ which cannot be defined in this way
in terms of time ordered products of factors without
any external derivative. Hence we proceed in an alternative,
recursive way: we give an explicit expression for the difference
\begin{equation}
T_{n+1}((\tilde\partial^\nu V)Wg\otimes W_1f_1\otimes...\otimes W_nf_n)
-T_{n+1}((\partial^\nu V)Wg\otimes W_1f_1\otimes...\otimes W_nf_n)
\label{diff}
\end{equation}
where $V,W,W_1,...,W_n\in\tilde {\cal P}_0$.
For this purpose we introduce some notations: by means of the Feynman
propagator $\Delta^F_{\chi,\psi}$
\begin{equation}
\int dx\,dy\,f(x)g(y)
i\Delta^F_{\chi,\psi}(x-y)\=d \langle\Omega,
T_2(\chi f\otimes\psi g)\Omega \rangle
,\quad\chi,\psi\in \tilde {\cal G},\>f,g\in {\cal D}(\RR^4),
\label{Feyprop}
\end{equation}
(where $\Omega$ denotes the Fock vacuum) we define
\begin{equation}
\delta^\mu_{\chi,\psi}\=d \partial^\mu\Delta^F_{\chi,\psi}
-\Delta^F_{\partial^\mu\chi,\psi},\quad\chi,\psi\in \tilde {\cal G}.
\label{delta2}
\end{equation}
Inserting the causal factorization of $T_2(...)(x,y)$ for $x\not=y$
we find that $\delta^\mu_{...}$ is a local distribution, hence it has
the form
\begin{equation}
\delta^\mu_{\chi,\psi}(z)=\sum_{a\in\NN_0^4}C^\mu_{\chi,\psi ;a}
\partial^a\delta (z),
\end{equation}
where the $C^\mu_{\chi,\psi ;a}\in\CC$ are constant numbers.
Then we define
\begin{eqnarray}
\Delta^\mu_{\chi,\psi}&:&
{\cal D}(\RR^4,\tilde {\cal P}_0)^{\times 2}
\longrightarrow {\cal D}(\RR^4,\tilde {\cal P}_0)\nonumber\\
\Delta^\mu_{\chi,\psi}(Vg,Wf)
&=&\sum_aC^\mu_{\chi,\psi;a}
(-1)^{|a|}\sum_{0\leq b\leq a}\nonumber\\
&&\frac{a!}{b!(a-b)!}
(\tilde\partial^b V)W(\partial^{(a-b)}g)f\label{Delta2}
\end{eqnarray}
where $|a|=a_0+a_1+a_2+a_3$ and $a!\equiv\prod_\mu a_\mu !$.
This formula is motivated by the identity
\begin{gather}
\int dx\,\d^a\delta (x-y)V(x)g(x)W(y)f(y)=\notag\\
\sum_{0\leq b\leq a}(-1)^{|a|}\frac{a!}{b!(a-b)!}
(\partial^b V)(y)W(y)(\partial^{(a-b)}g)(y)f(y),\label{ddelta}
\end{gather}
where $V(x)$ and $W(y)$ are here Wick polynomials (cf. (\ref{T_1})). The
subtle point in the definition (\ref{Delta2}) is that the derivative on $V$
on the r.h.s. is an external one. This results from the derivation of
the MWI in classical field theory \cite{DF2}. And, if the derivative
on $V$ would be an internal one, we would get wrong results, e.g. for
the BRST-transformation of the interacting gauge field in non-Abelian
models (\ref{BRST:A}).\footnote{On the heuristic level of the
Feyman rules this can be understood as follows (for simplicity we
assume $|a|=1$): one
shifts the derivative $\d$ from the difference (\ref{delta2})
of Feynman propagators $\delta^\mu\sim\d\delta (x-y)$
to $V(x)$, however the (time-ordered) contractions of the
legs of $V$ are already performed, i.e. $\d V$ must be an external
derivative. Thereby the term $VW(\d g)f$ is the boundary term.}
Note that $\Delta^\mu_{\chi,\psi}$ is not invariant with respect to
the exchange of its arguments.

We are now going to compute the difference (\ref{diff}) on a
{\it heuristic level} according to our prescription that the external
derivative acts after contracting. Let\footnote{In this calculation the
  indices $k$ of $\varphi_k$ and $\phi_k$ have nothing to do with the
  ones introduced in sect. 2.1.}
$V=\prod_{k=1}^m\varphi_k,\>W=\prod_{k=m+1}^p\varphi_k,\>
\varphi_k\in \tilde{\cal G},\>\varphi_k
=\tilde\d^{a_k}\mathrm{sym}(\phi_k)$.
We consider the sum of diagrams in
which $\phi_1,...,\phi_l\>(l\leq m)$ and $\phi_{m+1},...,\phi_q
\>(q\leq p)$ are contracted and $\phi_{l+1},...,\phi_m,
\phi_{q+1},...,\phi_p$ are not. By means of the Feynman rules
and the normalization condition\footnote{The following implication of
  {\bf (N3)} is used here: the Feynman propagators
$\Delta^F_{\varphi_j,\chi}$ which appear in (\ref{Fey})
depend on $(\varphi_j,\chi)$ only. This means that for
$(\varphi_j,\chi)=(\varphi_l,\psi)$ the undetermined parameters
(\ref{D^F}) in $\Delta^F_{\varphi_j,\chi}$ and $\Delta^F_{\varphi_l,
\psi}$ have the same values. Note additionally
$\Delta^F_{\varphi,\chi}=\pm\Delta^F_{\chi,\varphi}$ due to
(\ref{Feyprop}).} {\bf (N3)} we compute
the contribution of this sum of diagrams to the first $T$-product
in (\ref{diff}):
\begin{gather}
  T_{n+1}((\tilde\partial^\nu V)Wg\otimes W_1f_1\otimes...)=
\sum_{r_1,...,r_{l+q}}i^{l+q}\notag\\
\Bigl[\d^\nu_x\bigl(\Delta^F_{\varphi_1,\cdot}(x-x_{r_1})...
\Delta^F_{\varphi_l,\cdot}(x-x_{r_l})\bigr)
\Delta^F_{\varphi_{m+1},\cdot}(x-x_{r_{l+1}})...
\Delta^F_{\varphi_q,\cdot}(x-x_{r_{l+q-m}})\cdot\notag\\
:T(...)(x_1,...,x_n)\prod_{k=l+1}^m\d^{a_k}\phi_k(x)\cdot
\prod_{k=q+1}^p\d^{a_k}\phi_k(x):+\notag\\
\Delta^F_{\varphi_1,\cdot}(x-x_{r_1})...
\Delta^F_{\varphi_l,\cdot}(x-x_{r_l})
\Delta^F_{\varphi_{m+1},\cdot}(x-x_{r_{l+1}})...
\Delta^F_{\varphi_q,\cdot}(x-x_{r_{l+q-m}})\cdot\notag\\
:T(...)(x_1,...,x_n)\d^\nu_x\bigl(\d^{a_k}\prod_{k=l+1}^m\phi_k(x)\bigr)
\cdot\prod_{k=q+1}^p\d^{a_k}\phi_k(x):\Bigr]+... ,\label{Fey}
\end{gather}
where the double dots simply mean that the
$\phi_k(x),\>k=l+1,...,m,q+1,...,p$ are not contracted.
(Note that normal ordering is defined for monomials only, not for
polynomials.) With (\ref{Fey}) we obtain the following heuristic
result for the difference (\ref{diff})
\begin{gather}
  \sum_{r_1,...,r_{l+q}}i^{l+q}\sum_{t=1}^l
\Delta^F_{\varphi_1,\cdot}(x-x_{r_1})...\delta^{\nu}_
{\varphi_t,\cdot}(x-x_{r_t})...
\Delta^F_{\varphi_l,\cdot}(x-x_{r_l})\Delta^F_{\varphi_{m+1},\cdot}
(x-x_{r_{l+1}})...\notag\\
...\Delta^F_{\varphi_q,\cdot}(x-x_{r_{l+q-m}})
:T(...)(x_1,...,x_n)\prod_{k=l+1}^m\d^{a_k}\phi_k(x)\cdot
\prod_{k=q+1}^p\d^{a_k}\phi_k(x):+...\>.\label{Feyrules}
\end{gather}
We now require that this structure
is maintained in the process of renormalization:
\begin{eqnarray}
{\bf (\tilde N)}&&T_{n+1}((\tilde\partial^\nu V)Wg\otimes W_1f_1
\otimes...\otimes W_nf_n)=\nonumber\\
&&T_{n+1}((\partial^\nu V)Wg\otimes W_1f_1\otimes...\otimes W_nf_n)\nonumber\\
&&+i\sum_{m=1}^n\sum_{\chi,\psi\in\tilde {\cal G}}(\pm) T_n\Bigl(
\Delta^{\nu}_{\chi,\psi}\bigl(\frac{\partial V}
{\partial\chi}Wg,\frac{\partial W_m}{\partial\psi}f_m\bigr)\nonumber\\
&&\otimes W_1f_1\otimes...\hat m...\otimes W_nf_n\Bigr)\nonumber
\end{eqnarray}
where $V,W,W_1,...,W_n\in\tilde {\cal P}_0$,
the sign $(\pm)$ comes from permutations of Fermi operators and
$\hat m$ means that the corresponding factor is omitted. We now assume
that $(\tilde {\rm\bf N})$ holds true to lower orders $\leq n$.
Then, due to causal
factorization of time ordered products, we conclude that the condition
$(\tilde {\rm\bf N})$ is satisfied for
$\supp (g\otimes f_1\otimes ...\otimes f_n)\cap D_{n+1}=\emptyset$.
Hence $(\tilde {\rm\bf N})$ is in fact a normalization
condition. It can be satisfied by taking $(\tilde {\rm\bf N})$ as the
{\it definition} of the normalization of $T_{n+1}((\tilde\partial^\nu V)
Wg\otimes W_1f_1\otimes...\otimes W_nf_n)$. There is only one
non-trivial step in this procedure: the compatibility with {\bf (N3)}.
This is shown in sect. 3.1.

In models with anomalies, i.e. terms which violate the MWI
(see the next subsection),
the normalization condition $(\tilde {\rm\bf N})$ will be modified:
in order that (\ref{heur:tilde-d}) holds true
these anomalies must be taken into account in the difference (\ref{diff}),
they give an additional contribution to the r.h.s. of $(\tilde {\rm\bf N})$
(cf. sect. 5).

In particular the normalization condition $(\tilde {\rm\bf N})$ implies
\begin{equation}
\Delta^F_{\tilde\partial^a\varphi_j,\tilde\partial^b\varphi_l}=
(-1)^{|b|}\partial^a\partial^b \Delta^F_{\varphi_j,\varphi_l},\quad
\varphi_j,\varphi_l\in {\cal G},
\end{equation}
and hence
\begin{equation}
\delta^\mu_{\tilde\partial^a\varphi_j,\tilde\partial^b\varphi_l}=
(-1)^{|b|}\partial^a\partial^b \delta^\mu_{\varphi_j,\varphi_l},\quad
\varphi_j,\varphi_l\in {\cal G}.\label{delta2:tilde}
\end{equation}
By repeated application of $(\tilde {\rm\bf N})$ and $\langle\Omega ,T_1(Uh)
\Omega\rangle=0$ for $\tilde {\cal P}_0\ni U\not=
\lambda {\bf 1},\>\lambda\in\CC$,
one finds
\begin{gather}
\langle\Omega ,T_2\Bigl((\prod_{k=1}^r\tilde\partial^{a^{(k)}}\varphi_{j_k})g
\otimes (\prod_{k=1}^r\tilde\partial^{b^{(k)}}\varphi_{l_k})f\Bigr)
\Omega\rangle=\notag\\
\langle\Omega ,T_2\Bigl((\prod_{k=1}^r\partial^{a^{(k)}}\varphi_{j_k})g
\otimes (\prod_{k=1}^r\partial^{b^{(k)}}\varphi_{l_k})f\Bigr)\Omega\rangle
\end{gather}
for $r>1$, where $\varphi_m\in {\cal G}\quad\forall m,\>a^{(k)}
\equiv(a^{(k)}_0,a^{(k)}_1,a^{(k)}_2,a^{(k)}_3)$ and similar for $b^{(k)}$.

\subsection{The master Ward identity}

The MWI is an explicit formula for the difference
\begin{equation}
\d^\nu_x T(V,W_1,...,W_n)(x,x_1,...,x_n)- T(\d^\nu V,W_1,...,W_n)
(x,x_1,...,x_n)\label{dT-Td},
\end{equation}
where $V,W_1,...,W_n\in {\cal P}_0$. It may be regarded as the
postulate that the recursive definition $(\tilde {\rm\bf N})$
reproduces, in the case $W=1$ and $V,W_1,...,W_n\in {\cal P}_0$,
the direct definition (\ref{heur:tilde-d}) (see the Remark below).
However, this is a very technical and indirect way to the MWI.
We found it by the following, intuitive procedure:
the result of the Feynman rules for the difference
(\ref{dT-Td}) is obtained from (\ref{Feyrules}) by
choosing $\varphi_k\in\mathcal{G},\>\forall k$, and putting
$W=1$ (i.e. $p=q=m$). The MWI requires
that renormalization is done in such a way that this
heuristic result is essentially preserved:
\begin{eqnarray}
{\bf (N)}&&-T_{n+1}(V\partial^\nu g\otimes W_1f_1\otimes...\otimes W_nf_n)=
\nonumber\\
&&T_{n+1}((\partial^\nu V)g\otimes W_1f_1\otimes...\otimes W_nf_n)\nonumber\\
&&+i\sum_{m=1}^n\sum_{\chi,\psi\in {\cal G}}(\pm) T_n\Bigl(
\Delta^{\nu}_{\chi,\psi}
\bigl(\frac{\partial V}{\partial\chi}g,\frac{\partial W_m}{\partial
\psi}f_m\bigr)\nonumber\\
&&\otimes W_1f_1\otimes...\hat m...\otimes W_nf_n\Bigr)
\end{eqnarray}
where $V,W_1,...,W_n\in{\cal P}_0$ (not $\in\tilde {\cal P}_0$),
the sign $(\pm)$ is due to permutations of Fermi operators and
$\hat m$ means that the corresponding factor is omitted. We recall
that $\Delta^\mu$ contains external derivatives. To give the correct
formula for the difference (\ref{dT-Td}) one needs the external
derivative or an equivalent formalism (for a latter see \cite{DF2}
and the Remark at the end of sect. 2.1). Similarly to $(\tilde {\rm\bf
  N})$ the MWI {\bf (N)} presupposes the normalization condition
{\bf (N3)}, because {\bf (N3)} is used in (\ref{Fey})-(\ref{Feyrules}).

{\it Remark:}
Instead of requiring $(\tilde {\rm\bf N})$ and ({\bf N}), one can take
(\ref{heur:tilde-d}) and $(\tilde {\rm\bf N})$ as the primary
normalization conditions, because the latter two imply ({\bf N}).
This alternative and more compact formulation is the straightforward
way to formulate the quantum MWI \cite{DF2}, when departing from
classical field theory. However, the advantage of the present
procedure is that it explicitly distinguishes the 'weak'
normalization condition $(\tilde {\rm\bf N})$
(which only {\it defines} the normalization
of the time ordered products with external derivatives)
from the 'hard' one ({\bf N}) (which expresses deep symmetries).
This distinction plays an important role in our (incomplete)
proof of the MWI (sect. 3).

We now assume that {\bf (N)} holds true
to lower orders $\leq n$. Then, due to causal
factorization of time ordered products, we conclude that the condition
{\bf (N)} is satisfied for $\supp (g\otimes f_1\otimes ...\otimes f_n)
\cap D_{n+1}=\emptyset$. Hence {\bf (N)} is in fact a normalization
condition. The compatibility with {\bf (N0)-(N2)} is trivial and the
compatibility with {\bf (N3)} is proved in sect. 3.2.
The hard question is whether
{\bf (N)} can be satisfied by choosing suitable normalizations
(which are compatible with the other normalization conditions).
The answer depends on the model. We will see that the MWI implies that
there is no axial anomaly and no trace anomaly of the energy momentum
tensor. Hence it must be impossible to fulfil the MWI in these cases.
Generally we call any term that
violates the MWI (and cannot be removed by
an admissible, finite renormalization of the $T$-products) an {\it anomaly}.

If there is at most one contraction between $V$ and $W_1,...,W_n$
(i.e. we have $l=0$ or $l=1$
and of course $p=q=m$ in (\ref{Feyrules})) the expression
(\ref{Feyrules}) is well-defined and (re)normalization can be done
such that (\ref{Feyrules}) gives the contribution
of these diagrams to the difference (\ref{dT-Td}). In other words
one can fulfil the MWI {\bf (N)} for these
'tree-like' diagrams. The anomalies must come from 'loop-like' diagrams.
In sect. 5 we give a more general
formulation of the MWI which takes anomalies into account.

\section{Steps towards a proof of the master Ward identity}

We have to show that there exists a normalization of the $T$-products
which satisfies $(\tilde {\rm\bf N})$, ({\bf N}) and also {\bf (N0)-(N3)}).
The compatibility  of $(\tilde {\rm\bf N})$ and ({\bf N}) with {\bf (N0)-(N2)}
is obvious, but the compatibility with ({\bf N3}) requires some work
which is done in the next two subsections. The proof of
$(\tilde {\rm\bf N})$ is then easily completed (sect. 3.1).

But a general proof of ({\bf N}) is impossible, since it is
well-known that there exist anomalies in certain models. If solely massive
fields appear and if the scaling degrees (\ref{4.3b}) of the
individual C-number distributions appearing in {\bf (N)} are
not relatively lowered\footnote{We explain what we mean by this
  expression for the example of the identity $\d_\nu t_1^\nu=t_2
\quad (t_1, t_2\in {\cal D}'(\RR^k))$. According to (\ref{sd})
we naively expect ${\rm sd}(t_1)+1={\rm sd}(t_2)$. We say that the
scaling degree of $t_1$ (or $t_2$ resp.) is relatively lowered if
${\rm sd}(t_1)<{\rm sd}(t_2)-1$ (or ${\rm sd}(t_2)<{\rm sd}(t_1)+1$ resp.).

A relative pre-factor $m^a$ ($m=$ mass), $a>0$, indicates a relatively
lowered scaling degree. Let $\d_\nu t_1^\nu=m^a t_2$ and we
assume that $t_1$ and $t_2$
contain no global factor $m^b$ ($b\in\RR\setminus\{ 0\}$). Then,
for dimensional reasons, the scaling degree of $t_2$ is
relatively lowered: ${\rm sd}(t_2)={\rm sd}(t_1)+1-a$.},
we can give a constructive proof of {\bf (N)} (sect. 3.3).
More precisely we show that the so-called 'central solution' of Epstein
and Glaser, which is a distinguished extension $t^{(c)}\in {\cal D}'(\RR^k)$
of $t_0\in {\cal D}'(\RR^k\setminus\{ 0\})$, satisfies {\bf (N)} in this case.

To simplify the formulas we restrict this section to bosonic fields,
the inclusion of fermionic fields is obvious.

\subsection{Proof of $(\tilde {\rm\bf N})$}

The nontrivial part in the proof of $(\tilde {\rm\bf N})$
is the compatibility with {\bf (N3)}. The keys to show this
and the compatibility of {\bf (N)} with {\bf (N3)} (see sect. 3.2)
are the following two Lemmas:\\
{\bf Lemma 1:} Let $V\in \tilde {\cal P}_0,\>\varphi,\psi\in\tilde {\cal G}$
and $f,h\in {\cal D}(\RR^4)$. Then the following identities hold true
within ${\cal D}(\RR^4,\tilde {\cal P}_0)$:
\begin{eqnarray}
f\sum_{\varphi\in\tilde {\cal G}}\frac{\d(\d^a V)}{\d\varphi}
\Delta_{\varphi,\psi}\star h=f\sum_{\varphi\in\tilde {\cal G}}
\sum_{0\leq b\leq a}\frac{a !}{b !(a-b) !}\d^b
\frac{\d V}{\d\varphi}\d^{(a-b)}\Delta_{\varphi,\psi}\star h\label{identity1}\\
f\sum_{\varphi\in\tilde {\cal G}}\frac{\d(\tilde\d^a V)}{\d\varphi}
\Delta_{\varphi,\psi}\star h=f\sum_{\varphi\in\tilde {\cal G}}
\sum_{0\leq b\leq a}\frac{a !}{b !(a-b) !}\tilde\d^b
\frac{\d V}{\d\varphi}\d^{(a-b)}\Delta_{\varphi,\psi}\star h,\label{identity2}
\end{eqnarray}
where again $a!\equiv\prod_\mu a_\mu !$.\\
{\bf Proof:} We first prove (\ref{identity1}) for $|a|=1$, i.e.
\begin{equation}
f\sum_{\varphi\in\tilde {\cal G}}\frac{\d(\d^\mu V)}{\d\varphi}
\Delta_{\varphi,\psi}\star h=f\sum_{\varphi\in\tilde {\cal G}}
[\d^\mu\frac{\d V}{\d\varphi}\Delta_{\varphi,\psi}\star h+
\frac{\d V}{\d\varphi}\d^\mu \Delta_{\varphi,\psi}\star h].
\label{identity1.1}
\end{equation}
It suffices to consider the case in which $V$ is a monomial. The proof goes
by induction on the degree of this monomial. The case $V=1$ is trivial.
Let $V=\chi W,\>\chi\in \tilde {\cal G},\>W\in \tilde {\cal P}_0$.
By assumption $W$ satisfies (\ref{identity1.1}). Inserting now $V=\chi W$
into (\ref{identity1.1}) and using this assumption
most terms cancel and it remains to show
\begin{equation}
f\sum_{\varphi\in\tilde {\cal G}}\frac{\d(\d^\mu\chi)}{\d\varphi}W
\Delta_{\varphi,\psi}\star h=f\sum_{\varphi\in\tilde {\cal G}}
\frac{\d\chi}{\d\varphi}W\d^\mu \Delta_{\varphi,\psi}\star h.\label{xy}
\end{equation}
The l.h.s. is equal to $fW\Delta_{\d^\mu\chi,\psi}\star h$ and the r.h.s.
to $fW\d^\mu \Delta_{\chi,\psi}\star h$. Obviously these two expressions
agree.

To prove (\ref{identity1}) for arbitrary $|a|$ we proceed by induction on
$|a|$:
\begin{gather}
f\sum_{\varphi\in\tilde {\cal G}}\frac{\d(\d^\mu\d^a V)}{\d\varphi}
\Delta_{\varphi,\psi}\star h=\notag\\
f\sum_{\varphi\in\tilde {\cal G}}[(\d^\mu
\frac{\d(\d^a V)}{\d\varphi})\Delta_{\varphi,\psi}\star h+
\frac{\d(\d^a V)}{\d\varphi}\d^\mu\Delta_{\varphi,\psi}\star h]=\notag\\
f\sum_{\varphi\in\tilde {\cal G}}
\sum_{0\leq b\leq a}\frac{a!}{b!(a-b)!}[(\d^\mu\d^b
\frac{\d V}{\d\varphi})\d^{(a-b)}\Delta_{\varphi,\psi}\star h+(\d^b
\frac{\d V}{\d\varphi})\d^\mu\d^{(a-b)}\Delta_{\varphi,\psi}\star h]
=\notag\\
f\sum_{\varphi\in\tilde {\cal G}}
\sum_{0\leq b\leq (a+e_\mu)}\frac{(a+e_\mu)!}{b!(a+e_\mu-b)!}
(\d^b\frac{\d V}{\d\varphi})\d^{(a+e_\mu-b)}\Delta_{\varphi,\psi}\star h,
\end{gather}
where $e_\mu=(0,...,1,...,0)$ with $1$ at the $\mu$-th position.
First we have used (\ref{identity1.1}) (with $V$ replaced by $\d^a V$) and
in the second equality sign we have inserted (\ref{identity1})
(which is the inductive assumption) and $\d^\mu\otimes {\bf 1}$
applied to (\ref{identity1}) (cf. (\ref{space})).

The proof of the second identity (\ref{identity2}) is completely similar.
One simply has to replace the internal derivatives $\d^a\otimes {\bf 1}$
by external ones $\tilde\d^a\otimes {\bf 1}$. In particular the validity
of the equation corresponding to (\ref{xy}) relies on
$\Delta_{\tilde\d^\mu\chi,\psi}=\d^\mu \Delta_{\chi,\psi}$. $\w$

By means of Lemma 1 we will prove

{\bf Lemma 2:} Let $V,W\in \tilde {\cal P}_0,\>\chi,\psi,\kappa\in\tilde
{\cal G}$ and $f,g,h\in {\cal D}(\RR^4)$. Then
\begin{gather}
\sum_{\varphi\in\tilde {\cal G}}\frac{\d\Delta^\mu_{\chi,\psi}(Vg,Wf)}
{\d\varphi}\Delta_{\varphi,\kappa}\star h=\notag\\
\sum_{\varphi\in\tilde {\cal G}}[\Delta^\mu_{\chi,\psi}(\frac{\d V}
{\d\varphi}g\Delta_{\varphi,\kappa}\star h,Wf)+
\Delta^\mu_{\chi,\psi}(Vg,\frac{\d W}{\d\varphi}f
\Delta_{\varphi,\kappa}\star h)].\label{ableit:Delta2}
\end{gather}
{\bf Proof:} Using the explicit form (\ref{Delta2}) for $\Delta^\mu$
the l.h.s. of (\ref{ableit:Delta2}) is equal to
\begin{gather}
\sum_a\sum_{\varphi\in\tilde {\cal G}}C^\mu_{\chi,\psi; a}(-1)^{|a|}
\sum_{0\leq b\leq a}\frac{a!}{b!(a-b)!}[\frac{\d
(\tilde\d^{(a-b)}V)}{\d\varphi}W(\d^b g)f\notag\\
+(\tilde\d^{(a-b)}V)
\frac{\d W}{\d\varphi}(\d^b g)f]\Delta_{\varphi,\kappa}\star h.
\label{xxx}
\end{gather}
Again by means of (\ref{Delta2}) the r.h.s. of (\ref{ableit:Delta2})
can be written as
\begin{gather}
\sum_a\sum_{\varphi\in\tilde {\cal G}}C^\mu_{\chi,\psi; a}(-1)^{|a|}
\sum_{0\leq b\leq a}\frac{a!}{b!(a-b)!}
[\sum_{0\leq c\leq b}\frac{b!}{c!(b-c)!}\notag\\
(\tilde\d^{(a-b)}\frac{\d V}{\d\varphi})W(\d^c g)(\d^{(b-c)}
\Delta_{\varphi,\kappa}\star h)f+(\tilde\d^{(a-b)}V)\frac{\d W}{\d\varphi}
(\d^b g)f(\Delta_{\varphi,\kappa}\star h)].\label{xxxx}
\end{gather}
Due to (\ref{identity2}) the expressions (\ref{xxx}) and
(\ref{xxxx}) agree. $\quad\quad\w$

We now come to the proof of $(\tilde {\rm\bf N})$, i.e. we show that
there exists a normalization of the $T$-products which satisfies
{\bf (N0)}, {\bf (N1)}, {\bf (N2)}, {\bf (N3)} and $(\tilde {\rm\bf
  N})$.
Let the §T-products fulfil the first four of these normalization
conditions to all orders. In a double inductive procedure
we assume that $(\tilde {\rm\bf N})$ holds
to lower orders $\leq n$ and for all $T$-products to order $n+1$ of
sub-polynomials. More precisely, the second induction goes (for each
fixed $n$) with respect to the 'polynomial degree' $d$ which is the
sum of the degrees of the polynomials
$V_1,...,V_n\in\tilde\mathcal{P}_0$ in $T_n(V_1g_1\otimes
...\otimes V_ng_n)$: $d\=d |V_1|+...+|V_n|$. Note $|\d^a V|=|V|=
|\tilde\d^a V|$. By using {\bf (N3)} we want to show that
the commutators of the l.h.s. and of the r.h.s. of
$(\tilde {\rm\bf N})$ with $T_1(\kappa h)$ agree.
The commutator of the l.h.s. is equal to
\begin{eqnarray}
i\sum_{\varphi\in\tilde{\cal G}}\Bigl[T_{n+1}\Bigl([\frac{\d(\tilde\d^\nu V)}
{\d\varphi}W+(\tilde\d^\nu V)\frac{\d W}{\d\varphi}]g(\Delta_{\varphi,\kappa}
\star h)\otimes W_1f_1\otimes...\Bigr)\label{1}\\
+\sum_jT_{n+1}\Bigl((\tilde\partial^\nu V)Wg\otimes W_1f_1
\otimes...\otimes\frac{\d W_j}{\d\varphi}f_j(\Delta_{\varphi,\kappa}
\star h)\otimes ...\Bigr)\Bigr].\label{2}
\end{eqnarray}
To compute the commutator of the r.h.s. of $(\tilde {\rm\bf N})$ with
$T_1(\kappa h)$ we use again {\bf (N3)} and in addition Lemma 2. We obtain
\begin{eqnarray}
i\sum_{\varphi\in\tilde{\cal G}}\Bigl[T_{n+1}\Bigl([\frac{\d(\d^\nu V)}
{\d\varphi}W+(\d^\nu V)\frac{\d W}{\d\varphi}]g(\Delta_{\varphi,\kappa}
\star h)\otimes W_1f_1\otimes...\Bigr)\label{3}\\
+\sum_jT_{n+1}\Bigl(\Bigl[(\d^\nu V)Wg\otimes W_1f_1
\otimes...\otimes\frac{\d W_j}{\d\varphi}f_j(\Delta_{\varphi,\kappa}
\star h)\otimes ...\Bigr)\label{4}\\
+i\sum_{m=1}^n\sum_{\chi,\psi\in\tilde {\cal G}} T_n\Bigl(
\bigl[\Delta^{\nu}_{\chi,\psi}\bigl([\frac{\d^2 V}
{\d\varphi\d\chi}W+\frac{\d V}{\d\chi}\frac{\d W}{\d\varphi}]
g(\Delta_{\varphi,\kappa}\star h),\frac{\d W_m}{\d\psi}f_m\bigr)\label{5}\\
+\Delta^{\nu}_{\chi,\psi}\bigl(\frac{\partial V}
{\partial\chi}Wg,\frac{\d^2 W_m}{\d\varphi\d\psi}f_m
(\Delta_{\varphi,\kappa}\star h)\bigr)\bigr]
\otimes W_1f_1\otimes...\hat m...\Bigr)\label{6}\\
+i\sum_{m,j\,(m\not= j)}\sum_{\chi,\psi\in\tilde {\cal G}} T_n\Bigl(
\Delta^{\nu}_{\chi,\psi}\bigl(\frac{\partial V}
{\partial\chi}Wg,\frac{\partial W_m}{\partial\psi}f_m\bigr)
\otimes W_1f_1\otimes...\nonumber\\
...\hat m...\otimes \frac{\d W_j}{\d\varphi}f_j
(\Delta_{\varphi,\kappa}\star h)\otimes...\Bigr)\Bigr]\label{7}
\end{eqnarray}
Due to $(\tilde {\rm\bf N})$ for subpolynomials
we have the following equations:\\
(second term in (\ref{1}))$=$(second term in (\ref{3}))$+$
(second term in (\ref{5}))\\
and $\quad\quad\quad$ (\ref{2})=(\ref{4})+(\ref{6})+(\ref{7}).\\
To get the equality of (\ref{1})+(\ref{2}) and (\ref{3})+(\ref{4})+
(\ref{5})+(\ref{6})+(\ref{7}) it remains to show:\\
(first term in (\ref{1}))$=$(first term in (\ref{3}))$+$
(first term in (\ref{5})). $\quad\quad (*)$\\
To verify this we insert (\ref{identity1.1}) with $\d^\mu\otimes {\bf 1}$
replaced by $\tilde\d^\mu\otimes {\bf 1}$ into the first term in (\ref{1})
and the original (\ref{identity1.1}) into the first term in (\ref{3}).
The remaining terms in $(*)$ cancel by means of $(\tilde {\rm\bf N})$ for
subpolynomials.

From the just now proved result
we conclude that $(\tilde {\rm\bf N})$ can be violated by a C-number only:
\begin{gather}
T_{n+1}((\tilde\partial^\nu V)Wg\otimes W_1f_1\otimes...)-
T_{n+1}((\partial^\nu V)Wg\otimes W_1f_1\otimes...)\notag\\
-i\sum_{m=1}^n\sum_{\chi,\psi\in\tilde {\cal G}} T_n\Bigl(
\Delta^{\nu}_{\chi,\psi}\bigl(\frac{\partial V}
{\partial\chi}Wg,\frac{\partial W_m}{\partial\psi}f_m\bigr)
\otimes W_1f_1\otimes...\hat m...\Bigr)=\notag\\
\langle\Omega |T_{n+1}((\tilde\partial^\nu V)
Wg\otimes W_1f_1\otimes...)|\Omega\rangle-
\langle\Omega |T_{n+1}((\partial^\nu V)Wg\otimes
W_1f_1\otimes...)|\Omega\rangle\notag\\
-i\sum_{m=1}^n\sum_{\chi,\psi\in\tilde {\cal G}} \langle\Omega|T_n\Bigl(
\Delta^{\nu}_{\chi,\psi}\bigl(\frac{\partial V}
{\partial\chi}Wg,\frac{\partial W_m}{\partial\psi}f_m\bigr)
\otimes W_1f_1\otimes...\hat m...\Bigr)|\Omega\rangle\notag\\
\=d:\tilde a(g,f_1,...,f_n).\label{A}
\end{gather}
Due to causal factorization
of the $T$-products and the validity of $(\tilde {\rm\bf N})$ to lower
orders $\leq n$, the possible violation $\tilde a(g,f_1,...,f_n)$ of
$(\tilde {\rm\bf N})$ must be local
\begin{equation}
\tilde a(g,f_1,...,f_n)=\int dx\, dx_1...dx_n\,\sum_{|a|=0}^\omega
C_a\d^a\delta(x_1-x,...,x_n-x)g(x)f_1(x_1)...f_n(x_n),\label{B}
\end{equation}
with unknown constants $C_a$ and
\begin{equation}
\omega\=d {\rm sd}\bigl(\langle\Omega |T_{n+1}((\tilde\d^\nu V)W,W_1,...
,W_n)|\Omega\rangle\bigr)-4n.\label{C}
\end{equation}
After the finite renormalization
\begin{gather}
\langle\Omega |T_{n+1}((\tilde\partial^\nu V)Wg
\otimes W_1f_1\otimes...)|\Omega\rangle
\rightarrow\notag\\
 \langle\Omega |T_{n+1}((\tilde\partial^\nu V)Wg\otimes W_1f_1
\otimes...)|\Omega\rangle-\tilde a(g,f_1,...)\label{D1}
\end{gather}
$(\tilde {\rm\bf N})$ holds true. By construction (in particular (\ref{C}))
this renormalization respects {\bf (N0)}. From the
definition (\ref{A}) of $\tilde a(g,f_1,...)$ we see that (\ref{D1})
maintains {\bf (N1)}, {\bf (N2)} and the permutation symmetry of
$\langle\Omega |T_{n+1}((\tilde\d^\nu V)W,W_1,...)|\Omega\rangle$.
However, in general (\ref{D1}) violates {\bf (N3)}, namely in the
cases in which \break
$T_{n+1}((\tilde\d^\nu V)W,W_1,...)$ appears on the
r.h.s. of {\bf (N3)}.\footnote{The cases of {\bf (N3)} in which
$T_{n+1}((\tilde\d^\nu V)W,W_1,...)$ appears on the l.h.s. remain true,
because only the C-number part of $T_{n+1}((\tilde\d^\nu V)W,W_1,...)$
gets changed.} So we everywhere repair {\bf (N3)}  by a chain
of finite renormalizations of $T$-products
of order $n+1$ with polynomial degree
$d>|V|+|W|+|W_1|+...+|W_n|$.\footnote{The vacuum expectation values of
  these $T$-products remain unchanged; solely the operator parts get
renormalized.}
It is obvious that this can be done such that {\bf (N0)}, {\bf (N1)}
and {\bf (N2)} are preserved. The validity of $(\tilde {\rm\bf N})$
up to order $n+1$ and polynomial degree $|V|+|W|+|W_1|+...+|W_n|$
is not touched by these renormalizations. So the inductive step is
finished. $\quad\w$

In other words, the compatibility of the renormalization (\ref{D1})
with {\bf (N3)} follows from the following general observation:
{\bf (N3)} determines the operator-valued map $T_n$ completely
in terms of the $\CC$-valued map
$\langle\Omega|T_n(\cdot)|\Omega\rangle:$
${\cal D}(\RR^4,\tilde {\cal P}_0)^{\otimes n}\rightarrow\CC$.
However, {\bf (N3)} does not give any
relation among the vacuum expectation values of the $T$-products, they
may be arbitrarily given. Hence, renormalizations of the vacuum
expectation values of the $T$-products are not in conflict with
{\bf (N3)}. We will use this second way of argumentation in the following.

\subsection{Compatibility of the master Ward identity with (N3)}

We start with $T$-products which fulfil {\bf (N0)}, {\bf (N1)}, {\bf
  (N2)}, {\bf (N3)} and $(\tilde {\rm\bf N})$ to all orders.
We use the same double induction as in the preceding subsection:
we assume that ({\bf N}) holds
to lower orders $\leq n$ and for $T_{n+1}$ restricted to the elements
of $\mathcal{D}(\RR^4,\tilde\mathcal{P}_0)^{\otimes n+1}$ with a lower
polynomial degree.

By means of
{\bf (N3)} we are going to prove that the commutators of the
l.h.s. and of the r.h.s. of ({\bf N}) with $T_1(\kappa h)$ are equal.
For the l.h.s. it results
\begin{eqnarray}
-i\sum_{\varphi\in {\cal G}}\Bigl[T_{n+1}\bigl(\frac{\d V}{\d\varphi}
(\d^\nu g)(\Delta_{\varphi,\kappa}\star h)\otimes W_1f_1\otimes...\bigr)
\label{11}\\
+\sum_{l=1}^nT_{n+1}\bigl(V\d^\nu g\otimes W_1f_1\otimes...\otimes\frac{\d W_l}
{\d\varphi}f_l (\Delta_{\varphi,\kappa}\star h)\otimes...\bigr)\Bigr].
\label{12}
\end{eqnarray}
By using again {\bf (N3)} and in addition Lemma 2 we
compute the commutator of the r.h.s. and obtain
\begin{gather}
i\sum_{\varphi\in {\cal G}}\Bigl\{ T_{n+1}\bigl(\frac{\d (\d^\nu V)}
{\d\varphi}g(\Delta_{\varphi,\kappa}\star h)\otimes W_1f_1\otimes...\bigr)
\label{13}\\
+\sum_{l=1}^nT_{n+1}\bigl((\d^\nu V)
g\otimes W_1f_1\otimes...\otimes\frac{\d W_l}
{\d\varphi}f_l (\Delta_{\varphi,\kappa}\star h)\otimes...\bigr)\label{14}\\
+i\sum_{m=1}^n\sum_{\chi,\psi\in {\cal G}} \Bigl(T_n\bigl(\bigl[
\Delta^{\nu}_{\chi,\psi}\bigl(\frac{\d^2 V}{\d\varphi\d\chi}g
(\Delta_{\varphi,\kappa}\star h),\frac{\partial W_m}{\d\psi}f_m\bigr)
\label{15}\\
+\Delta^{\nu}_{\chi,\psi}\bigl(\frac{\d V}{\d\chi}g,\frac{\d^2 W_m}
{\d\varphi\d\psi}f_m (\Delta_{\varphi,\kappa}\star h)\bigr)\bigr]
\otimes W_1f_1\otimes...\hat m...\bigr)\label{16}\\
+\sum_{l\>(l\not= m)}T_n\bigl(\Delta^{\nu}_{\chi,\psi}
\bigl(\frac{\d V}{\d\chi}g,\frac{\d W_m}{\d\psi}f_m\bigr)\otimes W_1f_1
\otimes...\notag\\
...\hat m...\otimes\frac{\d W_l}{\d\varphi}f_l(\Delta_{\varphi,
\kappa}\star h)\otimes...\bigr)\Bigr)\Bigr\}.\label{17}
\end{gather}
The validity of ({\bf N}) for sub-polynomials implies\\
(\ref{12})=(\ref{14})+(\ref{16})+(\ref{17}).\\
It remains to prove\\
(\ref{11})=(\ref{13})+(\ref{15}).$\quad\quad (**)$\\
After inserting (\ref{identity1.1}) into (\ref{13}) this equation $(**)$
takes the form of ({\bf N}) for some sub-polynomials, which holds true
by the inductive assumption.$\quad\quad\w$

As in the case of $(\tilde {\rm\bf N})$ (\ref{A})-(\ref{C}) we conclude
that the operator identity {\bf (N)} can be violated by a local C-number
$a(g,f_1,...,f_n)$ only:
\begin{gather}
a(g,f_1,...,f_n)\=d \langle\Omega|
T_{n+1}(V\partial^\nu g\otimes W_1f_1\otimes...
)|\Omega\rangle+\notag\\
\langle\Omega|T_{n+1}((\partial^\nu V)g\otimes
W_1f_1\otimes...)|\Omega\rangle\nonumber\\
+i\sum_{m=1}^n\sum_{\chi,\psi\in {\cal G}} \langle\Omega|T_n\Bigl(
\Delta^{\nu}_{\chi,\psi}\bigl(\frac{\partial V}{\partial\chi}g,
\frac{\partial W_m}{\partial\psi}f_m\bigr)\otimes W_1f_1\otimes...
\hat m...\Bigr)|\Omega\rangle.\label{E}
\end{gather}
The aim is now to remove $a(g,f_1,...,f_n)$ by finite renormalizations
of the vacuum expectation values $\langle\Omega |T(...)|\Omega\rangle$ on the
r.h.s.. Such renormalizations are not in conflict with {\bf (N3)}
(see the end of the preceding subsection). So we have proved
the compatibility of {\bf N} with {\bf (N3)}.

We discuss the possibilities to remove $a(g,f_1,...,f_n)$:

(A) The finite renormalization
\begin{gather}
\langle\Omega|T_{n+1}((\partial^\nu V)g\otimes W_1f_1\otimes...)|\Omega\rangle
\rightarrow \notag\\
\langle\Omega|T_{n+1}((\partial^\nu V)g\otimes W_1f_1\otimes...)|\Omega\rangle
-a(g,f_1,...),\label{F}
\end{gather}
does this job and is compatible with {\bf (N0)-(N3)},
$(\tilde {\rm\bf N})$ and permutation symmetry. However,
this procedure works only if $\d^\nu V\not= 0$ and if
${\rm sd}\bigl(\langle\Omega|T_{n+1}(\partial^\nu V,W_1,...$\break
$...,W_n)|
\Omega\rangle\bigr)=
{\rm sd}\bigl(\langle\Omega|T_{n+1}(V,W_1,...,W_n)|\Omega\rangle\bigr)+1$.
In case that the latter does not hold one gets in conflict with
{\bf(N0)}. This happens e.g. for the axial and pseudo-scalar
triangle-diagram, see (\ref{axan}). In many important applications
of the MWI $V$ corresponds to a conserved current (i.e.
$\d^\nu V=0$), for example $V=\bar\psi\gamma\psi$, or if $V$ is
the free ghost current (cf. sect. 4.2 for both) or the free BRST-current
(sect. 4.4).

(B) If (\ref{F}) does not work one tries to satisfy {\bf (N)} by
renormalizing also $\langle\Omega|T_{n+1}(V,W_1,...)|\Omega\rangle$ and
eventually $\langle\Omega|T_n\Bigl(
\Delta^{\nu}_{\chi,\psi}\bigl(...\bigr),W_1...\Bigr)|\Omega\rangle$.
This method does not ensure success. In detail one proceeds in the
following way:\\
(B1) $a(g,f_1,...)$ has the form (\ref{B}) with $\omega\=d {\rm sd}\bigl(
\langle\Omega |T_{n+1}(V,W_1,...,W_n)|\Omega\rangle\bigr)+1-4n$.
Using symmetry
properties (e.g. Poincare covariance, permutation symmetries) of the
r.h.s. of (\ref{E}) the constants $C_a$ (we use the notation of (\ref{B}))
can be strongly restricted.\\
(B2) One works out the freedoms of normalization (\ref{4.3c}) of
$\langle\Omega|T_{n+1}(V,W_1,...)|\Omega\rangle$, $\langle\Omega|T_{n+1}
((\partial^\nu V),W_1...)|\Omega\rangle$ and
eventual $\langle\Omega|T_n\Bigl(\Delta^{\nu}(...),W_1...\Bigr)|\Omega\rangle$
(the second is only
available if $\d^\nu V\not= 0$) which respect {\bf (N0)-(N2)},
$(\tilde {\rm\bf N})$ and permutation symmetry. Renormalizations of
$\langle\Omega|T_n\Bigl(\Delta^{\nu}(...),W_1...\Bigr)|\Omega\rangle$ are also
restricted by the validity of the inductive assumption for {\bf (N)}.\\
(B3) One then tries to remove the remaining $a(g,f_1,...)$ by using the
freedoms which result from step (2).\\
Because the restricted $a(g,f_1,...)$ (step(1)) and the free normalization
polynomials (step (2)) depend strongly on $(V,W_1,...,W_n)$, one
has to treat each combination $(V,W_1,...,W_n)$ separately and this
gives quite a lot of work.
This method was used in \cite{DHKS} to prove 'perturbative gauge invariance'
(which is equation (\ref{N6}) with $j_1=...=j_n=0$) for $SU(N)$-Yang-Mills
theories. To restrict $a(g,f_1,...)$ sufficiently a weak assumption about
the infrared behaviour was necessary. (However, if this assumption would
not hold, the Green's functions would not exist.)

\subsection{Proof of the master Ward identity for solely massive
fields and not relatively lowered scaling degree}

We return to the end of sect. 2.2 and set
\begin{equation}
\omega (t_0)\=d {\rm sd}(t_0)-4(n-1).
\end{equation}
A possible extension $t\in {\cal D}'(\RR^{4(n-1)})$ of $t_0\in
{\cal D}'(\RR^{4(n-1)}\setminus \{0\})$ which respects {\bf (N0)} is given
by (cf. \cite{BF},\cite{P})
\begin{equation}
(t^{(w)},h)\=d (t_0,h^{(w)}),\quad\quad\forall h\in {\cal D}(\RR^{4(n-1)})
\label{extension}
\end{equation}
where
\begin{equation}
h^{(w)}(x)\=d h(x)-w(x)\sum_{|a|=0}^{\omega (t_0)}\frac{x^a}{a!}
(\d^a h)(0),\label{w-sub}
\end{equation}
and $w\in {\cal D}(\RR^{4(n-1)})$ is such that
there exists a neighbourhood ${\cal U}$ of $0\in \RR^{4(n-1)}$ with
$w\vert_{\cal U}\equiv 1$. A change of $w$ alters the normalization of
$t^{(w)}$. For $\omega (t_0)<0$ we have $h^{(w)}=h$ in agreement with the
fact that the extension is unique in that case. Because there is no Lorentz
invariant $w\in {\cal D}(\RR^{4(n-1)})$, the extension $t^{(w)}$ is not
Lorentz covariant (in general)
and one has to perform a finite renormalization
(\ref{4.3c}) to restore this symmetry (see the second papers of
\cite{St2} and \cite{DHKS}, as well as
\cite{BPP}). To avoid this one is tempted to choose $w\equiv 1$.
But $h^{(w\equiv 1)}$ is not a test-function. However, if all fields
are massive the infrared behavior is harmless and Epstein and Glaser
\cite{EG} have shown that one may indeed choose $w\equiv 1$ in this case.
The extension $t^{(c)}\=d t^{(w\equiv 1)}$ is called 'central solution'
(or better 'central extension' in our framework) and it was pointed out
that it preserves nearly all symmetries \cite{EG}, \cite{DKS}, \cite{S}
\footnote{Epstein and Glaser have proved that one may choose $w\equiv 1$
for the method of distribution splitting. In this footnote we show how
their result applies to our extension procedure (\ref{extension}). From
Epstein and Glaser \cite{EG} we know $t_0\in {\cal S}'(\RR^{4(n-1)}\setminus
\{0\})$ and hence $t^{(w)}\in {\cal S}'(\RR^{4(n-1)})$, so we may use
Fourier transformation. Epstein and Glaser have proved
that in the massive case the Fourier transformation $\hat t^{(w)}(p)$
(and therefore any extension (\ref{4.3c})) is analytic in a neighbourhood
of $p=0$. Then they define the central extension $t^{(c)}$ by
\begin{equation}
\d^a\hat t^{(c)}(0)=0\quad\quad\forall |a|\leq \omega (t_0).\label{central}
\end{equation}
The Fourier transformation of $t^{(w)}$ (\ref{extension}) reads \cite{P}
\begin{equation}
\hat t^{(w)}(p)=\hat t_0(p)-\sum_{|a|=0}^{\omega (t_0)}\frac{p^a}{a!}
\d^a(\widehat{t_0w})(0).\label{hat:t}
\end{equation}
Note $\widehat{t_0w}=(2\pi)^{-\frac{n}{2}}\hat t_0\star \hat w\in
{\cal C}^\infty$, i.e. $\d^a(\widehat{t_0w})(0)$ exists. Using the
definition (\ref{central}) of the central extension we now find
\begin{equation}
\hat t^{(c)}(p)=\hat t^{(w)}(p)-\sum_{|a|=0}^{\omega (t_0)}\frac{p^a}{a!}
(\d^a\hat t^{(w)})(0)=\hat t_0(p)-\sum_{|a|=0}^{\omega (t_0)}\frac{p^a}{a!}
(\d^a\hat t_0)(0).\label{central1}
\end{equation}
We see that we may set $w\equiv 1$ in (\ref{hat:t}) and hence also in
(\ref{extension}), and that this choice is the central extension
(\ref{central}).}.

We are now going to show that the central extensions fulfil the
MWI provided the scaling degree is not relatively lowered for the
individual, contributing C-number distributions
(a precise explanation of the latter expression
is given below in (\ref{om1}), (\ref{om2})).
We define $\bar t^{\nu (c)},\,t^{(c)}$ and
$t^{\nu (c)}_{b;\chi,\psi}$ to be central extensions:
\begin{gather}
\int dx\, dx_1...dx_n\,\bar t^{\nu (c)}(x_1-x,...,x_n-x)g(x)f_1(x_1)...
f_n(x_n)\=d \notag\\
\langle\Omega|T_{n+1}((\partial^\nu V)g\otimes W_1f_1\otimes...
\otimes W_nf_n)|\Omega\rangle^{(c)}\\
\int dx\, dx_1...dx_n\,t^{(c)}(x_1-x,...,x_n-x)g(x)f_1(x_1)...
f_n(x_n)\=d \notag\\
\langle\Omega|T_{n+1}(Vg\otimes W_1f_1\otimes...
\otimes W_nf_n)|\Omega\rangle^{(c)}\\
\sum_b\int dx\, dx_1...dx_n\,t^{\nu (c)}_{b;\chi,\psi}(x_1-x_m,
...\hat m...,x_n-x_m)\cdot\notag\\
(\d^b\delta) (x_m-x)g(x)f_1(x_1)...f_n(x_n)\=d\notag\\
\langle\Omega|T_n\Bigl(\Delta^{\nu}_{\chi,\psi}
\bigl(\frac{\partial V}{\partial\chi}g,\frac{\partial W_m}{\partial
\psi}f_m\bigr)\otimes W_1f_1\otimes...\hat m...\otimes W_nf_n\Bigr)
|\Omega\rangle^{(c)},\label{t_b}
\end{gather}
where we have taken the definition (\ref{Delta2}) of $\Delta^\mu$ into
account\footnote{From (\ref{Delta2}) we see that $\langle\Omega|T_n\Bigl(
\Delta^\mu\bigl(...\bigr)\otimes...\hat m...\Bigr)|\Omega\rangle^{(c)}$
is of the form
\begin{gather}
\sum_b\tilde t_b(f_1\otimes...\otimes (\d^b g)f_m\otimes...\otimes f_n)=
\notag\\
\sum_b\int dx\, dx_1...dx_n\, t_b(x_1-x_m,...\hat m...,x_n-x_m)
(\d^b\delta)(x_m-x)g(x)f_1(x_1)...f_n(x_n),
\end{gather}
$\tilde t_b\in {\cal D}'(\RR^{4n}),\> t_b\in {\cal D}'(\RR^{4(n-1)})$.}.
The corresponding non-extended distributions are
\begin{equation}
\bar t^\nu_0,\quad t_0\in {\cal D}'(\RR^{4n}\setminus \{0\})
\quad {\rm and}\quad t^\nu_{b;\chi,\psi;0}
\in {\cal D}'(\RR^{4(n-1)}\setminus \{0\}).\label{t_0}
\end{equation}
In the preceding subsection we have learnt that the validity of
{\bf(N3)} reduces the proof of the MWI to the vacuum sector
(see (\ref{E})). So we only have to show
\begin{gather}
-\d^\nu t^{(c)}(y_1,...,y_n)=\bar t^{\nu (c)}(y_1,...,y_n)\notag\\
+i\sum_{m=1}^n\sum_{\chi,\psi\in {\cal G}} t^{\nu (c)}_{b;\chi,\psi}
(y_1-y_m,...\hat m...,y_n-y_m)\d^b\delta (y_m),\label{Beh}
\end{gather}
where
\begin{equation}
\d^\nu\=d \d^\nu_1+...+\d^\nu_n.
\end{equation}
By causal factorization and induction
we know that this equation is fulfilled by the corresponding
non-extended distributions (\ref{t_0}). Setting $y\equiv
(y_1,...,y_n)$ we obtain
\begin{gather}
-\Bigl(\d^\nu t^{(c)}(y),h(y)\Bigr)=
\Bigl(t_0(y),[\d^\nu h(y)-\sum_{|a|=0}^{\omega (t_0)}\frac{y^a}{a!}
(\d^a\d^\nu h)(0)]\Bigr)=\notag\\
\Bigl(t_0(y),\d^\nu [h(y)
-\sum_{|a|=0}^{\omega (t_0)+1}\frac{y^a}{a!}
(\d^a h)(0)]\Bigr)=\notag\\
\Bigl(\bar t^\nu_0(y),[h(y)
-\sum_{|a|=0}^{\omega (t_0)+1}\frac{y^a}{a!}
(\d^a h)(0)]\Bigr)\notag\\
+i\sum_{m=1}^n\sum_{\chi,\psi\in {\cal G}} \Bigl(t^\nu_{b;\chi,\psi;0}
(y_1-y_m,...\hat m...,y_n-y_m)\d^b\delta (y_m),[h(y)\notag\\
-\sum_{|a|=0}^{\omega (t_0)+1}\frac{y^a}{a!}
(\d^a h)(0)]\Bigr).\label{master1}
\end{gather}
If the scaling degree is not relatively lowered, more precisely if
\begin{equation}
\omega (\bar t^\nu_0)=\omega(t_0)+1\label{om1}
\end{equation}
and
\begin{equation}
\omega (t^\nu_{b;\chi,\psi;0})=\omega(t_0)+1-|b|,\quad\forall b,\quad
\forall \chi,\psi\in {\cal G},\label{om2}
\end{equation}
then the terms in the final expression of (\ref{master1}) are the central
extensions. For the $\bar t^\nu_0$-term this is obvious. To verify this
statement for the $t^\nu_{b;\chi,\psi;0}$-terms (we omit the indices
$\nu,\chi,\psi$ in the following) it suffices to consider the term
$m=n$ and test-functions of the form $h(y_1,...,y_n)=h_1(z_1,...,z_{n-1})
h_2(z_n)$ where
\begin{equation}
z\equiv (z_1,...,z_n)=(y_1-y_n,...,y_{n-1}-y_n,y_n)\=d :Ay,\quad\quad
A\in SL(n,\RR).\nonumber
\end{equation}
Then we have
\begin{gather}
h(y)-\sum_{|a|=0}^{\omega (t_{b;0})+|b|}\frac{y^a}{a!}
(\d^a h)(0)=\notag\\
(h_1\otimes h_2)(z)-\sum_{|a|=0}^{\omega (t_{b;0})+|b|}\frac{y^a}{a!}
((A^T\d)^a (h_1\otimes h_2))(0)=\notag\\
(h_1\otimes h_2)(z)-\sum_{|a|=0}^{\omega (t_{b;0})+|b|}\frac{z^a}{a!}
(\d^a (h_1\otimes h_2))(0)\notag
\end{gather}
where $A^T$ denotes the transposed matrix and we have used
$y^a\cdot (A^T\d)^a=\break (Ay)^a\cdot\d^a$.
We set $a=(\bar a, a_n)$ and $z\equiv (\bar z,z_n)$.
Then the last term in (\ref{master1})
can be transformed in the following way:
\begin{gather}
\Bigl(t_{b;0}(y_1-y_n,...,y_{n-1}-y_n)\d^b\delta (y_n),
[h(y)-\sum_{|a|=0}^{\omega (t_{b;0})+|b|}\frac{y^a}{a!}
(\d^a h)(0)]\Bigr)_y=\notag\\
\Bigl(t_{b;0}(\bar z)\d^b\delta (z_n),[h_1(\bar z)h_2(z_n)
-\sum_{|\bar a|+|a_n|=0}^{\omega (t_{b;0})+|b|}\frac{
\bar z^{\bar a}z_n^{a_n}}{\bar a!a_n!}(\d^{\bar a}h_1)(0)
(\d^{a_n}h_2)(0)]\Bigr)_z=\notag\\
(-1)^{|b|}(\d^b h_2)(0)\Bigl(t_{b;0}(\bar z),[h_1(\bar z)
-\sum_{|\bar a|=0}^{\omega (t_{b;0})}\frac{
\bar z^{\bar a}}{\bar a!}(\d^{\bar a}h_1)(0)]\Bigr)_{\bar z}=\notag\\
\Bigl(t^{(c)}_b(\bar z)\d^b\delta (z_n),h_1(\bar z) h_2(z_n)\Bigr)_z=
\notag\\
\Bigl(t^{(c)}_b(y_1-y_n,...,y_{n-1}-y_n)\d^b\delta (y_n),h(y)\Bigr)_y.
\end{gather}
Summing up we find the assertion (\ref{Beh}) if (\ref{om1}) and (\ref{om2})
hold true, otherwise
we have over-subtracted extensions. Note that this proof works
also for $\bar t^\nu_0=0$ and hence $\bar t^{\nu (c)}=0$. Obviously
this method fails for extensions $t^{(w)}$ (\ref{extension}) with
$w\in {\cal D}(\RR^{4(n-1)})$, because additional
terms $\sim \d^\nu w$ appear in (\ref{master1}).\footnote{For the
method of distribution splitting the central solution in momentum space
can be obtained by a dispersion integral \cite{EG},\cite{DKS},\cite{S}.
In \cite{DKS} this dispersion integral has been used to prove gauge
invariance of QED. The present proof (\ref{master1}) is a kind of $x$-space
version of that procedure, which yields a more general result.
In addition, it has the advantage that it is not
necessary to treat the cases of different external legs individually.}
$\quad\w$

In case of the axial anomaly we set $j_A^\mu\=d\psq\gamma^\mu\gamma^5\psi,\,
j^\mu\=d\psq\gamma^\mu\psi$ and $j_\pi\=d i\psq\gamma^5\psi$, and have
\begin{eqnarray}
t^{\mu\lambda\tau (c)}(x,x_1,x_2)=\langle\Omega|T_3(j^\mu_A,j^\lambda,j^\tau)
(x,x_1,x_2)|\Omega\rangle^{(c)},\nonumber\\
\bar t^{\nu\mu\lambda\tau (c)}(x,x_1,x_2)=
2mg^{\mu\nu}\langle\Omega|T_3(j_\pi,j^\lambda,j^\tau)
(x,x_1,x_2)|\Omega\rangle^{(c)}\label{axan}
\end{eqnarray}
for the $AVV$-triangle diagram. The corresponding distributions for
the $AAA$-triangle are obtained by replacing $j^\lambda,j^\tau$ by
 $j^\lambda_A,j^\tau_A$. All $t_{b;...}$-distributions vanish.
One finds $\omega(t^{(c)})=1$ and $\omega(\bar t^{(c)})=0<
\omega(t^{(c)})+1$.\footnote{According to power counting one expects
$\omega(\bar t^{(c)})=1$, but the ($\omega=1$)-terms are proportional
to the spinor trace
${\rm tr}(\gamma^5p_{1\mu}\gamma^\mu\gamma^\lambda p_{2\nu}\gamma^\nu
\gamma^\tau p_{3\rho}\gamma^\rho)=0$.}
Hence, the present proof (\ref{master1}) does not apply.

\section{Applications of the master Ward identity}

The main success of the MWI are its many, important and far-reaching
consequences.

\subsection{Field equation}

Let us consider the pair $(\varphi,\chi)$ of symbols (corresponding to
massive or massless free fields which fulfill the Klein-Gordon or
wave equation) that is studied in appendix A\footnote{For simplicity
we choose $\epsilon =1$ in (\ref{D}).}
and let $W_1,...,W_n\in {\cal P}_0$. We assume that
$W_1,...,W_n$ contain only zeroth and first (internal) derivatives of
$\chi$. By applying twice the MWI
and using the explicit expressions (\ref{delta:0-0})-(\ref{delta:1-1})
for $\delta^\mu$ we obtain
\begin{gather}
T_{n+1}(\varphi(\w +m^2)g\otimes W_1f_1\otimes...\otimes W_nf_n)=\notag\\
-T_{n+1}((\d_\mu\varphi)\d^\mu g\otimes W_1f_1\otimes...\otimes W_nf_n)
+m^2 T_{n+1}(\varphi g\otimes W_1f_1\otimes...\otimes W_nf_n)\notag\\
-i\sum_{m=1}^n\sum_{\psi\in {\cal G}}(\pm) T_n\Bigl(
\Delta_{\mu\,\varphi,\psi}
\bigl(\d^\mu g,\frac{\partial W_m}{\partial\psi}f_m\bigr)
\otimes W_1f_1\otimes...\hat m...\otimes W_nf_n\Bigr)=\notag\\
i\sum_{m=1}^n\sum_{\psi\in {\cal G}}(\pm) T_n\Bigl(
\Delta^\mu_{\d_\mu\varphi,\psi}
\bigl(g,\frac{\partial W_m}{\partial\psi}f_m\bigr)
\otimes W_1f_1\otimes...\hat m...\otimes W_nf_n\Bigr)\notag\\
-i\sum_{m=1}^n(\pm) C T_n\Bigl(
\frac{\partial W_m}{\partial(\d^\mu\chi)}(\d^\mu g)f_m
\otimes W_1f_1\otimes...\hat m...\otimes W_nf_n\Bigr)=\notag\\
i\sum_{m=1}^n(\pm) T_n\Bigl(
\frac{\partial W_m}{\partial\chi} gf_m
\otimes W_1f_1\otimes...\hat m...\otimes W_nf_n\Bigr)\notag\\
+i\sum_{m=1}^n(\pm) T_n\Bigl(
\frac{\partial W_m}{\partial(\d^\mu\chi)}(\d^\mu g)f_m
\otimes W_1f_1\otimes...\hat m...\otimes W_nf_n\Bigr).\label{N4}
\end{gather}
This is the normalization condition {\bf (N4)} of \cite{DF} and
\cite{BDF}. It is equivalent to
\begin{gather}
T_{n+1}(\varphi g\otimes W_1f_1\otimes...\otimes W_nf_n)=\notag\\
i\sum_{l=1}^n\sum_{\psi\in {\cal G}}T_n(W_1f_1\otimes...\otimes\frac{\d W_l}
{\d\psi}f_l\Delta^F_{\psi,\varphi}\star g\otimes ...\otimes W_nf_n)+...\>,
\label{N4'}
\end{gather}
where the dots stand for the terms in which $\varphi g$ is not contracted.
We see from this formula (\ref{N4'}) that the normalization condition
{\bf (N4)} can always be satisfied without getting in conflict with
{\bf (N0)-(N3)}, even if anomalies are present.
Note that the final result (on the
r.h.s. in (\ref{N4})) is independent from the
normalization constant $C$ which appears in the intermediate formula.
This must be so, because the Feynman propagators $\Delta^F_{\psi,\varphi}$
in (\ref{N4'}) do not contain this constant.

Generalizing Bogoliubov's idea \cite{BS} we define the interacting
field $\Lambda_{g{\cal L}}$ belonging to
$\Lambda\in \mathcal{D}(\RR^4,\tilde{\cal P}_0)$ and to the
interaction ${\cal L}\in {\cal P}_0$ in terms of the $T$-products by
\begin{eqnarray}
\Lambda_{g{\cal L}}\=d S(g{\cal L})^{-1}\frac{d}{id\lambda}
\vert_{\lambda =0}S(g{\cal L}+\lambda \Lambda)=\nonumber\\
\sum_{n=0}^\infty\frac{i^n}{n!}
R_{n+1}(({\cal L}g)^{\otimes n};\Lambda)=
T_1(\Lambda)+{\cal O}(g),\label{intfield}
\end{eqnarray}
where the 'totally retarded products' $R_{n+1}(...)$ (also called
'$R$-product') are defined by
\begin{gather}
R_{n+1}(\Lambda_1\otimes...\otimes\Lambda_n;\Lambda)\=d
\sum_{I\subset \{1,...,n\}}(-1)^{|I|}
\bar T(\otimes_{l\in I}\Lambda_l)T((\otimes_{j\in I^c}\Lambda_j)
\otimes \Lambda)\label{R}
\end{gather}
and we have used (\ref{S}) and (\ref{S:inverse}). Similarly to the
$S$-matrix (\ref{S}), the interacting fields are formal power series.
In the particular case $\Lambda =Wf,\>W\in{\cal P}_0,\> f\in
\mathcal{D}(\RR^4)$ we write $W_{g{\cal L}}(f)$ instead of
$(Wf)_{g{\cal L}}$. Following \cite{DF} the condition (\ref{N4})
can easily be translated into an identity for
$R_{n+1}(W_1f_1\otimes...\otimes W_nf_n;\varphi(\w +m^2) f)$.
The latter implies the field equation
\begin{equation}
(\w +m^2)\varphi_{g{\cal L}}=-g\Bigl(\frac{\d{\cal L}}{\d\chi}
\Bigr)_{g{\cal L}}+\d^\mu [g\Bigl(\frac{\d{\cal L}}{\d(\d^\mu\chi)}
\Bigr)_{g{\cal L}}],\label{fieldeq}
\end{equation}
where $g$ is a test function.

The calculation (\ref{N4}) can be carried over to external derivatives
by using ($\tilde {\bf N}$) instead of {\bf (N)}. More precisely let
$W,W_1,...,W_n\in {\cal P}_0$ and let us assume that $W_1,...,W_n$ contain
only zeroth and first (internal) derivatives of $\chi$. With that we obtain
\begin{gather}
T_{n+1}(((\tilde{\w}+m^2)\varphi)Wg\otimes W_1f_1\otimes...\otimes W_nf_n)=
\notag\\
i\sum_{m=1}^n(\pm) T_n\Bigl(W\frac{\partial W_m}{\partial\chi}gf_m
\otimes W_1f_1\otimes...\hat m...\otimes W_nf_n\Bigr)\notag\\
+i\sum_{m=1}^n(\pm) T_n\Bigl(W\frac{\d W_m}{\d(\d_\mu\chi)}(\d_\mu g)f_m
\otimes W_1f_1\otimes...\hat m...\otimes W_nf_n\Bigr)\notag\\
+i\sum_{m=1}^n(\pm) T_n\Bigl((\tilde\d_\mu W)\frac{\d W_m}{\d(\d_\mu\chi)}
gf_m\otimes W_1f_1\otimes...\hat m...\otimes W_nf_n\Bigr)\label{N4b}
\end{gather}
by proceeding analogously to (\ref{N4}), i.e. we have twice applied
($\tilde {\bf N}$). In the special case that no
derivatives of $\chi$ are present, the last two
terms on the r.h.s. vanish.

\subsection{Charge- and ghost-number conservation}

We consider massive or massless spinors $\psi,\psq\in {\cal P}_0$
fulfilling the Dirac equation and in particular the matter
current $j_\mu\=d\psq\gamma_\mu\psi$ (which is conserved).
We assume $W_1,...,W_n\in {\cal P}_0$ and that no derivatives of $\psi$
and $\psq$ are present.
{\bf Charge conservation} is expressed by the following Ward identity
{\bf (N5) (charge)} which is an immediate consequence of the master
Ward identity {\bf (N)}
\begin{gather}
-T_{n+1}(j_\mu\d^\mu g\otimes W_1f_1\otimes...\otimes W_nf_n)=
\notag\\
i\sum_{m=1}^n\Bigl[(\pm) T_n\Bigl(
\Delta^\mu_{\gamma_\mu\psi,\psq}
\bigl(\frac{\d j_\mu}{\d(\gamma_\mu\psi)}g,\frac{\d W_m}{\d\psq}f_m\bigr)
\otimes W_1f_1\otimes...\hat m...\otimes W_nf_n\Bigr)\notag\\
+(\pm) T_n\Bigl(\Delta^\mu_{\psq\gamma_\mu,\psi}
\bigl(\frac{\d j_\mu}{\d(\psq\gamma_\mu)}g,\frac{\d W_m}{\d\psi}f_m\bigr)
\otimes W_1f_1\otimes...\hat m...\otimes W_nf_n\Bigr)\Bigr]=\notag\\
\sum_{m=1}^n T_n\Bigl(W_1f_1\otimes ...\otimes
(\psq\frac{\d W_m}{\d\psq}-\psi\frac{\d W_m}{\d\psi})gf_m\otimes
...\otimes W_nf_n\Bigr).\label{N5}
\end{gather}
In the second step we have used the
formulas (\ref{delta:psi-psq})-(\ref{delta:psq-psi}) for $\delta^\mu$.
Each monomial $W$ is an eigenvector of the operator
$(\psq\frac{\d}{\d\psq}-\psi\frac{\d}{\d\psi})$ with eigenvalue:
(number of $\psq$ in $W$) minus (number of $\psi$ in $W$),
which we call 'spinor charge'.
That this Ward identity can be satisfied by choosing suitable
normalizations which are compatible with {\bf (N0)-(N4)}
has been proved in \cite{DF} for the case that $W_1,...,W_n$ are
sub-monomials of the QED-interaction ${\cal L}=A^\mu\psq\gamma_\mu\psi$.\\
\\
We turn to models which contain pairs $(\tilde u_a,\,u_a)$ of
massive or massless, scalar, but {\bf fermionic ghost fields}, e.g.
non-Abelian gauge theories (see appendix A for the
anti-commutators and Feynman propagators of the free ghost fields
$\tilde u_a,\,u_a\in {\cal P}_0$.) The free ghost current
\begin{equation}
k^\mu=i\sum_a [u_a\d^\mu\tilde u_a-\d^\mu u_a\tilde u_a]\label{ghost-current}
\end{equation}
is conserved, because $u_a,\,\tilde u_a$ satisfy the
Klein-Gordon or wave equation. Let
$W_1,...,W_n\in {\cal P}_0$ and we assume that only zeroth and first
(internal) derivatives of $u_a$ and $\tilde u_a$
appear in $W_1,...,W_n$. Similarly
to (\ref{N5}) the MWI {\bf (N)} implies the following
Ward identity {\bf (N5) (ghost)}:
\begin{gather}
-T_{n+1}(k_\mu\d^\mu h\otimes W_1f_1\otimes...\otimes W_nf_n)=
\notag\\
\sum_{m=1}^n T_n\Bigl(W_1f_1\otimes ...\otimes
\Bigl[\Bigl(u_a\frac{\d W_m}{\d u_a}-C_{u_a} \d^\mu u_a\frac{\d W_m}{\d
(\d^\mu u_a)}\notag\\
-\tilde u_a\frac{\d W_m}{\d \tilde u_a}+C_{u_a} \d^\mu \tilde
u_a\frac{\d W_m}{\d (\d^\mu \tilde u_a)}\Bigr)hf_m\notag\\
+(1+C_{u_a})\Bigl((\tilde\d^\mu u_a)\frac{\d W_m}{\d (\d^\mu u_a)}hf_m
+u_a\frac{\d W_m}{\d (\d^\mu u_a)}(\d^\mu h)f_m\notag\\
-(\tilde\d^\mu \tilde u_a)\frac{\d W_m}{\d (\d^\mu \tilde u_a)}hf_m
-\tilde u_a\frac{\d W_m}{\d (\d^\mu \tilde u_a)}(\d^\mu h)f_m
\Bigr)\Bigr]\otimes ...\otimes W_nf_n\Bigr),\label{N5:ghost}
\end{gather}
where the normalization constant $C$ appearing in (\ref{delta:0-1}),
(\ref{delta:1-1}) is specified by a lower index $u_a$. Every monomial $W$
is an eigenvector of the operator
\begin{equation}
\Theta_g\=d
u_a\frac{\d}{\d u_a}+(\d^\mu u_a)\frac{\d}{\d (\d^\mu u_a)}
-\tilde u_a\frac{\d}{\d \tilde u_a}-(\d^\mu \tilde
u_a)\frac{\d}{\d (\d^\mu \tilde u_a)}
\end{equation}
and the eigenvalue is the ghost number $g(W)$:
\begin{equation}
\Theta_g W=g(W) W,\quad\quad g(W)\in\ZZ.\label{ghostnr}
\end{equation}
The identity (\ref{N5:ghost})
expresses {\bf ghost number conservation} correctly if and only if
\begin{equation}
C_{u_a}=-1,\quad\quad\forall a.\label{C_u}
\end{equation}
With this normalization {\bf (N5) (ghost)} takes the form
\begin{gather}
-T_{n+1}(k_\mu\d^\mu h\otimes W_1f_1\otimes...\otimes W_nf_n)=
\notag\\
\sum_{m=1}^n g(W_m)T_n\Bigl(W_1f_1\otimes ...\otimes W_mf_mh\otimes...
\otimes W_nf_n\Bigr)\label{N5:ghost'}
\end{gather}
for monomials $W_1,...,W_m\in {\cal P}_0$.
That the normalization condition {\bf (N5) (ghost)} (with $C_{u_a}=-1$)
has common solutions with {\bf (N0)-(N4)} has been proved in \cite{BDF}
by using the method of \cite{DF} appendix B. (A slight restriction on
$W_1,...,W_n$ is used in that proof).

{\it Remark}: The (free) ghost charge $Q_g$ is defined by
\begin{equation}
Q_g\=d\int_{x^0={\rm const.}}d^3 x\,T_1(k^0(x)).\label{Q_g}
\end{equation}
{\bf (N5) (ghost)} implies the identity
\begin{equation}
[Q_g,T_n\Bigl(W_1f_1\otimes ...\otimes W_nf_n\Bigr)]=
\Bigl(\sum_{m=1}^n g(W_m)\Bigr)T_n\Bigl(W_1f_1\otimes ...
\otimes W_nf_n\Bigr)\label{[Q_g,T]}
\end{equation}
as can be seen by a suitable choice of the test-function $h$ in
(\ref{N5:ghost'}). For the details of this conclusion as well as for the
existence of $Q_g$ see the corresponding procedure (\ref{Q_0})-(\ref{j-W})
for the free BRST-current.

\subsection{Non-Abelian matter currents}

The aim of this subsection is to derive the identity (\ref{IZ}) from the
MWI. Let
\begin{equation}
j^\mu_a\=d \psq_\alpha\gamma^\mu\frac{(\lambda_a)_{\alpha\beta}}{2}
\psi_\beta
\end{equation}
(we use matrix notation for the spinor structur) and
\begin{equation}
[\lambda_a,\lambda_b]=2if_{abc}\lambda_c,
\end{equation}
where $(f_{abc})_{a,b,c}$ are the structure constants of some Lie algebra.
We assume that the masses of the spinor fields are colour independent
\begin{equation}
(i\gamma_\mu\d^\mu -m)\psi_\alpha =0,\quad\quad\forall\alpha,
\end{equation}
which implies
\begin{equation}
\d_\mu j^\mu_a=0.\label{div(j_a)}
\end{equation}
We denote by $(A_a)_a$ the gauge fields and by $(u_a,\tilde u_a)_a$ the
corresponding fermionic ghost fields, and consider an interaction of the
form
\begin{equation}
{\cal L}=j^\mu_a A_{a\mu}+{\cal L}_1(A,u,\tilde u),
\end{equation}
where ${\cal L}_1(A,u,\tilde u)$ is a polynomial in the symbols $A,u,\tilde u$
and internal derivatives thereof. QCD fits in this framework: the quark
fields $\psi_\alpha$ are in the fundamental representation of $SU(3)$.

To apply the MWI we need
\begin{equation}
i\sum_{\chi,\varphi}\Delta^\mu_{\chi,\varphi}
\Bigl(\frac{\d j_{a\mu}}{\d\chi}f,
\frac{\d j_{b}^\nu}{\d\varphi}h\Bigr)=\frac{fh}{4}\psq\gamma^\nu
[\lambda_a,\lambda_b]\psi =ifh f_{abc}j_c^\nu\label{j-j}
\end{equation}
((\ref{delta:psi-psq}) and (\ref{delta:psq-psi}) are used), and by
contracting with $A_{b\nu}$ we obtain $i\sum\Delta^\mu_{\chi,\varphi}
\Bigl(\frac{\d j_{a\mu}}{\d\chi}f,
\frac{\d {\cal L}}{\d\varphi}g\Bigr)$. So the MWI
for $T(g{\cal L}\otimes...\otimes g{\cal L}\otimes j^\mu_a\d_\mu f)$ implies
\begin{equation}
-R_{n+1}((g{\cal L})^{\otimes n};j^\mu_a\d_\mu f)=in R_{n+1}
((g{\cal L})^{\otimes (n-1)};f_{abc}A_{b\nu}j_c^\nu fg),\label{R(j)}
\end{equation}
and hence
\begin{equation}
j^\mu_{a\,g{\cal L}}(\d_\mu f)=(f_{abc}A_{b\nu}j_c^\nu)_{g{\cal L}}(fg),
\label{div(j_gL)}
\end{equation}
which corresponds to the covariant conservation of the interacting
classical current.

To formulate (\ref{IZ}) we need the time-ordered product
$T_{g{\cal L}}(W_1f_1\otimes...\otimes W_mf_m)$ of the interacting fields
$W_{1\,g{\cal L}}(f_1),...,W_{m\,g{\cal L}}(f_m)$, which is defined
by generalizing (\ref{intfield}) (cf. \cite{BS}, \cite{EG})
\begin{gather}
T_{g{\cal L}}(W_1f_1\otimes...\otimes W_mf_m)\=d\notag\\
S(g{\cal L})\frac{d^m}{i^md\lambda_1...d\lambda_m}\vert_{\lambda_1=
...=\lambda_m=0}S(g{\cal L}+\sum_{l=1}^m\lambda_l W_lf_l)=\notag\\
\sum_{n=0}^\infty\frac{i^n}{n!}R_{n,m}((g{\cal L})^{\otimes n};
W_1f_1\otimes...\otimes W_mf_m)\label{T(intfield)}
\end{gather}
with\footnote{The connection to the notation (\ref{R}) reads:
$R_{n,1}\equiv R_{n+1}$.}
\begin{gather}
R_{n,m}(g_1V_1\otimes...\otimes g_nV_n;W_1f_1\otimes...\otimes W_mf_m)\=d
\notag\\
\sum_{I\subset\{1,...,n\}}(-1)^{|I|}\bar{T}(\otimes_{l\in I}g_lV_l)
T((\otimes_{j\in I^c}g_jV_j)\otimes (\otimes_{k=1}^m f_kW_k)).
\end{gather}
By using (\ref{div(j_a)}) and (\ref{j-j}) the MWI yields
\begin{gather}
-R_{n,2}((g{\cal L})^{\otimes n};j^\mu_a\d_\mu f\otimes j^\nu_bh)=
R_{n,1}((g{\cal L})^{\otimes n};if_{abc}j_c^\nu fh)\notag\\
+inR_{n-1,2}((g{\cal L})^{\otimes (n-1)};f_{acd}A_{c\tau}j_d^\tau fg
\otimes j^\nu_bh)\label{R(jj)}
\end{gather}
which gives
\begin{equation}
-T_{g{\cal L}}(j^\mu_a\d_\mu f\otimes j^\nu_bh)=
if_{abc}j^\nu_{c\,g{\cal L}}(fh)
-T_{g{\cal L}}(f_{acd}A_{c\tau}j_d^\tau fg\otimes j^\nu_bh).
\end{equation}
Due to (\ref{div(j_gL)}) this is the formulation of (\ref{IZ}) in the
framework of causal perturbation theory. In the simple case that the
gauge fields $A_a$ are external fields (which implies
${\cal L}_1(A,u,\tilde u)\equiv 0$) and the
spinor fields are massive ($m>0$), the proof of
sect. 3.3 applies, i.e. the central extensions fulfil
the MWI. (Note that no factor $m$ appears in
(\ref{R(j)}) and (\ref{R(jj)}), which indicates that the scaling
degree is not lowered. )

\subsection{The master BRST-identity}

We consider free gauge fields $A^\mu_a,\>a=1,...,N$,
with mass $m_a\geq 0$ in {\it Feynman gauge} and the corresponding
free ghost fields $\tilde u_a,\,u_a$ with the
same mass $m_a$. For each fixed value of $a$ and $\mu$ the field $A^\mu_a$
is quantized as a real scalar field satisfying the
Klein-Gordon or wave equation, i.e. in the formalism of appendix A we
set $\varphi =A^\mu_a=\chi,\>\epsilon=1$. The free ghost fields
fulfil the same algebraic relations as in sect. 4.2 and in
appendix A. For each massive gauge field $A^\mu_a$,
$m_a>0$, we introduce a free, real scalar field $\phi_a$ with the
same mass $m_a$, which is quantized
with a minus sign in the commutator, i.e. we have $\varphi =\phi_a
=\chi,\>\epsilon=-1$ in the formalism of appendix A. (For the
Fock space representation of these free fields see e.g. \cite{S-wiley}.)
There is no obstacle to include spinor fields in our treatment of
BRST-symmetry (sects. 4.4 and 4.5), see \cite{DF}, the last paper of
of \cite{DHKS}, \cite{DS} and \cite{G}.

The free BRST-current (cf. \cite{K},\cite{DS})
\begin{equation}
j^\mu \=d\sum_a[(\d_\tau A^\tau_a+m_a\phi_a)\d^\mu u_a-
\d^\mu(\d_\tau A^\tau_a+m_a\phi_a) u_a]
\label{BRST-current}
\end{equation}
is conserved, because $\d_\tau A^\tau_a,\,u_a$ and $\phi_a$
fulfill the Klein-Gordon equation with the same mass $m_a$.
We will see that the corresponding charge
\begin{equation}
Q_0\=d\int_{x^0={\rm const.}}d^3 x\,T_1(j^0)(x),\label{Q_0}
\end{equation}
is the generator of the BRST-transformation of the free fields and
Wick monomials. $Q_0$ is nilpotent,
\begin{equation}
2Q_0^2=[Q_0,Q_0]_+=0,\label{Q_0^2=0}
\end{equation}
because $[(\d_\tau A^\tau_a+m_a\phi_a),(\d_\rho A^\rho_b+m_b\phi_b)]=0$.
Without the scalar fields $\phi_a$ the charge $Q_0$ would not be nilpotent,
if some gauge fields are massive. So, a main purpose of the the scalar
fields $\phi_a$ is to restore the nilpotency of $Q_0$.
(For a rigorous definition of $Q_0$, with 4-dimensional smearing with
a test function and taking a suitable limit, see \cite{DF} where a method
of Requardt \cite{R} is used.)

To obtain the master BRST-identity (i.e. the
(anti)commutator of $Q_0$ with arbitrary $T$-products (\ref{MBRST}))
we will compute
\begin{equation}
T_{n+1}(j_\mu\d^\mu g\otimes W_1f_1\otimes...\otimes W_nf_n),
\quad\quad W_1,...,W_n\in {\cal P}_0,\label{divT(j)}
\end{equation}
by means of the MWI {\bf (N)}. Thereby we assume that $W_1,...,W_n$
have an even or odd ghost number (no mixture). From this result we
shall get $[Q_0,T(W_1,...,W_n)]_\mp$ in the following way:
let ${\cal O}$ be an open double cone
with ${\rm supp}\>f_j\subset {\cal O},\>\forall j=1,...,n$. Following
\cite{DF} (appendix B) we choose $g$
to be equal to 1 on a neighbourhood of $\overline{\cal O}$ and
decompose $\partial^{\mu} g=b^{\mu}-a^{\mu}$
such that $\supp a^{\mu}\cap (\overline{V}_-+{\cal O})=\emptyset$ and
$\supp b^{\mu}\cap (\overline{V}_++{\cal O})=\emptyset$. Then we apply
causal factorization of the $T$-products:
\begin{gather}
-T_{n+1}(j_\mu(\d^\mu g)\otimes W_1f_1\otimes...\otimes W_nf_n)=\notag\\
T_1(j_\mu a^\mu)T_n(W_1f_1\otimes...\otimes W_nf_n)
\mp T_n(W_1f_1\otimes...\otimes W_nf_n)T_1(j_\mu b^\mu)=\notag\\
[T_1(j_\mu a^\mu),T_n(W_1f_1\otimes...\otimes W_nf_n)]_\mp
\mp T_n(W_1f_1\otimes...\otimes W_nf_n)T_1(j_\mu \d^\mu g).\label{j-W}
\end{gather}
The last term on the r.h.s. vanishes because of $\d^\mu j_\mu=0$. Since
$T_n(W_1f_1\otimes...\otimes W_nf_n)$ is localized in ${\cal O}$,
we may vary $a^\mu$ in the spatial
complement of $\overline{\cal O}$ without affecting $[T_1(j_\mu
a^\mu),T_n(W_1f_1\otimes...)]_\mp$. In this way and by using
$\d^\mu j_\mu=0$ we find
\begin{equation}
[T_1(j_\mu a^\mu),T_n(W_1f_1\otimes...\otimes W_nf_n)]_\mp
=[Q_0,T_n(W_1f_1\otimes...\otimes W_nf_n)]_\mp\label{QW}
\end{equation}
(see \cite{DF}, appendix B for details of this conclusion).

We start the computation of (\ref{divT(j)})
with the simplest case: $n=1$. We assume that the symbols
in $W$ carry at most a first (internal) derivative (no higher derivatives)
and give the calculation in detail
\begin{equation}
-T_2(j_\mu(\d^\mu g)\otimes Wf)=
i\sum_{\chi,\psi\in {\cal G}} T_1\Bigl(\Delta^\mu_{\chi,\psi}
\bigl(\frac{\d j_\mu}{\d\chi}g,\frac{\d W}{\d\psi}f\bigr)\Bigr).\label{j1}
\end{equation}
The explicit results for the $\Delta^\mu$ with a non-vanishing
contribution are listed in appendix B.
Thereby $C_{A_a}\>(C_{1A_a}$ resp.), $C_{\phi_a}$ and $C_{u_a}$ mean the
normalization constants $C\>\> (C_1$ resp.)
in the cases $\varphi =A^\mu_a=\chi,\>
\varphi =\phi_a=\chi$ and $\varphi=\tilde u_a,\>\chi =u_a$.
In the present context they may depend on the colour index $a$.
Inserting (\ref{j2})-(\ref{j9d}) into (\ref{j1}) we obtain
\begin{gather}
-T_2(j_\mu(\d^\mu g)\otimes Wf)=T_1\Bigl(s_0(W)gf\notag\\
+\Bigl[...\Bigr](\d^\nu g)f+\Bigl[...\Bigr](\d^\nu\d^\sigma g)f+
\Bigl[...\Bigr](\w g)f\Bigr)\label{j10}
\end{gather}
by means of $T_1((\tilde\d^a V)Wg)=T_1((\d^a V)Wg)$, where
\begin{gather}
  s_0(W)\=d (\d^\mu u_a)
\frac{\d W}{\d A^\mu_a}
+(\d^\sigma\d^\mu u_a)\frac{\d W}{\d (\d^\sigma A^\mu_a)}
-(\d_\tau A^\tau_a+m_a\phi_a)\frac{\d W}{\d\tilde u_a}\notag\\
-(\d_\nu(\d_\tau A^\tau_a+m_a\phi_a))\frac{\d W}{\d(\d_\nu\tilde u_a)}
+m_au_a\frac{\d W}{\d\phi_a}+m_a(\d_\mu u_a)
\frac{\d W}{\d(\d_\mu\phi_a)}.\label{s_0}
\end{gather}
(The terms which are not written out depend on the normalization constants
$C_{A_a},C_{\phi_a},C_{u_a}$ and $C_{1A_a}$.) Using
$gf=f$ and $(\d^a g)f=0,\forall |a|\geq 1$ we end up with
\begin{equation}
[Q_0,T_1(Wf)]_\mp = T_1(s_0(W)f),\label{Q-W}
\end{equation}
where we have the anti-commutator iff $W$ has an odd ghost number.
The normalization constants $C_{A_a},C_{\phi_a},C_{u_a}$ and $C_{1A_a}$
have dropped out on the r.h.s., as it must be since they do not appear
on the l.h.s. of (\ref{Q-W}).
The result (\ref{Q-W}) is the well-known free
BRST-transformation of a Wick polynomial $T_1(W)\rightarrow T_1(s_0(W))$
(cf. \cite{DHKS}) which we have obtained here with quite a
lot of calculations. Note that in our framework $s_0$ is a derivation
$s_0: {\cal P}_0\rightarrow {\cal P}_0$.

However, the advantage of the present method is that it can be used to
compute commutators of $Q_0$ with $T$-products of higher orders.
For $n=2$ in (\ref{divT(j)}) we obtain
\begin{gather}
-T_3(j_\mu(\d^\mu g)\otimes W_1f_1\otimes W_2f_2)=
i\sum_{\chi,\psi\in {\cal G}}\Bigl[T_2\Bigl(\Delta^\mu_{\chi,\psi}
\bigl(\frac{\d j_\mu}{\d\chi}g,\frac{\d W_1}{\d\psi}f_1\bigr)
\otimes W_2f_2\Bigr)\notag\\
+(\pm) [(W_1,f_1)\leftrightarrow (W_2,f_2)]\Bigr]\label{j11}
\end{gather}
where $(\pm)$ is still a sign coming from permutations of Fermi operators.
We insert the expressions (\ref{j2})-(\ref{j9d}) for the various
$\Delta^\mu$. For given $f_1,f_2$ we then choose $g$ as in
(\ref{j-W}), hence $gf_j=f_j$ and $(\d^a g)f_j=0,\forall |a|\geq 1$.
It results
\begin{gather}
[Q_0,T_2(W_1f_1\otimes W_2f_2)]_\mp =i\Bigl[T_2\Bigl(
\Bigl[\bigl(\frac{1}{4}(\d^\mu u_a)+\frac{3}{4}(\tilde\d^\mu u_a)\bigr)
\frac{\d W_1}{\d A^\mu_a}\notag\\
+[(C_{A_a}+\frac{1}{2}+2C_{1A_a})(\tilde\d^\mu\d_\mu u_a)-
(\frac{1}{2}+2C_{1A_a})
\tilde{\w}u_a+C_{A_a}m_a^2 u_a]\frac{\d W_1}{\d (\d_\nu A^\nu_a)}\notag\\
+[-C_{1A_a}(\tilde\d^\sigma\d^\nu u_a)+(\frac{1}{2}+2C_{1A_a})
(\tilde\d^\nu\d^\sigma u_a)+
(\frac{1}{2}-C_{1A_a})(\tilde\d^\nu\tilde\d^\sigma u_a)]
\frac{\d W_1}{\d (\d^\sigma A^\nu_a)}\notag\\
-(\d_\tau A^\tau_a+m_a\phi_a)\frac{\d W_1}{\d \tilde u_a}
+[-(1+C_{u_a})(\tilde\d_\nu(\d_\tau A^\tau_a+m_a\phi_a)\notag\\
+C_{u_a}(\d_\nu(\d_\tau A^\tau_a+m_a\phi_a))]
\frac{\d W_1}{\d (\d_\nu\tilde u_a)}\notag\\
+m_au_a\frac{\d W_1}{\d\phi_a}+m_a[(1+C_{\phi_a})\tilde\d_\nu u_a-
C_{\phi_a}\d_\nu u_a]\frac{\d W_1}{\d(\d_\nu\phi_a)}\Bigr]f_1
\otimes W_2f_2\Bigr)\notag\\
+(\pm) [(W_1,f_1)\leftrightarrow (W_2,f_2)]\Bigr].\label{j12}
\end{gather}
To simplify this expression we insert the value $C_{u_a}=-1$ which is required
from ghost number conservation (\ref{N5:ghost})-(\ref{C_u}).
By means of $(\tilde {\rm\bf N})$ we replace the external derivatives by
internal ones
\begin{eqnarray}
[Q_0,T_2(W_1f_1\otimes W_2f_2)]_\mp =\Bigl[T_2\Bigl(s_0(W_1)f_1
\otimes W_2f_2\Bigr)\nonumber\\
+T_1\Bigl(G^{(1)}(W_1f_1, W_2f_2)\Bigr)\Bigr]
+(\pm)\Bigl[ (W_1,f_1)\leftrightarrow (W_2,f_2)\Bigr],\label{j13}
\end{eqnarray}
where
\begin{gather}
G^{(1)}(W_1f_1, W_2f_2)\=d
-\sum_{\psi\in {\cal G}}\Bigl[\frac{3}{4}
\Delta^\mu_{u_a,\psi}\bigl(\frac{\d W_1}{\d A^\mu_a}f_1,
\frac{\d W_2}{\d\psi}f_2\bigr)\notag\\
+C_{A_a}\Delta^\mu_{\d_\mu u_a,\psi}
\bigl(\frac{\d W_1}{\d (\d_\nu A^\nu_a)}
f_1,\frac{\d W_2}{\d\psi}f_2\bigr)\notag\\
-(\frac{1}{2}+2C_{1A_a})\Delta^\mu_{\tilde\d_\mu u_a,\psi}
\bigl(\frac{\d W_1}{\d (\d_\nu A^\nu_a)}f_1,\frac{\d W_2}{\d\psi}f_2\bigr)
\notag\\
+(\frac{1}{2}-2C_{1A_a})\Delta^{\sigma}_{\d^\nu u_a,\psi}\bigl(\frac{\d W_1}
{\d (\d^\sigma A^\nu_a)}f_1,\frac{\d W_2}{\d\psi}f_2\bigr)\notag\\
+(\frac{1}{2}+2C_{1A_a})\Delta^{\nu}_{\d^\sigma u_a,\psi}\bigl(\frac{\d W_1}
{\d (\d^\sigma A^\nu_a)}f_1,\frac{\d W_2}{\d\psi}f_2\bigr)\notag\\
+(\frac{1}{2}-C_{1A_a})\Delta^{\nu}_{\tilde\d^\sigma u_a,\psi}
\bigl(\frac{\d W_1}{\d (\d^\sigma A^\nu_a)}f_1,
\frac{\d W_2}{\d\psi}f_2\bigr)\notag\\
+m_a(1+C_{\phi_a})\Delta^{\nu}_{u_a,\psi}\bigl(\frac{\d W_1}
{\d (\d^\nu\phi_a)}f_1,\frac{\d W_2}{\d\psi}f_2\bigr)\Bigr].\label{j14}
\end{gather}
Note that $G^{(1)}(\cdot,\cdot)$ is not invariant
with respect to the exchange
of the two arguments. Now we assume that $s_0(W_j)$ is a divergence,
i.e. that there exists a (Lorentz) vector $(W'_{j\nu})_{\nu=0,...,3},
\> W'_{j\nu}\in\mathcal{P}_0$ with
\begin{equation}
s_0(W_j)=i\d^\nu W'_{j\nu},\quad j=1,2.\label{div}
\end{equation}
By means of the MWI {\bf (N)} we shift this derivative to the
test-function
\begin{gather}
[Q_0,T_2(W_1f_1\otimes W_2f_2)]_\mp =\Bigl[-iT_2\Bigl(W'_{1\nu}\d^\nu f_1
\otimes W_2f_2\Bigr)\notag\\
+T_1\Bigl(G((W_1,W'_{1})f_1, W_2f_2)\Bigr)\Bigr]
+(\pm)\Bigl[ (W_1,W'_{1},f_1)\leftrightarrow (W_2,
W'_{2},f_2)\Bigr],\label{j15}
\end{gather}
where
\begin{equation}
G((W_1,W'_{1})f_1, W_2f_2)=G^{(1)}(W_1f_1, W_2f_2)+
G^{(2)}(W'_{1}f_1, W_2f_2),\label{j16}
\end{equation}
with
\begin{equation}
G^{(2)}(W'_{1}f_1, W_2f_2)\=d
\sum_{\chi,\psi\in {\cal G}} \Delta^{\nu}_{\chi,\psi}
\bigl(\frac{\d W'_{1\nu}}{\d\chi}f_1,\frac{\d W_2}{\d\psi}f_2\bigr).
\label{j16a}
\end{equation}
If we only know that $s_0(W_1)$ is a divergence, our final result reads
\begin{gather}
[Q_0,T_2(W_1f_1\otimes W_2f_2)]_\mp =\notag\\
-iT_2\Bigl(W'_{1\nu}\d^\nu f_1
\otimes W_2f_2\Bigr)+T_1(G((W_1,W'_{1\nu})f_1, W_2f_2))\notag\\
+(\pm) \Bigl[T_2(s_0(W_2)f_2\otimes W_1f_1)+
T_1(G^{(1)}(W_2f_2, W_1f_1))\Bigr]
\label{j17}
\end{gather}
instead of (\ref{j15}).

The $(n=2)$-calculation generalizes to higher orders $n\geq 2$ in a
straightforward way: let $W_1,...,W_k,V_1,...,V_{n-k}\in{\cal P}_0$
with $s_0(W_j)=i\d^\nu W'_{j\nu},\>\forall j=1,...,k$ and $f_j,h_i\in
{\cal D}(\RR^4)$. For simplicity we assume that each polynomial
$W_1,...,W_k,V_1,...,V_{n-k}$ has an even ghost number,
otherwise some additional, obvious signs appear in the following formula.
Setting $m\=d n-k$ we obtain
\begin{gather}
[Q_0,T_n(W_1f_1\otimes ...\otimes W_kf_k\otimes V_1h_1\otimes ...
\otimes V_mh_m)]=\notag\\
=-i\sum_{l=1}^kT_n(W_1f_1\otimes ...\otimes W'_{l\nu}\d^\nu f_l\otimes ...
\otimes W_kf_k\otimes V_1h_1\otimes ...)\notag\\
+\sum_{l=1}^mT_n(W_1f_1\otimes ...\otimes V_1h_1\otimes
...\otimes  s_0(V_l)h_l\otimes ...\otimes V_mh_m)\notag\\
+\sum_{l,r=1\>(l\not= r)}^k T_{n-1}(G((W_l,W'_{l})f_l, W_rf_r)\otimes
W_1f_1\otimes ...\hat l...\hat r...\otimes W_kf_k\otimes V_1h_1\otimes ...)
\notag\\
+\sum_{l=1}^k\sum_{r=1}^m[ T_{n-1}(G((W_l,W'_{l})f_l, V_rh_r)\otimes
W_1f_1\otimes ...\hat l...\otimes V_1h_1\otimes ...\hat r...)\notag\\
+ T_{n-1}(G^{(1)}(V_rh_r,W_lf_l)\otimes
W_1f_1\otimes ...\hat l...\otimes V_1h_1\otimes ...\hat r...)]\notag\\
+\sum_{l,r=1\>(l\not= r)}^m T_{n-1}(G^{(1)}(V_lh_l,V_rh_r)\otimes
W_1f_1\otimes ...\otimes V_1h_1\otimes ...\hat l...\hat r...\otimes
V_mh_m),\label{j19}
\end{gather}
where $\hat l$ or $\hat r$ means that the corresponding factor is omitted.
We call this equation the {\bf 'master BRST-identity'}. It is a consequence
of the master Ward identity. Hence, the master BRST-identity is also
a normalization condition. We point out that the $G$-terms (i.e. the
terms in the last four lines) are explicitly known. We are not aware
of any reference which gives a general formula for these terms, as it
is done here.

So far we have not spoken about the interaction ${\cal L}\equiv
{\cal L}_0$; the master BRST-identity is a condition on $T$-products of
arbitrary factors. Now we require that the interaction
is $s_0$-invariant
in some sense. The requirement $s_0{\cal L}_0=0$ is too restrictive, it is
not satisfied for physically relevant models. So we impose the weaker
condition that $s_0{\cal L}_0$ is a divergence:
\begin{equation}
s_0 {\cal L}_0=i\d^\nu {\cal L}_{1\nu}.\label{L:div}
\end{equation}
The requirements that

(a) the master BRST-identity becomes particularly simple, and

(b) can be satisfied to all orders\\
for $T$-products involving the interaction are good criterions (among others)
to restrict ${\cal L}_0$ further. We will make (a) explicit
by the formula
\begin{equation}
  G(({\cal L}_0,{\cal L}_1)f,{\cal L}_0g)+(f\leftrightarrow g)=0.
\label{G(L)=0}
\end{equation}
(b) means that anomalies (in the master BRST-identity) may not
occur or must cancel. It is a hard job to work this out.
For example it is well-known that in weak interactions the axial
anomalies cancel only if the numbers of generations for leptons and quarks
agree.

For an interaction $\mathcal{L}$ fulfilling (\ref{L:div}) and
(\ref{G(L)=0}) the validity of the master
BRST-identity for $[Q_0,T_n(\mathcal{L},...,\mathcal{L})]$ $\forall
n\in\NN$ implies
\begin{equation}
[Q_0,S]=0, \label{[Q,S]}
\end{equation}
where $S$ is the $S$-matrix in the adiabatic limit,
\begin{equation}
  S\=d\lim_{g\to 1}S(g\mathcal{L}),\label{adlim(S)}
\end{equation}
provided this limit exists \cite{EG1}. Hence, $S$ induces a
well-defined operator on the physical Hilbert space ${\cal H}_{\rm phys}=
\frac {{\rm ker}\>Q_0}{{\rm ran}\>Q_0}$, which is unitary if
${\cal L}_0^*={\cal L}_0$ and {\bf (N2)} is satisfied \cite{EG1},
\cite{DS}, \cite{DSchroer}, \cite{G}.

Having determined the interaction by using
(\ref{L:div}), (\ref{G(L)=0}) and other (quite obvious) requirements,
we will show that the validity of the master BRST-identity and
of the ghost number conservation {\bf (N5) (ghost)} suffices for
a local construction of observables in non-Abelian gauge theories.
This is a generalization of the corresponding construction for QED
in \cite{DF}. In particular we will obtain an explicit formula for
the computation of the nonlinear term in the BRST-transformation
of an arbitrary interacting field.

\subsection{Local construction of observables in gauge theories}

For massive gauge fields the procedure is more involved. So we first
treat massless gauge fields and afterwards give the modifications for the
massive case.

\subsubsection{Massless gauge fields: determination of the interaction}

Since we are considering solely massless
fields ($m_a=0,\>\forall a$), the scalar fields
$\phi_a$ are superfluous. So we set $\phi_a\equiv 0,\>\forall a$.

First we {\bf determine the interaction} ${\cal L}_0$ by the following
requirements (cf. \cite{stora}, \cite{D1}, \cite{G}, \cite{DSchroer}
and \cite{S-wiley}):

(A) There exist ${\cal L}_j\in ({\cal P}_0)^{4^j},\>
j=0,1,...,M$ which
satisfy the ladder equations
\begin{equation}
s_0{\cal L}_j^{\mu_1...\mu_j}=i\d_{\mu_{j+1}}{\cal L}_{j+1}^
{\mu_1...\mu_j\mu_{j+1}},\quad j=0,1,...,M-1,\quad s_0{\cal L}_M=0
\label{leiter}
\end{equation}

(B) ${\cal L}_j$ is a polynomial in the gauge field $A_a^\mu$
and in the fermionic ghost fields $u_a,
\tilde u_a,\> a=1,...,N,$ and internal derivatives of these symbols;
each monomial has at least three factors.

(C) ${\cal L}_j$ has UV-dimension $\leq 4$.

(D) ${\cal L}_j$ is a Lorentz tensor of rank $j$.

(E) ${\cal L}_j$ has ghost number $j$:
\begin{equation}
g({\cal L}_j)=j\label{g(L)}
\end{equation}
(cf. (\ref{ghostnr})). Thereby we take into account that
$s_0$ increases the ghost number by $1$. We conclude that the ladder
(\ref{leiter}) stops at $M\leq 3$ for trilinear terms.

(F) unitarity (for  ${\cal L}_0$ only): ${\cal L}_0^*={\cal L}_0$.\\
\\
Following \cite{stora}
we make the most general ansatz for ${\cal L}_j,\>j=0,1,2,3$
which satisfies (B)-(F) and insert it into (\ref{leiter}).
The calculation excludes quadrilinear terms in ${\cal L}_0$. Using
$F^{\nu\mu}\=d\d^\nu A^\mu-\d^\mu A^\nu$ the most general solution
for ${\cal L}_0$ reads
\begin{equation}
{\cal L}_0=g_0f_{abc}[\frac{1}{2}A_{a\mu}A_{b\nu}F^{\nu\mu}_c-
u_a\d^\mu\tilde u_bA_{c\mu}]-is_0K_1+\d_\nu K_2^\nu,\label{L_0}
\end{equation}
where $f_{abc}$ must be totally antisymmetric and $g_0\in\RR$
is a constant. This implies that the
colour index takes at least $N\geq 3$ values. The $K_j$ are trilinear
polynomials with ghost number $(j-2)$. We assume that the colour tensor
in $K_j, j=1,2$ is also totally antisymmetric (i.e.
$K_j=h^j_{abc}\varphi^{(1)}_a\varphi^{(2)}_b\varphi^{(3)}_c$
with a totally antisymmetric $h^j$). Then one finds
\begin{equation}
K_1=g_0h^1_{abc}u_a\tilde u_b\tilde u_c,\quad\quad
K_2^\mu =g_0h^2_{abc}A^\mu_au_b\tilde u_c.\label{K_1,2}
\end{equation}
The most general solutions for ${\cal L}_j,j\geq 1$ contain
trilinear terms only. Assuming again that solely totally antisymmetric
colour tensors appear, we obtain
\begin{gather}
{\cal L}_1^\nu=g_0f_{abc}[A_{a\mu}u_bF^{\nu\mu}_c-\frac{1}{2}u_au_b\d^\nu
\tilde u_c]-is_0K_2^\nu +g_0h^3_{abc}\d_\mu (u_aA^\nu_bA^\mu_c),\notag\\
{\cal L}_2^{\nu\mu}=g_0\frac{1}{2}f_{abc}u_au_bF^{\nu\mu}_c
-ig_0s_0(h^3_{abc}u_aA^\nu_bA^\mu_c)\notag\\
+g_0h^4_{abc}[u_au_b\d^\nu A^\mu_c+2u_a\d^\nu u_bA^\mu_c-
g^{\nu\mu}(u_au_b\d_\lambda A^\lambda_c-2u_a\d_\lambda u_b A^\lambda_c)]
\notag\\
{\cal L}_3^{\nu\mu\lambda}=g_0h^4_{abc}[g^{\nu\mu}u_au_b\d^\lambda u_c
+g^{\lambda\nu}u_au_b\d^\mu u_c],\label{L_1,2,3}
\end{gather}
where $h^3,h^4$ are totally antisymmetric. Note that the divergence
$\d_\mu$ of the $h^4$-term in ${\cal L}_2^{\nu\mu}$ vanishes.
To simplify the formulas we choose
\begin{equation}
h^j_{abc}=0,\quad\forall j=1,2,3,4.\label{h=0}
\end{equation}
$h^4=0$ is equivalent to ${\cal L}_2^{\mu\nu}=-{\cal L}_2^{\nu\mu}$ and
also to ${\cal L}_3=0$.

The requirements (A)-(F) used so far do not involve $T$-products,
they are of first order perturbation theory. We now restrict
$\mathcal{L}_0$ further by (\ref{G(L)=0}), which can be interpreted as
a requirement for second order tree diagrams, see (\ref{G=T}). More
precisely, we will work with the generalization of (\ref{G(L)=0})
to the ladder (\ref{leiter}): in order that
the master BRST-identity (\ref{j19}) implies the important equation
({\bf 'generalized perturbative gauge invariance'}\footnote{In
\cite{BDF} this identity is called the normalization condition
{\bf (N6)}, in \cite{DSchroer} it is called 'generalized (free
perturbative operator) gauge invariance'.
The importance and usefulness of this identity has also been pointed out
in the earlier paper \cite{D1}.
The particular case $j_1=...=j_n=0$ is the 'perturbative gauge invariance'
(or 'free perturbative operator gauge invariance') which has been proved
in \cite{DHKS} for $SU(N)$-Yang-Mills theories. In \cite{D} it has been shown
that this perturbative gauge invariance implies the usual
Slavnov-Taylor identities.})
\begin{gather}
[Q_0,T_n({\cal L}_{j_1}f_1\otimes...\otimes {\cal L}_{j_n}f_n)]_\mp=
\notag\\
-i\sum_{l=1}^n(-1)^{j_1+...+j_{l-1}}
T_n({\cal L}_{j_1}f_1\otimes...\otimes {\cal L}_{j_l+1}^\nu
\d_\nu f_l\otimes ...\otimes {\cal L}_{j_n}f_n),\label{N6}
\end{gather}
we require
\begin{equation}
G(({\cal L}_j,{\cal L}_{j+1})f, {\cal L}_kg)
+(-1)^{jk}G(({\cal L}_k,{\cal L}_{k+1})g, {\cal L}_jf)=0,\quad\quad
j,k=0,1,2,3,\label{G=0}
\end{equation}
$\forall f,g\in {\cal D}(\RR^4)$,
where we set ${\cal L}_4\equiv 0$. Or, with the simplification (\ref{h=0})
(which will always be used in the following),
$j$ and $k$ run only through the values $j,k=0,1,2$. In the present
case of solely massless fields this requirement can be fulfilled.
It restricts the interaction ${\cal L}_0$ further and
determines the normalization constant $C_{A_a}$. Namely, using the
simplification (\ref{h=0}), one finds by explicit calculation that
the requirement (\ref{G=0}) holds true if and only if
\begin{equation}
C_{A_a}=-\frac{1}{2},\quad\quad\forall a,\label{C_A}
\end{equation}
and the $f_{abc}$ fulfil the {\it Jacobi identity}
\cite{stora}.\footnote{The Jacobi
identity and (\ref{C_A}) are required even from the particular case
$G(({\cal L}_0,{\cal L}_1)f, {\cal L}_0g)+(f\leftrightarrow g)=0$.
This was demonstrated in \cite{stora} by reversing the calculation in
\cite{DHKS}. The computation of the l.h.s. of (\ref{G=0}) is lengthy.
The straightforward way uses the definitions (\ref{j14}) and (\ref{j16a})
of $G^{(1)}$ and $G^{(2)}$. To shorten the calculation one may choose
$C_{1A_a}=0=C_{1u_a}$, because the terms $\sim C_{1A_a},C_{1u_a}$
must drop out. This follows from the fact that ${\cal L}_0$ (\ref{L_0}),
${\cal L}_1$ and ${\cal L}_2$ (\ref{L_1,2,3}) do not contain symbols
with second or higher derivatives and, hence, the r.h.s. in
\begin{gather}
T_1\Bigl( G(({\cal L}_j,{\cal L}_{j+1})f, {\cal L}_kg)
+(-1)^{jk}\{(j,f)\leftrightarrow (k,g)\}\Bigr) =\notag\\
[ Q_0,T_2({\cal L}_jf\otimes {\cal L}_kg) ]\vert_{\rm 4-legs} +i\Bigl(
T_2({\cal L}_{j+1}^\nu\d_\nu f\otimes {\cal L}_k g)+
(-1)^{jk}\{(j,f)\leftrightarrow (k,g)\}\Bigr)\vert_{\rm 4-legs}\label{G=T}
\end{gather}
(cf. (\ref{j15}), $...\vert_{\rm 4-legs}$ expresses that we
mean the terms with 4 free field operators only)
does not contain the constants $C_{1A_a}$ and $C_{1u_a}$
(according to the definition(\ref{2-1})). However, even
with this simplification, it seems to be faster to compute the r.h.s.
of (\ref{G=T}) (by using the techniques of \cite{DHKS}, \cite{DS}),
instead of the straightforward computation of the l.h.s.. Thereby,
the {\it $T$-products of second order must fulfill the normalization
  condition} {\bf (N3)}, because the MWI presupposes this condition.
Note that the derivation of (\ref{G=T}) uses the MWI for
tree diagrams only and, hence, (\ref{G=T}) holds surely true.}
Hence, the $f_{abc}$ are structure constants of some Lie algebra.
The total antisymmetry of $f_{abc}$ implies that this Lie algebra is
isomorphic to a direct sum of Abelian and simple compact Lie algebras,
see e.g. \cite{S-wiley}. We point out, that the Lie
algebraic structure is not put in, it is a consequence of our requirements.

{\it Remarks}: (1) If we do not use the simplification (\ref{h=0}), but
assume that $K_1$ and $K_2$ are built up from the same colour tensor
$f_{abc}$ as the first two terms in (\ref{L_0}),
\begin{equation}
K_1=-i\beta_1 g_0f_{abc}u_a \tilde u_b\tilde u_c,\quad
K_2^\mu=\beta_2 g_0f_{abc}A^\mu_au_b\tilde u_c,\quad
\beta_1,\beta_2\in \RR\label{K_1,K_2}
\end{equation}
(instead of (\ref{K_1,2})), we can determine the parameters $\beta_1$
and $\beta_2$ from the particular case $(j,k)=(0,0)$ of the
requirement (\ref{G=0}): additionally to (\ref{C_A}) and the Jacobi
identity this condition yields
\begin{equation}
(\beta_1,\beta_2)\in \{(0,0),\,({1\over 2},1),\,(-{1\over 2},0),\,(1,1)\}
\label{beta}
\end{equation}
by a generalization of the calculation (\ref{C_A}),
where $C_{u_a}=-1$ (\ref{C_u}) is used. Note that the
first two solutions in (\ref{beta}),
and also the latter two, are obtained
from each other by replacing $u$ by $\gamma\tilde u$ and $\tilde u$ by
$(-{1\over\gamma}u)$ in ${\cal L}_0$, $\gamma\in i\RR\setminus\{0\}$
arbitrary.

(2) In \cite{SQED} it was found how the quadrilinear interactions
of the usual Lagrangian formalism appear in our framework.
Namely, from (\ref{L_0}), (\ref{K_1,K_2}) and {\bf (N3)} it results
\begin{gather}
  T_2({\cal L}_0,{\cal L}_0)(x_1,x_2)=f_{abr}f_{cds}\notag\\
\cdot\Bigl(\Delta^F_{\d^\nu A^\mu_r,\d^\tau A^\rho_s}(x_1-x_2)
[:A_{a\,\mu}(x_1)A_{b\,\nu}(x_1)A_{c\,\rho}(x_2)A_{d\,\tau}
(x_2):\notag\\
+(\beta_2-2\beta_1)^2g_{\nu\mu}g_{\tau\rho}:u_a(x_1)
\tilde u_b(x_1)u_c(x_2)\tilde u_d(x_2):]\notag\\
+\{\Delta^F_{\d^\mu\tilde u_r,\d^\nu u_s}(x_1-x_2)
\beta_2(\beta_2-1):A_{a\,\mu}(x_1)u_b(x_1)A_{c\,\nu}(x_2)
\tilde u_d(x_2):\notag\\
+(x_1\longleftrightarrow x_2)\}\Bigr)+...,
\end{gather}
where the terms which are denoted by the dots have no contributions
of the form
\begin{equation}
  \delta (x_1-x_2):B_1(x_1)B_2(x_1)B_3(x_2)B_4(x_2):\label{T_2:delta}
\end{equation}
($B_1,...,B_4$ are free field operators), they are
disconnected terms, tree terms with
propagators $\Delta^F(x_1-x_2)$ or $\d^\nu\Delta^F(x_1-x_2)$, or loop
terms. The terms of the form (\ref{T_2:delta}) correspond to the
quadrilinear terms of the interaction Lagrangian of the conventional
theory. So, $C_{A_a}=-\frac{1}{2}$ (\ref{C_A}) yields the
usual 4-gluon coupling (cf. the first paper of \cite{DHKS}) and,
if $\beta_2-2\beta_1\not=0$, a 4-ghost coupling. However, there is no
$AuA\tilde u$-coupling coming from $C_{u_a}=-1$,
because $\beta_2(\beta_2-1)=0$.

(3) By using {\bf (N4)} (\ref{N4'}) we obtain for the interacting $F$-field
($F^{\mu\nu}\=d\d^\mu A^\nu -\d^\nu A^\mu$)
\begin{equation}
F^{\mu\nu}_{a\,g{\cal L}}(x)=\d^\mu A^\nu_{a\,
g{\cal L}}(x)-\d^\nu A^\mu_{a\,g{\cal L}}(x)-2C_{A_a}g_0g(x)f_{abc}
(A^\mu_b A_c^\nu)_{g{\cal L}}(x).\label{F_gL}
\end{equation}
We see that the nonlinear term is due to the non-vanishing of $C_{A_a}$
and that it agrees with the usual nonlinear term iff
$C_{A_a}=-\frac{1}{2}$, in agreement with (\ref{C_A}).

(4) In sect. 4.5.3 we will see that our local construction of
observables works also if one replaces (\ref{G=0}) by the weaker
condition (\ref{G=N}) (with the specifications
(\ref{T_2^N})-(\ref{N_{j,k}}) and (\ref{G3=0})-(\ref{G4=0})), i.e.
one allows to introduce additional 4-legs couplings 'by hand'. This
relaxed version of (\ref{G=0}) has solutions for arbitrary
$(\beta_1,\beta_2)\in\RR^2$ (\ref{K_1,K_2}) (at least in
the case $(j,k)=(0,0)$), see e.g. \cite{D1}.

\subsubsection{Massless gauge fields: local construction of observables}

In \cite{DF} a general {\bf local construction of
observables in gauge theories and of the physical Hilbert space}
(in which the observables are faithfully represented) is given.
This construction relies on some assumptions which can be fulfilled
in QED \cite{DF}. We are now going to generalize the latter result
to the class of interactions we have selected in the preceding
subsection, which includes non-Abelian gauge theories.
Thereby we assume that ghost number conservation {\bf (N5) (ghost)}
and certain cases of the master BRST-identity (\ref{j19}) are
satisfied.

As in \cite{DF} we start with the {\it local} algebra
of interacting gauge and ghost fields (\ref{intfield})
\begin{equation}
{\cal F}({\cal O})\=d\bigvee\{W_{g{\cal L}}(f)\>|\>f\in {\cal D}({\cal O}),
\>W=A^\mu,\,u,\,\tilde u,...\}\label{F(O)}
\end{equation}
(the dots stand for polynomials in $A^\mu,\,u$ and $\tilde u$),
where ${\cal O}$ is a double cone and $g(x)=1$,
$\forall x\in {\cal O}$. In \cite{BF} the crucial observation has been
made that a change of the switching function $g$ outside of ${\cal O}$,
transforms all interacting fields $\in {\cal F}({\cal O})$ by the same
unitary transformation\footnote{An alternative proof of this fact,
which applies also to classical field theory, is given in the second
paper of \cite{DF1}.}. Therefore, the algebraic properties of
${\cal F}({\cal O})$ are independent of the adiabatic limit $g(x)
\rightarrow 1,\>\forall x$.
Hence, we may avoid this limit, which saves
us from infrared divergences. It seems that a consistent
perturbative construction of massless non-Abelian gauge theories
can be done only {\it locally},
i.e. without performing the adiabatic limit, due to the confinement.

The field algebra ${\cal F}({\cal O})$ contains unphysical fields.
The central problem in gauge theories is to eliminate the latter, i.e.
to select the observables, and, in a second step, to construct (physical)
states on the algebra of observables. We proceed as in \cite{DF}:
roughly speaking we will define the observables to be the BRST-invariant
fields. Thereby, we will define the BRST-transformation $\tilde s$
as the ($\ZZ_2$-graded) commutator with the (modified)
interacting BRST-charge $Q_{g{\cal L}}$ of Kugo and Ojima \cite{KO}.
But in contrast to this reference we do not perform the
adiabatic limit. The latter causes the complication that $Q_{g{\cal L}}$
does not agree with its zeroth order contribution $Q_0$: it is a
non-trivial formal power series (cf. \cite{DF}, \cite{DSchroer}).

The current belonging to $Q_{g\cal L}$ is the interacting BRST-current
$\tilde j_{g\cal L}^\mu$. From our experience made in QED \cite{DF} we
know that $\tilde j_{g\cal L}^\mu$ should have the following properties:\\
(a) to zeroth order it agrees with the free BRST current $j^\mu$
(\ref{BRST-current}),\\
(b) it is conserved up to (first) derivatives of the coupling 'constant'
$g$: $\d_\mu \tilde j_{g\cal L}^\mu(x)\sim (\d g)(x)$.\\
Unfortunately, (b) does not hold true for the interacting field
$j_{g{\cal L}}$ (\ref{intfield})-(\ref{R}) where
$j_{g{\cal L}}$ is constructed in terms of $T$-products
satisfying the MWI {\bf (N)}: from (\ref{j1})-(\ref{j10})
we get
\begin{equation}
-T_2(j^\mu\d_\mu f\otimes {\cal L}_0g)=-iT_1({\cal L}_1^\nu f\d_\nu g)+
iT_1({\cal M}^\nu (\d_\nu f)g),\label{MWI-J_BRS}
\end{equation}
where
\begin{equation}
{\cal M}^\nu\=d -{\cal L}^\nu_1+(3C_{1A_a}+\frac{1}{2})\d_\mu u_a
\frac{\d {\cal L}_0}{\d(\d_\mu A_{a\nu})}+\frac{3}{4}u_a
\frac{\d {\cal L}_0}{\d A_{a\nu}}.
\end{equation}
However, the wanted conservation property can be achieved by a change of
the normalization of $T_{n+1}(j^\mu,{\cal L}_{j_1},...,{\cal L}_{j_n})$:
motivated by
\begin{equation}
s_0(k^\mu)=j^\mu\label{s(k)=j}
\end{equation}
(where $k^\mu$ is the (free) ghost current (\ref{ghost-current}))
and generalized perturbative gauge invariance (\ref{N6}) we define
\begin{gather}
\tilde T_{n+1}(j^\mu f\otimes {\cal L}_{j_1}f_1\otimes...
\otimes {\cal L}_{j_n}f_n)\=d [Q_0,T_{n+1}(k^\mu f\otimes {\cal L}_{j_1}f_1
\otimes...\otimes {\cal L}_{j_n}f_n)]_\mp \notag\\
+i\sum_{l=1}^n (-1)^{j_1+...+j_{l-1}}
T_{n+1}(k^\mu f\otimes {\cal L}_{j_1}f_1
\otimes...\otimes {\cal L}_{j_l+1}^\nu\d_\nu f_l\otimes
...\otimes {\cal L}_{j_n}f_n).\label{T:tilde}
\end{gather}
By means of (\ref{s(k)=j}) and (\ref{N6}) we find that $\tilde T_{n+1}
(j^\mu ,{\cal L}_{j_1},...,{\cal L}_{j_n})$ factorizes causally
(\ref{caus}), e.g. $\tilde T_{n+1}(j, {\cal L}_{...},...)=
\tilde T_{l+1}(j, {\cal L}_{...},...)T_{n-l}({\cal L}_{...},...)$.
In addition it is symmetrical in ${\cal L}_{j_1},...,{\cal L}_{j_n}$ and
fulfills the normalization conditions {\bf (N1), (N2)} and {\bf (N0)}.
Hence, $T_{n+1}(j^\mu,{\cal L}_{j_1},...)\rightarrow
\tilde T_{n+1}(j^\mu,{\cal L}_{j_1},...)$ is an admissible finite
renormalization of $T$-products (solely the extension to $D_{n+1}$
is changed), which however violates {\bf (N3)} and the MWI {\bf (N)}.
We point out that the requirement $\d_\mu \tilde j_{g\cal L}^\mu(x)
\sim (\d g)(x)$ is not compatible with {\bf (N3)}, which can be seen
by an explicit calculation of the first order tree diagrams of
$\tilde j_{g\cal L}^\mu$.\footnote{The first order tree diagrams of
$\tilde j_{g\cal L}^\mu$ read:
\begin{gather}
\tilde j_{g\cal L}^{\mu\>(1)}\vert_{\rm tree}(x)=g_0\int d^4x_1\,g(x_1)
\{-:\d^\mu u_a(x)A_{\nu\,b}(x_1)F_c^{\nu\lambda}(x_1):
f_{abc}\d_\lambda D^{\rm ret}(x-x_1)+\label{c1}\\
+:u_a(x)A_{\nu\,b}(x_1)F_c^{\nu\lambda}(x_1):f_{abc}
[\d^\mu\d_\lambda D^{\rm ret}(x-x_1)+C_3g^\mu_\lambda
\delta (x-x_1)]+\label{c2}\\
+:\d^\mu u_a(x)u_b(x_1)\d^\lambda\tilde u_c(x_1):
f_{abc}\d_\lambda D^{\rm ret}(x-x_1)-\label{c3}\\
-:u_a(x)u_b(x_1)\d^\lambda\tilde u_c(x_1):f_{abc}
[\d^\mu\d_\lambda D^{\rm ret}(x-x_1)+C_2g^\mu_\lambda
\delta (x-x_1)]+\label{c4}\\
+:\d^\mu\d_\nu A_a^\nu (x)A_{\lambda\,b}(x_1)u_c(x_1):f_{abc}
\d^\lambda D^{\rm ret}(x-x_1)-\label{c5}\\
-:\d_\nu A_a^\nu (x)A_{\lambda\,b}(x_1)u_c(x_1):f_{abc}
[\d^\mu\d^\lambda D^{\rm ret}(x-x_1)+C_1g^{\mu\lambda}
\delta (x-x_1)],\}\label{c6}
\end{gather}
($D^{\rm ret}$ is the retarded Green's function of the wave operator)
where we have used the simplification (\ref{h=0}) and $C_1,C_2$ and
$C_3$ are undetermined normalization constants. The requirement
$\d_\mu \tilde j_{g\cal L}^\mu(x)\sim (\d g)(x)$ fixes the latter
uniquely:
\begin{equation}
  C_1=-1,\quad\quad C_2=-\frac{1}{2},\quad\quad C_3=0.\label{c7}
\end{equation}
If {\bf (N3)} (or equivalently the causal Wick expansion \cite{EG},
\cite{BF}) holds true, the propagators in (\ref{c2}) and (\ref{c4})
are both equal to
\begin{equation}
  i[...]=\langle\Omega |R_2(A_\lambda;\d^\mu\d_\tau A^\tau)
(x_1;x)|\Omega \rangle.\label{c8}
\end{equation}
But this contradicts $C_2\not= C_3$ (\ref{c7}).}
To compute the divergence of (\ref{T:tilde}) with respect to $j^\mu$
we first apply {\bf (N5)(ghost)} (\ref{N5:ghost'}) to all terms on the
r.h.s. (where we use $g({\cal L}_j)=j$) and afterwards generalized
gauge invariance (\ref{N6}). In this way we obtain
\begin{gather}
\tilde T_{n+1}(j^\mu \d_\mu f\otimes {\cal L}_{j_1}f_1\otimes...
\otimes {\cal L}_{j_n}f_n)=\notag\\
i\sum_{l=1}^n (-1)^{j_1+...+j_{l-1}}\Bigl(
T_n({\cal L}_{j_1}f_1\otimes...
...\otimes {\cal L}_{j_l+1}^\nu f\d_\nu f_l\otimes...\otimes
{\cal L}_{j_n}f_n)\notag\\
-j_lT_n({\cal L}_{j_1}f_1\otimes...\otimes
{\cal L}_{j_l+1}^\nu (\d_\nu f) f_l\otimes...\otimes
{\cal L}_{j_n}f_n)\Bigr).\label{divT(jL)}
\end{gather}
Conversely, this current conservation identity (\ref{divT(jL)}) implies
the generalized perturbative gauge invariance (\ref{N6}) by
proceeding similarly to (\ref{j-W}).
We denote by $\tilde R_{n+1}({\cal L},...,{\cal L};j)$
(where ${\cal L}\equiv {\cal L}_0$) the $R$-product
(\ref{R}) which is constructed in terms of $T_k({\cal L},...,{\cal L})$
and $\tilde T_{k+1}(j,{\cal L},...,{\cal L}),\>1\leq k\leq n$. Then,
the identity (\ref{divT(jL)}) implies
\begin{equation}
\tilde R_{n+1}\bigl( ({\cal L}g)^{\otimes n};j^\mu\d_\mu f\bigr)=
in R_n \bigl( ({\cal L}g)^{\otimes n-1};{\cal L}_1^\nu f\d_\nu g\bigr).
\label{divR(jL)}
\end{equation}
Analogously to (\ref{intfield}) we define
\begin{equation}
\tilde j_{g{\cal L}}^\mu (f)\=d\sum_{n=0}^\infty\frac{i^n}{n!}
\tilde R_{n+1}\bigl( ({\cal L}g)^{\otimes n};j^\mu f\bigr),
\label{j:tilde}
\end{equation}
and this interacting BRST-current has the wanted conservation property
\begin{equation}
\tilde j_{g{\cal L}}^\mu (\d_\mu f)=-{\cal L}_{1\,g{\cal L}}^\nu
(f\d_\nu g),\label{div(j)}
\end{equation}
which agrees precisely with the corresponding result for QED
(formula (5.12) of the first paper of \cite{DF}).
The difference $(\tilde j_{g\cal L}^\mu - j_{g\cal L}^\mu)$
(where $j_{g\cal L}^\mu$ still denotes the interacting field
constructed in terms of $T$-products satisfying also the MWI and
{\bf (N3)}) is immediately obtained by applying the master BRST-identity:
\begin{equation}
\tilde j_{g\cal L}^\mu (f)- j_{g\cal L}^\mu (f)=
i\bigl( G^{(1)}(k^\mu f,{\cal L}g)\bigr)_{g\cal L}+
i\bigl( G(({\cal L},{\cal L}_1)g,k^\mu f)\bigr)_{g\cal L}.\label{tildej-j}
\end{equation}

The interacting BRST-charge operator is now defined by
\begin{equation}
Q_{g{\cal L}}\=d \int d^4x\,h^\mu (x)\tilde j_{\mu\,g{\cal L}}(x),\label{Q_gL}
\end{equation}
where ${\cal L}\equiv {\cal L}_0$ and $h^\mu$ is a suitable test function
(see \cite{DF} and subsection 4.5.4). $Q_{g{\cal L}}$ is a formal power
series and the construction is such that the relations
\begin{equation}
Q_{g{\cal L}}=Q_{g{\cal L}}^*,\quad\quad (Q_{g{\cal L}})_0=Q_0,\label{Q_{int}}
\end{equation}
hold true (where $(...)_0$ means the zeroth order) and that $Q_{g{\cal L}}$
is nilpotent
\begin{equation}
(Q_{g{\cal L}})^2=0.\label{Q^2=0}
\end{equation}
The latter property is proved in subsection 4.5.4
by using current conservation (\ref{div(j)}) and generalized
gauge invariance (\ref{N6}). We point out that the conservation of the
BRST-current (\ref{div(j)}) and the construction of the nilpotent
BRST-charge (\ref{Q_gL})-(\ref{Q^2=0}) use the master BRST-identity
for $T_n({\cal L}_0,...,{\cal L}_0)$ and $T_n({\cal L}_1,{\cal L}_0,
...,{\cal L}_0)$ $(\forall n\in\NN)$ only.

The BRST-transformation $\tilde s$ of the interacting fields
$W_{g{\cal L}}(f),\>f\in {\cal D}({\cal O})$,
is then defined by the commutator with $Q_{g{\cal L}}$ (or anti-commutator
if $W$ has an odd ghost number)
\begin{equation}
\tilde s(W_{g{\cal L}}(f))\=d [Q_{g{\cal L}},W_{g{\cal L}}(f)]_\mp,
\quad\quad f\in {\cal D}({\cal O}).
\label{BRST-trafo}
\end{equation}
We extend $\tilde s$ to a graded derivation ${\cal F}({\cal O})
\rightarrow {\cal F}({\cal O})$. The local
observables are selected by the definition \cite{DF}
\begin{equation}
{\cal A}({\cal O})\=d \frac {{\rm ker}\>\tilde s}{{\rm ran}\>\tilde s}.
\label{obs}
\end{equation}

Following \cite{DF} we look for states on ${\cal A}({\cal O})$
which take values
in $\tilde{\CC}$ (by which we mean the formal power series with
coefficients in $\CC$). Thereby, $a\in \tilde{\CC}$ is called
positive, if there
exists $b\in \tilde{\CC}$ with $a=b^*b$, where $*$ means complex
conjugation. In \cite{DF} it is shown that ${\cal A}({\cal O})$
can be naturally represented on the cohomology of $Q_{g{\cal L}}$
\begin{equation}
{\cal H}_{\rm phys}\=d \frac {{\rm ker}\>Q_{g{\cal L}}}
{{\rm ran}\>Q_{g{\cal L}}}
\end{equation}
and that the induced inner product on ${\cal H}_{\rm phys}$ is
positive definite in the just mentioned sense.
Hence ${\cal H}_{\rm phys}$ is a pre Hilbert space and its elements
are interpreted as physical states.

The master BRST-identity (\ref{j19}) yields also an explicit formula
for the {\bf BRST-transformation of the interacting fields}
$\in {\cal F}({\cal O})$, by which we will see
that the definition (\ref{BRST-trafo}) agrees with the usual
BRST-transformation. (In addition this ensures the existence of
non-trivial observables.) For this purpose we note that the
master BRST-identity and the requirement (\ref{G=0}) imply the following
relation for the $R$-products (\ref{R}):
\begin{gather}
[Q_0,R_{n+1}\bigl(({\cal L}_0g)^{\otimes n}; Wf\bigr)]_\mp =\notag\\
-inR_{n+1}\bigl(({\cal L}_0g)^{\otimes (n-1)}\otimes {\cal L}_1^\nu
\d_\nu g; Wf\bigr)
-iR_{n+1}\bigl(({\cal L}_0g)^{\otimes n};W'_\nu\d^\nu f\bigr)\notag\\
+nR_n\Bigl(({\cal L}_0g)^{\otimes (n-1)};
[G(({\cal L}_0,{\cal L}_1)g,Wf)+G((W,W')f,{\cal L}_0g)]
\Bigr),\label{N6b}
\end{gather}
where we have assumed $s_0(W)=i\d^\nu W'_\nu$.
In subsection 4.5.4 it will be shown that this identity implies the
BRST-transformation formula
\begin{gather}
\tilde s(W_{g{\cal L}}(f))=[Q_{g{\cal L}},W_{g{\cal L}}(f)]_\mp=
-iW^{\prime\nu}_{g{\cal L}}(\d_\nu f)\notag\\
+i\bigl(G(({\cal L},{\cal L}_1)g,Wf)+G((W,W')f,{\cal L}g)
\bigr)_{g{\cal L}},
\quad\quad f\in {\cal D}({\cal O}),\label{trafo:intfield}
\end{gather}
where ${\cal L}\equiv {\cal L}_0$. The term in the second line
is the nonlinear part of the BRST-transformation. In case that $W$ is
a single symbol we find that $[G(({\cal L},{\cal L}_1)g,Wf)+
G((W,W')f,{\cal L}g)]$ is quadratic in the symbols (because
${\cal L}$ and ${\cal L}_1$ are trilinear), in agreement with the usual
BRST-transformation. (To prove
(\ref{trafo:intfield}) it will be shown that the terms
$[(Q_{g{\cal L}}-Q_0),W_{g{\cal L}}(f)]$ cancel out with
the terms $-inR_{n+1}\bigl(({\cal L}_0g)^{\otimes (n-1)}
\otimes {\cal L}_1^\nu\d_\nu g; Wf\bigr)$.)

If we do not assume that $s_0 W$ is a divergence we end up with
\begin{equation}
\tilde s(W_{g{\cal L}}(f))=(s_0W)_{g{\cal L}}(f)
+i\bigl( G^{(1)}(Wf,{\cal L}g)+G(({\cal L},{\cal L}_1)g,Wf)
\bigr)_{g{\cal L}},
\end{equation}
instead of (\ref{trafo:intfield}). We choose $W=k^\mu$, compare with
(\ref{tildej-j}) and find
\begin{equation}
[Q_{g{\cal L}},k^\mu_{g{\cal L}}]=\tilde j^\mu_{g{\cal L}}.
\end{equation}
Introducing the interacting ghost charge
\begin{equation}
Q^u_{g{\cal L}}\=d \int d^4x\,h^\mu (x)
k_{\mu\,g{\cal L}}(x),
\end{equation}
where $h^\mu\in\mathcal{D}(\RR^4)$ is chosen in precisely the
same way as in $Q_{g{\cal L}}$ (see sect. 4.5.4), it results
\begin{equation}
[Q_{g{\cal L}},Q^u_{g{\cal L}}]=Q_{g{\cal L}}
\end{equation}
as in \cite{KO}.

{\bf Examples}: In most of the following examples the computation of
$G^{(1)}$ and $G^{(2)}$ gives less work than it seems, because only very
few terms contribute. We use the values $C_{u_a}=-1$ (\ref{C_u}) and
$C_{A_a}=-\frac{1}{2}$ (\ref{C_A}) without further mentioning it.

(1) BRST-transformation of $A^\mu_{a\,{g{\cal L}}}(h)$:
\begin{eqnarray}
G^{(1)}({\cal L}_0g,A^\mu_ah)=0,\quad
G^{(1)}(A^\mu_ah,{\cal L}_0g)=-\frac{3}{4}g_0f_{abc}
A^\mu_b u_c hg,\nonumber\\
G^{(2)}({\cal L}_1g,A^\mu_a h)=\frac{3}{4}g_0f_{abc}A^\mu_b u_c gh,
\quad G^{(2)}(g^{\mu\nu}u_a h,{\cal L}_0g)=g_0f_{abc}A^\mu_b u_c hg.
\end{eqnarray}
Therefore,
\begin{equation}
\tilde s(A^\mu_{a\,{g{\cal L}}}(h))=
-iu_{a\,{g{\cal L}}}(\d^\mu h)+ig_0(f_{abc}A^\mu_b u_c)_{g{\cal L}}(gh).
\label{BRST:A}
\end{equation}
Taking $g\vert_{{\rm supp}\>h}=1$ into account the last term takes
the usual form $ig_0(f_{abc}A^\mu_b u_c)_{g{\cal L}}(h)$. We see
that $G^{(1)}(A^\mu_ah,{\cal L}_0g)\not= 0$ gives a non-vanishing
contribution to the non-linear term in (\ref{BRST:A}). This shows that
the distinction of internal and external derivatives and in particular
the appearance of the external derivative in the definition of
$\Delta^\mu$ (\ref{Delta2}) is crucial to obtain
the correct BRST-transformation.

(2) BRST-transformation of $u_{a\,{g{\cal L}}}(h)$:
\begin{eqnarray}
G^{(1)}({\cal L}_0g,u_ah)=0,\quad
G^{(1)}(u_ah,{\cal L}_0g)=0,\nonumber\\
G^{(2)}({\cal L}_1g,u_a h)=-\frac{1}{2}g_0f_{abc}u_b u_c gh,
\quad G(0 h,{\cal L}_0g)=0.
\end{eqnarray}
Hence,
\begin{equation}
\tilde s(u_{a\,{g{\cal L}}}(h))=
-\frac{i}{2}g_0(f_{abc}u_b u_c)_{g{\cal L}}(h).
\label{BRST:u}
\end{equation}

(3) BRST-transformation of $\tilde u_{a\,{g{\cal L}}}(h)$:
\begin{eqnarray}
G^{(1)}({\cal L}_0g,\tilde u_ah)=0,\quad
G^{(1)}(\tilde u_ah,{\cal L}_0g)=0,\nonumber\\
G^{(2)}({\cal L}_1g,\tilde u_a h)=0,
\quad G^{(2)}(-\d_\nu A^\nu_a h,{\cal L}_0g)=0,
\end{eqnarray}
where we have used $g^{\nu\mu}\frac{\d{\cal L}_0}{\d (\d^\nu A^\mu_a)}=0$.
So we obtain
\begin{equation}
\tilde s(\tilde u_{a\,{g{\cal L}}}(h))=
iA^\nu_{a\,{g{\cal L}}}(\d_\nu h).
\label{BRST:tilde-u}
\end{equation}

(4) BRST-transformation of $F^{\mu\nu}_{a\,{g{\cal L}}}(h)$:
\begin{gather}
G^{(1)}({\cal L}_0g,F^{\mu\nu}_ah)=0,\notag\\
G^{(1)}(F^{\mu\nu}_ah,{\cal L}_0g)=(\frac{1}{2}+3C_{1A_a})g_0f_{abc}
A^\nu_bu_c(\d^\mu h)g-(\mu\leftrightarrow\nu),\notag\\
G^{(2)}({\cal L}_1g,F^{\mu\nu}_a h)=(\frac{1}{2}+3C_{1A_a})g_0f_{abc}
[\tilde\d^\mu (A^\nu_bu_c)gh+A^\nu_bu_c(\d^\mu g)h]\notag\\
-(\mu\leftrightarrow\nu)+g_0f_{abc}F_b^{\mu\nu}u_c gh,\quad\quad
G^{(2)}(0 h,{\cal L}_0g)=0.
\end{gather}
Now we apply the relation (\ref{heur:tilde-d}) (which holds obviously also
for the $R$-products; cf. the remark at the end of sect. 2.4)
\begin{equation}
R_{n+1}({\cal L}_0g\otimes...;f_{abc}\tilde\d^\mu (A^\nu_bu_c)gh)=
-R_{n+1}({\cal L}_0g\otimes...;f_{abc}A^\nu_bu_c\d^\mu (gh)).
\end{equation}
Hence, inserting these formulas into (\ref{trafo:intfield}), the
terms $\sim (\frac{1}{2}+3C_{1A_a})$ cancel and it remains
\begin{equation}
\tilde s(F^{\mu\nu}_{a\,{g{\cal L}}}(h))=
ig_0(f_{abc}F^{\mu\nu}_b u_c)_{g{\cal L}}(h).
\label{BRST:F}
\end{equation}

(5) BRST-transformation of $(\sum_aF^{\mu\nu}_aF^{\rho\tau}_a)_
{g{\cal L}}(h)$ (we do not write $\sum_a$ but always perform this sum):
\begin{gather}
G^{(1)}({\cal L}_0g,F^{\mu\nu}_aF^{\rho\tau}_ah)=0,\notag\\
G^{(1)}(F^{\mu\nu}_aF^{\rho\tau}_ah,{\cal L}_0g)=\{(\frac{1}{2}+3C_{1A_a})
g_0f_{abc}[F^{\rho\tau}_aA^\nu_bu_c(\d^\mu h)g
+(\tilde\d^\mu F^{\rho\tau}_a)A^\nu_bu_chg]\notag\\
-(\mu\leftrightarrow\nu)\}+\{(\mu,\nu)\leftrightarrow (\rho,\tau)\},\notag\\
G^{(2)}({\cal L}_1g,F^{\mu\nu}_aF^{\rho\tau}_a h)=
\{(\frac{1}{2}+3C_{1A_a})g_0f_{abc}[F^{\rho\tau}_a\tilde\d^\mu (A^\nu_bu_c)gh+
F^{\rho\tau}_aA^\nu_bu_c(\d^\mu g)h]\notag\\
-(\mu\leftrightarrow\nu)
+g_0f_{abc}F^{\rho\tau}_aF_b^{\mu\nu}u_c gh\}+\{(\mu,\nu)
\leftrightarrow (\rho,\tau)\},\quad\quad
G^{(2)}(0 h,{\cal L}_0g)=0.
\end{gather}
The term $\sim FFugh$ in $G^{(2)}({\cal L}_1 g,FF h)$ drops out
because $f_{abc}$ is totally antisymmetric.
Inserting these formulas into (\ref{trafo:intfield}), the
terms $\sim (\frac{1}{2}+3C_{1A_a})$ cancel again due to (\ref{heur:tilde-d}).
So we obtain
\begin{equation}
\tilde s((F^{\mu\nu}_aF^{\rho\tau}_a)_{g{\cal L}}(h))=0\label{BRST:FF}
\end{equation}
and, hence, the corresponding equivalence class (cf. (\ref{obs})) is
a non-trivial observable.

(6) Due to the requirement $G(({\cal L}_j,{\cal L}_{j+1})f,{\cal L}_kg)
+(-1)^{jk}\{(j,f)\leftrightarrow (k,g)\}=0$
(\ref{G=0}) we can easily write down the BRST-transformation of
${\cal L}_{j\,g{\cal L}},\>j=0,1,2$:
\begin{eqnarray}
\tilde s({\cal L}_{0\,g{\cal L}}(h))=-i{\cal L}^\nu_{1\,g{\cal L}}(\d_\nu h),
\nonumber\\
\tilde s({\cal L}^\nu_{1\,g{\cal L}}(h))=-i{\cal L}^{\nu\mu}
_{2\,g{\cal L}}(\d_\mu h),\nonumber\\
\tilde s({\cal L}^{\nu\mu}_{2\,g{\cal L}}(h))=0.\label{BRST:L}
\end{eqnarray}

{\it Remarks:} (1) Having determined the interaction ${\cal L}_0$ we can
explicitly write down the interacting field equations by means of
(\ref{fieldeq}): with the simplification (\ref{h=0}) they read
\begin{gather}
\w A_{\mu\,a\,g{\cal L}}=g_0f_{abc}\d^\nu[g
(A_{\mu\,b}A_{\nu\,c})_{g{\cal L}}]-\notag\\
-g_0gf_{abc}(A^\nu_b F_{\nu\mu\,c})_{g{\cal L}}+
g_0gf_{abc}(u_b\d_\mu\tilde u_c)_{g{\cal L}},\label{fieldeq:A}\\
\w u_{a\,g{\cal L}}=-g_0f_{abc}\d_\mu[g(A^\mu_bu_c)_{g{\cal L}}],
\label{fieldeq:u}\\
\w \tilde u_{a\,g{\cal L}}=-g_0gf_{abc}(A^\mu_b\d_\mu \tilde u_c)
_{g{\cal L}}.\label{fieldeq:tilde-u}
\end{gather}
They hold true everywhere, $g$ needs not to be constant. In the
classical limit $\hbar\rightarrow 0$ interacting fields factorize,
$(VW)_{g{\cal L}}(x)=V_{g{\cal L}}(x)W_{g{\cal L}}(x)$ (see
\cite{DF2}), and hence (\ref{fieldeq:A})-(\ref{fieldeq:tilde-u}) go over
into the usual Yang-Mills equations.

(2) One might think that the MWI agrees with the quantum Noether
condition (QNC) \cite{HS}\footnote{Note that the QNC in terms
of $T$-products (given in the first paper of \cite{HS})
does not agree with the QNC in terms of interacting fields (second
paper): they normalize the interacting BRST-current differently.
The second formulation fixes the BRST-current such that it is
conserved up to terms $\sim\d g$ where $g$ is the coupling 'constant',
in particular it yields the values (\ref{c7}) for the constants
$C_1, C_2$ and $C_3$ in (\ref{c1})-(\ref{c6}). But the
first formulation requires
\begin{equation}
  C_1=-1,\quad\quad C_2=0,\quad\quad C_3=1.\label{c9}
\end{equation}
} in the application to $\d_\mu T(j^\mu,\mathcal{L},...,\mathcal{L})$,
where $j^\mu$ is a conserved current and $\mathcal{L}$ the
interaction. But for $j$ being the BRST-current (\ref{BRST-current})
we explicitly see that the two conditions are {\it different}
and that the {\it QNC is less general}. Namely, the
QNC in terms of interacting fields is precisely the particular case
$j_1=...=j_n=0$ of (\ref{divT(jL)}). This equation is sufficient for
perturbative gauge invariance ((\ref{N6}) for $j_1=...=j_n=0$)
and, if ghost number conservation holds true, it is also necessary.
The QNC does not contain any information about the
$G$-terms in the master BRST-identity (\ref{j19}). In addition we
recall that (\ref{divT(jL)}) (and hence in particular the QNC
in terms of interacting fields)
is not compatible with the MWI (\ref{MWI-J_BRS}) and {\bf (N3)}.
Comparing (\ref{c9}) with (\ref{c8}) we find that this holds true also
for the QNC in terms of $T$-products.

In general note that the QNC
is formulated for $T(j^\mu,\mathcal{L},...,\mathcal{L})$
with $\d_\mu j^\mu =0$ and $\mathcal{L}$ the interaction, only,
whereas the MWI deals with $T$-products of arbitrary
factors. In particular $\d^\nu W$ in (\ref{[d,T]}) needs not to vanish
and there are important applications with $\d^\nu W\not= 0$, e.g.
the field equation (sect. 4.1), non-Abelian matter currents
(sect. 4.3), massive axial currents $j_A$ (\ref{axan}) (see sect.
5.1), etc..

\subsubsection{Massive gauge fields}

To simplify the notations we consider the most simple non-Abelian model,
namely three massive gauge fields, $m_a>0,\>a=1,2,3$ and no massless fields.
However, as far as anomalies are absent and there is a solution for
the below given requirements (A)-(F) and (\ref{G=N})-(\ref{N_{j,k}}),
(\ref{G3=0})-(\ref{G4=0}) on the interaction ${\cal L}$,
our method applies also to general
models with arbitrary numbers of massive and massless gauge fields and
spinor fields. We will find the well-known result that with the fields
$A^\mu_a,u_a,\tilde u_a$ and $\phi_a$ ($a=1,2,3$) only, a consistent
construction of the model is impossible, more precisely generalized
perturbative gauge invariance (\ref{N6}) for second order tree
diagrams cannot be satisfied \cite{DS}. We will solve this problem in
the usual way: besides the scalar fields $(\phi_a)_{a=1,2,3}$,
we introduce an additional
real (free) scalar field $H$, the 'Higgs field', with arbitrary
mass $m_H\geq 0$, which is quantized according to
\begin{equation}
(\w+m_H^2)H=0,\quad\quad H^*=H,\quad\quad \Delta_{H,H}=-D_{m_H}
\end{equation}
and $H$ commutes with all other free fields.

To determine the interaction
${\cal L}_0$ we require the same properties (A)-(F) as in the massless case.
The only modification is that ${\cal L}_j$ is now a polynomial in
$A^\mu_a,\,u_a,\,\tilde u_a,\phi_a,\>a=1,2,3$ and $H$ and internal
derivatives of these symbols (again we solely admit monomials
which have at least three factors).
Proceeding as in the massless case we find the following
particular solution of (A)-(F):
\begin{eqnarray}
{\cal L}_0&=&g_0\{f_{abc}[A_{a\,\mu }A_{b\,\nu }\d^{\nu }A_{c}^\mu
-u_{a}\d^{\mu }\tilde{u}_{b}A_{c\,\mu }]  \nonumber \\
&+&d_{abc}A_{a}^{\mu }\phi _{b}\d_{\mu }\phi _{c}
+e_{abc}A_a^\mu A_{b\,\mu}\phi_c
+h_{abc}\tilde{u}_{a}u_{b}\phi _{c}
\nonumber\\
&+&l_{ab}[\frac{1}{m_b}(-HA_{a}^{\mu }\d_{\mu }\phi _b+
(\d_\mu H)A_{a}^{\mu }\phi_b)
+A_{a}^\mu A_{b\,\mu }H\nonumber\\
&-&H\tilde{u}_{a}u_b
-\frac{m_H^2}{2m_a m_b}H\phi_{a}\phi _{b}]+pH^{3}+tH^4\}, \label{L_0:SU2}
\end{eqnarray}
\begin{eqnarray}
{\cal L}_1^\nu&=&g_0\{f_{abc}[A_{a\mu}u_bF^{\nu\mu}_c-\frac{1}{2}u_au_b\d^\nu
\tilde u_c]+2e_{abc}u_aA^\nu_b\phi_c\nonumber\\
&+&d_{abc}[u_a\phi_b\d^\nu\phi_c+m_c A^\nu_a\phi_bu_c]\nonumber\\
&+&l_{ab}[\frac{1}{m_b}u_a(\phi_b\d^\nu H-H\d^\nu\phi_b)+A^\nu_au_bH]\},
\label{L_1:SU2}
\end{eqnarray}
\begin{eqnarray}
{\cal L}_2^{\nu\mu}&=&g_0\frac{f_{abc}}{2}u_au_bF_c^{\nu\mu},\label{L_2:SU2}
\end{eqnarray}
\begin{equation}
{\cal L}_3=0, \label{L_3:SU2}
\end{equation}
where $f_{abc}\in\RR$ is totally antisymmetric, $l_{ab}\in\RR$ is
symmetric and
\begin{eqnarray}
d_{abc} &=&f_{abc}\frac{m_{b}^{2}+m_{c}^{2}-m_{a}^{2}}{2m_{b}m_{c}},\quad
e_{abc}=f_{abc}\frac{m_{b}^{2}-m_{a}^{2}}{2m_{c}},  \nonumber \\
h_{abc} &=&f_{abc}\frac{m_{a}^{2}+m_{c}^{2}-m_{b}^{2}}{2m_{c}},\quad\quad
p,t\in\RR.
\end{eqnarray}
The most general solution for ${\cal L}_0$ differs from the particular
solution (\ref{L_0:SU2}) by a coboundary $-is_0K_1$ and a divergence
$\d_\nu K_2^\nu$ as in (\ref{L_0}). In addition one has the freedom to
add terms with vanishing divergence to ${\cal L}_1$ and ${\cal L}_2$
as in (\ref{L_1,2,3}). It is a peculiarity of the present model, that the
total antisymmetry of $f_{abc}$ implies the Jacobi identity, so that we obtain
\begin{equation}
f_{abc}=\epsilon_{abc}\>(=\>{\rm structure\>constant\>of}\>SU(2))\label{f=eps}
\end{equation}
by absorbing a constant factor in $g_0$.
So far we could set $l_{ab}=0,\>p=0,\>t=0$, in other words the
Higgs field $H$ is not needed to satisfy (A)-(F).

Now we come to an interesting complication of the massive case: the
requirement $G(({\cal L}_j,{\cal L}_{j+1})f, {\cal L}_kg)
+(-1)^{jk}\{(j,f)\leftrightarrow (k,g)\}=0$
(\ref{G=0}) {\it cannot be satisfied!} To save generalized
perturbative gauge invariance (\ref{N6}) we require instead the
following weaker condition:
there exist $N_{j,k}\in\mathcal{P}_0,\> j,k=0,1,2$ such that
\begin{gather}
G(({\cal L}_j,{\cal L}_{j+1})f, {\cal L}_kg)+
(-1)^{jk}G(({\cal L}_k,{\cal L}_{k+1})g, {\cal L}_jf)+s_0(N_{j,k})fg
\notag\\
=-i[N^\nu_{j+1,k}(\d_\nu f)g+(-1)^{jk}N^\nu_{j,k+1}f(\d_\nu g)],\quad
j,k\in\{0,1,2\},\>\forall f,g\in {\cal D}(\RR^4),\label{G=N}
\end{gather}
where $N_{3,k}=0=N_{k,3},\>k=0,1,2$, and that the finite renormalization
\begin{equation}
T_2({\cal L}_j f\otimes {\cal L}_k g)\rightarrow
T_2^N({\cal L}_j f\otimes {\cal L}_k g)\=d
T_2({\cal L}_j f\otimes {\cal L}_k g)+T_1(N_{j,k}fg)\label{T_2^N}
\end{equation}
maintains the permutation symmetry of $T_2$ and the normalization
conditions {\bf (N1), (N2), (N0)}, and preserves the ghost number:
$[Q_g,T_2^N({\cal L}_j f\otimes {\cal L}_k g)]=
(j+k)T_2^N({\cal L}_j f\otimes {\cal L}_k g)$ where we take
$g({\cal L}_j)=j$ (\ref{g(L)}) into account (cf. (\ref{[Q_g,T]})).
This requirement can only be satisfied for $t=0$ in
(\ref{L_0:SU2}). Hence, the $G$-terms in (\ref{G=N}) are 4-legs
terms and, therefore, we may restrict the $N_{j,k}$ to be 4-legs
terms, too. In other words the $N_{j,k}$ are
sums of monomials of degree four in $A^\mu_a,u_a,
\tilde u_a,\phi_a,\>a=1,2,3$ and in $H$ (without any derivative),
which are Lorentz tensors of rank $(j+k)$ and satisfy
\begin{equation}
g(N_{j,k})=j+k,\quad\quad
N_{j,k}^*=-N_{j,k},\quad\quad N_{j,k}=(-1)^{jk}N_{k,j}.\label{N_{j,k}}
\end{equation}
These properties imply
\begin{equation}
N_{2,1}=0=N_{1,2},\quad\quad N_{2,2}=0,\quad\quad N_{1,1}^{\nu\mu}=
-N_{1,1}^{\mu\nu}.
\end{equation}
If (\ref{G=N})-(\ref{T_2^N}) is satisfied we indeed obtain
\begin{gather}
[Q_0,T^N_2({\cal L}_{j_1}f_1\otimes {\cal L}_{j_2}f_2)]_\mp=
-i\Bigl(T^N_2({\cal L}_{j_1+1}^\nu\d_\nu f_1\otimes {\cal L}_{j_2}f_2)
\notag\\
+(-1)^{j_1}T^N_2({\cal L}_{j_1}f_1\otimes {\cal L}_{j_2+1}^\nu
\d_\nu f_2)\Bigr),
\end{gather}
by using the master BRST-identity (\ref{j15}).

Turning to arbitrary orders $n\geq3$ we look for a sequence of $T$-products
$(T_n^N)_{n\in\NN}$ (in the sense of sect. 2.2)

- which satisfies the normalization conditions {\bf (N1), (N2)} and
{\bf (N0)},

- which agrees as far as possible with the given sequence $(T_n)_{n\in\NN}$
that satisfies all normalization conditions (also {\bf (N3)},
${\bf (\tilde N)}$ and the MWI {\bf (N)}),

- and for which $T_2^N({\cal L}_j,{\cal L}_k)$ is connected
with $T_2({\cal L}_j,{\cal L}_k)$ by (\ref{T_2^N}) for all $j,k$.\\
For this purpose let ${\cal B}=\{{\cal L}_0,{\cal L}_1,
{\cal L}_2,B_1,B_2,...\}$ be a (vector space)
basis of $\tilde {\cal P}_0$. Due to causality
(\ref{caus}) the renormalization terms $T_1(N_{j,k}fg)$ in (\ref{T_2^N})
propagate to higher orders. More precisely we define
\begin{gather}
T^N_{l+n}(B_{k_1}g_1\otimes ...\otimes B_{k_l}g_l\otimes {\cal L}_{j_1}f_1
\otimes...\otimes {\cal L}_{j_n}f_n)\=d\sum_{m=0}
^{[\frac{n}{2}]}\sum_{\pi\in {\cal S}_n}\frac{1}{2^m m! (n-2m)!}\notag\\
\eta^\pi (j_1,...,j_n)
T_{l+n-m}(B_{k_1}g_1\otimes ...\otimes B_{k_l}g_l\otimes
N_{j_{\pi 1},j_{\pi 2}}f_{\pi 1}f_{\pi 2}
\otimes...\notag\\
...\otimes N_{j_{\pi (2m-1)},j_{\pi 2m}}f_{\pi (2m-1)}f_{\pi 2m}\otimes
{\cal L}_{j_{\pi (2m+1)}}f_{\pi (2m+1)}\otimes...\otimes
{\cal L}_{j_{\pi n}}f_{\pi n}),\label{T^N}
\end{gather}
where $B_{k_1},...,B_{k_l}\in {\cal B}\setminus\{{\cal L}_0,{\cal L}_1,
{\cal L}_2\}$ and
$\eta^\pi (j_1,...,j_n)$ is the sign coming from the permutation
of Fermi-operators in
$({\cal L}_{j_1},...,{\cal L}_{j_n})\rightarrow ({\cal L}_{j_{\pi 1}}
,...,{\cal L}_{j_{\pi n}})$. We extend this definition to
${\cal D}(\RR^4,\tilde {\cal P}_0)^{\otimes (l+n)}$ by requiring linearity
and (permutation) symmetry. Obviously this $(T_n^N)_{n\in\NN}$
solves our requirements. The formula (\ref{T^N}) is a particular (simple)
case of Theorem 3.1 in \cite{Pi}, which is a precise formulation
of a formula given in \cite{BS}. For later purpose we mention that
the $T^N$-products (\ref{T^N}) satisfy {\bf (N5)(ghost)}, because the
$T$-products do so. In particular we have
\begin{eqnarray}
T^{N}_{n+1}( k^\mu\d_\mu g\otimes {\cal L}_{j_1}f_1\otimes...\otimes
{\cal L}_{j_n}f_n)=\nonumber\\
\sum_{l=1}^nj_lT^N({\cal L}_{j_1}f_1\otimes...\otimes {\cal L}_{j_l}f_lg
\otimes...\otimes {\cal L}_{j_n}f_n).\label{N5:ghost:T'^N}
\end{eqnarray}
But in general the $T^N$-products violate {\bf (N3)} and the MWI {\bf (N)}.

To obtain generalized perturbative gauge invariance to
orders $n\geq3$ we additionally require
\begin{equation}
G(({\cal L}_j,{\cal L}_{j+1})f,N_{k,l}g)+G^{(1)}(N_{k,l}g,{\cal L}_j f)
=0,\quad j,k,l\in\{0,1,2\},\quad \forall f,g\in {\cal D}(\RR^4),\label{G3=0}
\end{equation}
and
\begin{equation}
G^{(1)}(N_{k,l}f,N_{r,s}g)=0,\quad\quad k,l,r,s\in\{0,1,2\},\quad
\forall f,g\in {\cal D}(\RR^4).\label{G4=0}
\end{equation}
Then, applying the master BRST-identity (\ref{j19}) to $[Q_0,
T^N_n({\cal L}_{j_1}f_1\otimes...\otimes {\cal L}_{j_n}f_n)]_\mp$ and taking
(\ref{G=N}), (\ref{G3=0}) and (\ref{G4=0}) into account, we find the
wanted (modified) generalized perturbative gauge invariance\footnote{For
$j_1=...=j_n=0$ this is the formulation of pertubative
gauge invariance for massive fields in \cite{DS}, \cite{Sch} and \cite{G}.}
\begin{gather}
[Q_0,T^N_n({\cal L}_{j_1}f_1\otimes...\otimes {\cal L}_{j_n}f_n)]_\mp=
\notag\\
-i\sum_{l=1}^n(-1)^{j_1+...+j_{l-1}}
T^N_n({\cal L}_{j_1}f_1\otimes...\otimes {\cal L}_{j_l+1}^\nu
\d_\nu f_l\otimes ...\otimes {\cal L}_{j_n}f_n).\label{N6'}
\end{gather}

The particular case $j=k=l=0$ of the requirements
(\ref{G=N})-(\ref{T_2^N}) and (\ref{G3=0}) has been worked out
for general models in \cite{DS}, \cite{Sch} and \cite{G}. We
specialize these results to the present model:

{\bf (1)} The second order requirement (\ref{G=N})-(\ref{T_2^N})
is very restrictive: its restriction to $j=k=0$ is
satisfied if and only if the following relations (a)-(e) hold:\\
(a) the masses agree\footnote{For general models (\ref{m}) is replaced
by more complicated mass relations, see \cite{DS}, \cite{Sch}, \cite{G}.}
\begin{equation}
m\=d m_1=m_2=m_3.\label{m}
\end{equation}
(b) $f_{abc}$ satisfies the Jacobi identity. (In our simple model this is
already known (\ref{f=eps}), but in the general case the Jacobi
identity is obtained only at this stage here, similar to the
massless case.)\\
(c) The $H$-coupling parameters take the values
\begin{equation}
l_{ab}=\frac{\kappa m}{2}\delta_{ab},\quad\quad t=0,
\end{equation}
where $\kappa\in\{-1,1\}$ is an undetermined sign, and $p$ is still free.
In particular we see that the Higgs field (or another
enlargement/modification of the model) is indispensable.\\
(d) The constants $C_{A_a},\>C_{\phi_a}$ and $C_H$ have the values\footnote{We
recall that $C_{u_a}=-1$ has already been used in the derivation of the
master BRST-identity.}
\begin{equation}
C_{A_a}=-\frac{1}{2},\quad C_{\phi_a}=-1,\quad\forall a,\quad\quad
C_H=-1.
\end{equation}
(e) The polynomial $N_{0,0}$ reads
\begin{equation}
N_{0,0}=ig_0^2\{\frac{m_H^2}{16m^2}[(\sum_{a=1}^3\phi_a^2)^2+
2H^2\sum_{a=1}^3\phi_a^2]+\lambda H^4\}\delta(x_1-x_2),\label{N}
\end{equation}
where $\lambda\in\RR$ is a constant which is undetermined so far.

{\bf (2)} The third oder requirement (\ref{G3=0})
fixes the remaining free parameters
$p$ and $\lambda$ (which are the parameters of the $H$-self-couplings):
the particular case $j=k=l=0$ of
(\ref{G3=0}) holds true if and only if
\begin{equation}
p=\frac{m_H^2}{4m},\quad\quad\lambda=-\frac{m_H^2}{16m^2}.
\end{equation}
\\

The $C_{A},\> C_u,\>C_{\phi}$- and $C_H$-terms in the tree-diagram part
of $T_2^N({\cal L}_0,{\cal L}_0)$ and $N_{0,0}$ correspond to
the {\it quadrilinear terms in the interaction Lagrangian
of the conventional theory} (the latter are also of order $g_0^2$),
cf. Remark (2) in sect. 4.5.1. With this
identification our resulting interaction agrees precisely with the
$SU(2)$ Higgs-Kibble model, which is usually obtained by the Higgs
mechanism. Here we have derived it in a completely different way
(cf. \cite{DS}, \cite{Sch}, \cite{G} and \cite{DSchroer}).

By inserting the explicit expressions (\ref{N}) into the
definition (\ref{j14}) of $G^{(1)}(.,.)$ we verify that the
fourth order requirement (\ref{G4=0}) holds true for
$k=l=r=s=0$. We strongly presume that the requirements
(\ref{G=N})-(\ref{T_2^N}), (\ref{G3=0}) and (\ref{G4=0}) can be
fulfilled for all values of $j,k,l,r,s$.\footnote{Additionally we
  expect that these requirements determine $N_{1,0}, N_{1,1}$
and $N_{2,0}$ uniquely, similarly to $N_{0,0}$.} In the following we
assume that this conjecture holds true and that the master
BRST-identity is fulfilled. In particular we will use the modified
generalized perturbative gauge invariance (\ref{N6'}).

From the time ordered products $(T_n^N)_n$ we obtain the corresponding
anti-chronological products $(\bar T_n^N)_n$ by (\ref{T:bar}).
In terms of the $T^N$- and $\bar T^N$-products we construct the
totally retarded products $(R^N_{n+1})_n$ (\ref{R}). The generating
functional of the latter is the interacting field
$\Lambda^N_{g{\cal L}}$ or $W^N_{g{\cal L}}(f)$ (\ref{intfield}).
Similarly to the original $R$-products,
the $R^N$-products have retarded support
\begin{equation}
{\rm supp}\>R^N_{n+1}(...)(x_1,...,x_n;x)\subset\{(y_1,...,y_n,y)|
y_l\in y+\bar V_-,\forall l\}.\label{supp(R)}
\end{equation}
The proof of this support property uses only the causal or
anti-causal factorization of the $T^N$- and $\bar T^N$-products
(see \cite{EG}). The replacement
$\Lambda_{g{\cal L}}\rightarrow \Lambda^N_{g{\cal L}}(f)$ is a
finite renormalization of the interacting field.

The field equation for
$\varphi_{g{\cal L}}^N$ (where $\varphi\in\mathcal{P}_0$
corresponds to a free field without any derivative) differs from
the one of $\varphi_{g{\cal L}}$: instead of (\ref{fieldeq}) the master
Ward identity implies
\begin{gather}
(\w +m^2)\varphi_{g{\cal L}}^N(x)=-g(x)\Bigl(\frac{\d{\cal L}}{\d\chi}
\Bigr)_{g{\cal L}}^N(x)\notag\\
-\frac{1}{2}(g(x))^2\Bigl(\frac{\d N_{0,0}}
{\d\chi}\Bigr)_{g{\cal L}}^N(x)
+\d^\mu_x [g(x)\Bigl(\frac{\d{\cal L}}{\d(\d^\mu\chi)}
\Bigr)_{g{\cal L}}^N (x)],\label{fieldeq-N}
\end{gather}
where we use that $N_{0,0}$ contains no derivatives. The additional term
corresponds to the contribution to the Euler-Lagrange equation
of that quadrilinear terms (in the conventional Lagrangian)
which belong to $N_{0,0}$. The contributions of the quadrilinear terms
belonging to the $C_{A},\> C_u,\>C_{\phi}$- and $C_H$-terms are
contained in the first term on the r.h.s. of (\ref{fieldeq-N}).
For example the contribution of the $g_0^2A^4$-coupling
(which belongs to the $C_A$-terms) is contained
in the $g_0AF$-term in (\ref{fieldeq:A}), because $F_{g{\cal L}}$
(\ref{F_gL}) has a nonlinear term $\sim g_0(AA)_{g{\cal L}}$.
The latter is indeed $\sim C_A$ in our framework.

The construction of the interacting BRST-current $\tilde j_{g{\cal L}}$
(\ref{j:tilde}) must be modified correspondingly. We define
\begin{gather}
\tilde T^N_{n+1}(j^\mu f\otimes {\cal L}_{j_1}f_1\otimes...
\otimes {\cal L}_{j_n}f_n)\=d [Q_0,T^{N}_{n+1}(k^\mu f\otimes
{\cal L}_{j_1}f_1\otimes...\otimes {\cal L}_{j_n}f_n)]_\mp \notag\\
+i\sum_{l=1}^n (-1)^{j_1+...+j_{l-1}}
T^N_{n+1}(k^\mu f\otimes {\cal L}_{j_1}f_1
\otimes...\otimes {\cal L}_{j_l+1}^\nu\d_\nu f_l\otimes
...\otimes {\cal L}_{j_n}f_n).\label{T:tilde'}
\end{gather}
$\tilde T^N(j, {\cal L}_{j_1},...)$ has the same properties as
$\tilde T(j, {\cal L}_{j_1},...)$ (\ref{T:tilde}), in particular it
satisfies causality (\ref{caus}) and the normalization condition {\bf (N0)}.
Hence, $T(j, {\cal L}_{j_1},...)\rightarrow\tilde T^N(j, {\cal L}_{j_1},...)$
is a change of normalization. The divergence identity (\ref{divT(jL)})
holds true also for $(\tilde T^N_{n+1},T^N_n)$, because $T^{N}
(k^\mu,{\cal L}_{j_1},...)$ fulfills {\bf (N5)(ghost)} (\ref{N5:ghost:T'^N})
and $T^N({\cal L}_{j_1},...)$ satisfies generalized perturbative gauge
invariance (\ref{N6'}). Let $\tilde R^N_{n+1}
({\cal L},...,{\cal L};j)$ be the $R$-product (\ref{R}) which is
constructed in terms of $T^N_k({\cal L},...,{\cal L})$
and $\tilde T^N_{k+1}(j,{\cal L},...,{\cal L}),\>1\leq k\leq n$.
Then we define $\tilde j^{N\,\mu}_{g{\cal L}}(f)$ similarly to
(\ref{j:tilde}), replacing $\tilde R_{n+1}(...)$ by $\tilde R^N_{n+1}(...)$.
Analogously to (\ref{div(j)}) we then find that this interacting
BRST-current is conserved up to terms $\sim\d g$, more precisely
\begin{equation}
\tilde j^{N\,\mu}_{g{\cal L}} (\d_\mu f)=-{\cal L}_{1\,g{\cal L}}^{N\,\nu}
(f\d_\nu g).\label{div(j^N)}
\end{equation}
The interacting BRST-charge is now defined by
\begin{equation}
Q^N_{g{\cal L}}\=d \int d^4x\,h^\mu (x)\tilde j^N_{\mu\,g{\cal L}}(x)
\label{Q^N}
\end{equation}
instead of (\ref{Q_gL}). As in the massless case the construction can
be done such that $Q^N_{g{\cal L}}$ fulfills (\ref{Q_{int}}) and
is nilpotent (\ref{Q^2=0}) (see subsection 4.5.4).

We turn to the BRST-transformation of the interacting fields
$\in {\cal F}({\cal O})$ (\ref{F(O)}). Let $s_0
(W)=i\d^\nu W'_\nu$. The formula (\ref{N6b}) is violated by the non-vanishing
of $G(({\cal L}_0,{\cal L}_1)f, {\cal L}_0g)+(f\leftrightarrow g)$.
It must be modified:
\begin{gather}
[Q_0,R^N_{n+1}\bigl(({\cal L}_0g)^{\otimes n}; Wf\bigr)]_\mp =\notag\\
-inR^N_{n+1}\bigl(({\cal L}_0g)^{\otimes (n-1)}\otimes {\cal L}_1^\nu
\d_\nu g; Wf\bigr)
-iR^N_{n+1}\bigl(({\cal L}_0g)^{\otimes n};W'_\nu\d^\nu f\bigr)\notag\\
+nR^N_n\Bigl(({\cal L}_0g)^{\otimes (n-1)};
[G(({\cal L}_0,{\cal L}_1)g,Wf)+G((W,W')f,{\cal L}_0g)]
\Bigr),\label{N6a'}
\end{gather}
where for simplicity we assume that $W'\in {\cal P}_0$ does not contain
any derivative of $\phi_a$ or $H$.\footnote{For
$W\in [\{{\cal L}_0,{\cal L}_1,{\cal L}_2\}]$ (the $[...]$-bracket
denotes the linear span) this assumption is not needed:
$[G(...)+G(...)]$ vanishes and (\ref{N6a'}) follows immediately from
(\ref{N6'}).} This assumption ensures
\begin{equation}
G((W,W')f,N_{0,0}g)=0,\quad\quad\forall f,g\in {\cal D}(\RR^4),
\label{G(W,N)=0}
\end{equation}
as can be seen by inserting the explicit expression (\ref{N}) for
$N_{0,0}$ into the definition (\ref{j16}) of $G(.,.)$. Moreover,
one finds
\begin{equation}
G^{(1)}(N_{0,0}g,Wf)=0,\quad\quad\forall f,g\in {\cal D}(\RR^4),
\label{G1(N,W)=0}
\end{equation}
by inserting (\ref{N}) into the definition (\ref{j14}) of $G^{(1)}$.
Analogously to the derivation of (\ref{N6'}),
the proof of (\ref{N6a'}) is a straightforward application of the
master BRST-identity to $[Q_0,T^{N}_{l+1}\bigl(
Wf\otimes ({\cal L}_0g)^{\otimes l}\bigr)]$, which uses (\ref{G=N}),
(\ref{G3=0}), (\ref{G4=0}), (\ref{G(W,N)=0}) and (\ref{G1(N,W)=0}).
The resulting formula is then translated into an identity for the
$R$-products (\ref{R}). As in the massless case, (\ref{N6a'})
yields the BRST-transformation
\begin{gather}
\tilde s(W^N_{g{\cal L}}(f))\=d [Q^N_{g{\cal L}},W^N_{g{\cal L}}(f)]_\mp=
-iW^{\prime\nu\,N}_{g{\cal L}}(\d_\nu f)\notag\\
+i\bigl(G(({\cal L},{\cal L}_1)g,Wf)+G((W,W')f,{\cal L}g)
\bigr)^N_{g{\cal L}},\quad f\in {\cal D}({\cal O})\label{trafo:intfield'}
\end{gather}
(see subsection 4.5.4).

\subsubsection{The interacting BRST-charge}

In this subsection we summarize the construction of the interacting
BRST-charge given in \cite{DF}. With that we prove the properties
(\ref{Q_{int}}) and the nilpotency. Finally we show that the identity
(\ref{N6b}) ((\ref{N6a'}) resp.) implies the BRST-transformation formula
(\ref{trafo:intfield}) ((\ref{trafo:intfield'}) resp.) for the
interacting fields. We deal with massive
gauge fields, however, the massless case is included: in all formulas
it is allowed to set $m=0$, $\phi\equiv 0$ and $H\equiv 0$ (which replaces
$T^N$ by $T$ etc.).

We assume that the double cone ${\cal O}$ (\ref{F(O)}) is
centered at the origin. Let $r$ be the diameter of ${\cal O}$. The question
is, how to choose $g$ and $h^\mu$ such that $Q^N_{g{\cal L}}$ (\ref{Q^N})
satisfies (\ref{Q_{int}}) and is nilpotent. As explained in
\cite{DF} it is hard to avoid a volume divergence in
$\int d^4x\,h^\mu (x)\tilde j^N_{\mu\,g{\cal L}}(x)$, at
least for massless fields. To get rid of
this problem we proceed as in \cite{DF}: we embed our double cone
${\cal O}$ isometrically into
the cylinder  $\RR\times C_L$ (the first factor denotes the time
axis), where $C_L$ is a cube of
length $L,\quad L\gg r$, with metallic boundary
conditions for each free gauge fields $A_a$ ($a=1,...,N$),
and Diriclet boundary
conditions for the free ghost fields $u_a,\>\tilde u_a$ and the
free bosonic scalar fields $\phi_a$ and $H$. If we choose
the compactification length $L$ big enough, the physical
properties of the local algebra ${\cal F}({\cal O})$ are unchanged.

Following \cite{DF} we choose the switching function
$g$ to fulfil
\begin{equation}
g(x)=1\quad\forall x\in {\cal O}\cup
\{(x_0,\vec x)|\,|x_0|<\epsilon\}\quad (r\gg\epsilon>0)\label{g}
\end{equation}
on $\RR\times C_L$ and to have compact support in
timelike directions. In addition let $h^\mu$ be such that
\begin{equation}
h^{\mu}(x)\=d\delta^{\mu}_{0}h(x_0),
\quad\mathrm{where}\quad h\in {\cal D}([-\epsilon,\epsilon],\RR),
\quad\int dx_0\,h(x_0)=1.\label{h}
\end{equation}
Then
\begin{equation}
Q^N_{g{\cal L}}\=d \int dx_0\,h(x_0)\int_{C_L} d^3x\,\tilde
j^N_{0\,g{\cal L}}(x)\label{Q^N:comp}
\end{equation}
is well-defined, because $(x_0,\vec x)
\rightarrow h(x_0)$ is an admissible test function on $\RR\times C_L$.
$(Q^N_{g{\cal L}})_0=Q_0$ holds true, since $(\tilde
j^N_{\mu\,g{\cal L}})_0=j_\mu$ is conserved.
From the conservation of the interacting current, $\d^\mu\tilde
j^N_{\mu\,g{\cal L}}(x)=0$ for $x\in [-\epsilon,\epsilon]\times C_L$, we
conclude that $Q^N_{g{\cal L}}$ is independent from the choice of $h$.
By {\bf (N2)} and by the fact that $h$ and $g$ are real-valued we obtain
$Q^{N\>*}_{g{\cal L}}=Q^N_{g{\cal L}}$.

It remains to prove the nilpotency. For this purpose we first show
\begin{equation}
Q^N_{g{\cal L}}=Q_0+{\cal L}_{1\,g{\cal L}}^{N\,\nu}(H\d_\nu g)\label{Q^N:L_1}
\end{equation}
where
\begin{equation}
H(x)\equiv \tilde H(x_0)\=d\int_{-\infty}^{x_0}dt\,[-h(t)+h(t-a)]\label{H}
\end{equation}
and $a\in\RR$ is such that the support of $(x_0,\vec x)
\rightarrow h(x_0-a)$ is earlier than the support of $g$:
\begin{equation}
x_0<y_0,\quad\forall x_0\in{\rm supp}\>h(\cdot -a)\>\wedge\>\forall y_0\>
{\rm with}\>\exists \vec y\in C_L\> {\rm with}\> (y_0,\vec y)\in
{\rm supp}\>g.\label{a}
\end{equation}
In particular we will need
\begin{gather}
H(y)\d g(y)=\d g(y),\quad\quad\forall y\in (({\rm supp}\>H
\cup {\cal O})+\bar V_-),\label{Hdg}\\
{\cal O}\cap (\supp (H\d g)+\bar V_-)=\emptyset\label{O-Hdg}
\end{gather}
and
\begin{equation}
\d H(y)\d g(y)=0.\label{dHdg}
\end{equation}

{\bf Proof of (\ref{Q^N:L_1})}: From our definitions we immediately obtain
\begin{equation}
Q^N_{g{\cal L}}=\int_{\RR\times C_L}d^4x\,\tilde j^{N\,\mu}_{g{\cal L}}(x)
[-\d_\mu H(x)+g_{\mu 0}h(x_0-a)].\label{Q^N'}
\end{equation}
Due to (\ref{a}) and the retarded support of the $R$-products (\ref{supp(R)})
we have
\begin{equation}
{\rm supp}\>(\tilde j^N_{\mu\,g{\cal L}}-j_\mu)\cap {\rm supp}\>((x_0,\vec x)
\rightarrow h(x_0-a))=\emptyset.
\end{equation}
So the contribution of $g_{\mu 0}h(x_0-a)$ to (\ref{Q^N'}) is $Q_0$, and by
inserting (\ref{div(j^N)}) into the $\d_\mu H$-term we obtain the assertion
(\ref{Q^N:L_1}). $\quad\w$

The formula (\ref{Q^N:L_1}) manifestly shows that $Q^N_{g{\cal L}}$
converges to $Q_0$ in the adiabatic limit $g(x)\rightarrow 1,\>\forall x$,
provided this limit exists. For pure massive theories this limit exists
indeed in the strong sense \cite{EG1}. So, in the adiabatic limit
of a pure massive gauge theory, one has the simplification that the
BRST-cohomology is given in terms of $Q_0$ (cf.
(\ref{[Q,S]})-(\ref{adlim(S)}) and \cite{DSchroer}).\\
\\
In the following proofs we will use Proposition 2 of \cite{DF}, which
is a formula for the (anti-)commutator of two interacting
fields:\footnote{This formula, the retarded support of the
$R$-products (\ref{supp(R)}) and some further, quite obvious requirements
can be viewed as the defining properties of the retarded products.
They determine a direct construction of the $R_{n+1}\>(n\in\NN_0)$
by induction on $n$. If wanted, the $T$-products can then be obtained
by reversing (\ref{R}), see \cite{SD}, the second paper of
\cite{DF1} and \cite{DF3}.}
\begin{equation}
[W^N_{g{\cal L}}(f),V^N_{g{\cal L}}(h)]_\mp =
R^N_{g{\cal L}}(W,V)(f,h)\mp R^N_{g{\cal L}}(V,W)(h,f),\label{comm:intfield}
\end{equation}
where
\begin{equation}
R^N_{g{\cal L}}(W,V)(f,h)\=d
-\sum_{n=0}^\infty\frac{i^n}{n!}R^N_{n+2}\bigl(
(g{\cal L})^{\otimes n}\otimes Wf;Vh\bigl).\label{R(W,V)}
\end{equation}
Due to the retarded support (\ref{supp(R)}) of the $R^N$-products we
call $R^N_{g{\cal L}}(W,V)(f,h)$ the retarded part (and $\mp
R^N_{g{\cal L}}(V,W)(h,f)$ the advanced part) of
$[W^N_{g{\cal L}}(f),V^N_{g{\cal L}}(h)]_\mp$.\\
\\
{\bf Proof of} $\mathbf{(Q^N_{g{\cal L}})^2=0}$:
Generalized perturbative gauge invariance (\ref{N6'}) implies the relation
\begin{gather}
[Q_0,R^N_{n+1}\bigl(({\cal L}g)^{\otimes n};{\cal L}_1^\mu f\bigr)]_+=
-inR^N_{n+1}\bigl(({\cal L}g)^{\otimes (n-1)}\otimes
{\cal L}_1^\rho\d_\rho g;{\cal L}_1^\mu f\bigr)\notag\\
-iR^N_{n+1}\bigl(({\cal L}g)^{\otimes n};
{\cal L}_2^{\mu\rho}\d_\rho f\bigr).\label{[Q_0,R^N]}
\end{gather}
Due to ${\cal L}_2^{\mu\rho}=-{\cal L}_2^{\rho\mu}$ we may require
\begin{equation}
T^N_l({\cal L},...,{\cal L},{\cal L}_2^{\mu\rho})=-
T^N_l({\cal L},...,{\cal L},{\cal L}_2^{\rho\mu}),
\quad\quad\forall l\in\NN.\label{mu-rho}
\end{equation}
This is an additional normalization condition, which is compatible with
the other normalization conditions. Similarly to {\bf (N2)}, it can
be satisfied by antisymmetrizing in $\mu\leftrightarrow\rho$ a
$T^N_l({\cal L},...,{\cal L};{\cal L}_2^{\mu\rho})$ which satisfies the
other normalization conditions. The corresponding $R^N_{n+1}({\cal L},...,
{\cal L};{\cal L}_2^{\mu\rho})$ is then also antisymmetric in
$\mu\leftrightarrow\rho$ and, hence, we have ${\cal L}_{2\,g{\cal L}}
^{N\,\rho\mu}=-{\cal L}_{2\,g{\cal L}}^{N\,\mu\rho}$.

By means of (\ref{Q^N:L_1}) and $Q_0^2=0$ we find
\begin{equation}
2(Q^N_{g{\cal L}})^2=[Q^N_{g{\cal L}},Q^N_{g{\cal L}}]_+=
2[Q_0,{\cal L}_{1\,g{\cal L}}^{N\,\mu}(H\d_\mu g)]_+ +
[{\cal L}_{1\,g{\cal L}}^{N\,\nu}(H\d_\nu g),
{\cal L}_{1\,g{\cal L}}^{N\,\mu}(H\d_\mu g)]_+ .\label{Q^2=...}
\end{equation}
Using (\ref{[Q_0,R^N]}) and (\ref{R(W,V)}) we obtain
\begin{equation}
2[Q_0,{\cal L}_{1\,g{\cal L}}^{N\,\mu}(H\d_\mu g)]_+=
-2i{\cal L}_{2\,g{\cal L}}^{N\,\mu\rho}(\d_\rho (H\d_\mu g))-
2R^N_{g{\cal L}}({\cal L}_{1}^{\nu},{\cal L}_{1}^{\mu})
(\d_\nu g,H\d_\mu g).\label{Q_0Q_n}
\end{equation}
In the first term the $(\d_\rho H)(\d_\mu g)$-part vanishes by (\ref{dHdg})
and the $H\d_\rho\d_\mu g$-part is zero because of (\ref{mu-rho}).
In the remaining second term we first take (\ref{Hdg}) into account
and then apply (\ref{comm:intfield})
\begin{gather}
2[Q_0,{\cal L}_{1\,g{\cal L}}^{N\,\mu}(H\d_\mu g)]_+=
-2R^N_{g{\cal L}}({\cal L}_{1}^{\nu},{\cal L}_{1}^{\mu})
(H\d_\nu g,H\d_\mu g)\notag\\
=-[{\cal L}_{1\,g{\cal L}}^{N\,\nu}(H\d_\nu g),
{\cal L}_{1\,g{\cal L}}^{N\,\mu}(H\d_\mu g)]_+.
\end{gather}
Inserting this into (\ref{Q^2=...}) we see that $Q^N_{g{\cal L}}$ is
in fact nilpotent. $\quad\w$\\
\\
{\bf Proof of the formula}
\begin{equation}
[Q^N_{g{\cal L}},W^N_{g{\cal L}}(f)]_\mp =
-iW^{\prime\nu\,N}_{g{\cal L}}(\d_\nu f)+i\bigl(
G(({\cal L},{\cal L}_1)g,Wf)+G((W,W')f,{\cal L}g)\bigr)^N_{g{\cal L}}
\label{trafo:intfield''}
\end{equation}
{\bf for the BRST-transformation of the interacting fields}:
let $f\in {\cal D}({\cal O})$. According to (\ref{Q^N:L_1}) we have to
compute two terms:
\begin{equation}
[Q^N_{g{\cal L}},W^N_{g{\cal L}}(f)]_\mp =
[Q_0,W^N_{g{\cal L}}(f)]_\mp
+[{\cal L}_{1\,g{\cal L}}^{N\,\nu}(H\d_\nu g),W^N_{g{\cal L}}(f)]_\mp.
\label{[Q^N,W^N]}
\end{equation}
For the first one the identity (\ref{N6a'}) gives
\begin{gather}
[Q_0,W^N_{g{\cal L}}(f)]_\mp =
-iW^{\prime\nu\,N}_{g{\cal L}}(\d_\nu f)\notag\\
+i\bigl(G(({\cal L},{\cal L}_1)g,Wf)+G((W,W')f,{\cal L}g)
\bigr)^N_{g{\cal L}}
-R^N_{g{\cal L}}({\cal L}_{1}^{\nu},W)(\d_\nu g,f),
\label{[Q_0,W^N]}
\end{gather}
where we have used (\ref{R(W,V)}).

We turn to the other term in (\ref{[Q^N,W^N]}) and
apply (\ref{comm:intfield}):
\begin{equation}
[{\cal L}_{1\,g{\cal L}}^{N\,\nu}(H\d_\nu g),W^N_{g{\cal L}}(f)]_\mp =
R^N_{g{\cal L}}({\cal L}_{1}^{\nu},W)(H\d_\nu g,f)\mp
R^N_{g{\cal L}}(W,{\cal L}_{1}^{\nu})(f,H\d_\nu g).
\label{R-R}
\end{equation}
The second term vanishes due to the support properties
(\ref{supp(R)}) and (\ref{O-Hdg}). Because of (\ref{Hdg}) we may omit
the factor $H$ in the first term. Hence, we find that
$[{\cal L}_{1\,g{\cal L}}^{N\,\nu}(H\d_\nu g),W^N_{g{\cal L}}(f)]_\mp$
cancels out with the last term in (\ref{[Q_0,W^N]}) and it remains the
assertion (\ref{trafo:intfield''}). $\quad\w$\\
\\
We summarize: the following conditions on an interaction ${\cal L}_0$
are sufficient for our local construction of observables.\\
- ${\cal L}_0$ fulfills the conditions (A)-(F) given at the beginning
of sect. 4.5.1. or 4.5.3 resp..\\
- There exist sums $N_{j,k}\in {\cal P}_0$ of monomials of degree
four which satisfy (\ref{G=N})-(\ref{N_{j,k}}) and (\ref{G3=0})-
(\ref{G4=0}).\\
- There is no restriction coming from ghost number conservation
{\bf (N5) (ghost)}. In the relevant cases, this normalization
condition has common solutions with {\bf (N0)-(N3)}, see
\cite{BDF} appendix B.1.\\
- Simultaneously with {\bf (N0)-(N3)} and {\bf (N5) (ghost)},
the master BRST-identity can be fulfilled for
\begin{equation}
  [Q_0,T_n({\cal L}_0,...,{\cal L}_0)]\quad\quad\mathrm{and}\quad\quad
[Q_0,T_n({\cal L}_1,{\cal L}_0,...,{\cal L}_0)]\label{vs1}
\end{equation}
to all orders $n\in\NN$.

The conditions mentioned so far suffice for the construction of the
BRST-charge $Q^N_{g{\cal L}}$. In our derivation of the
BRST-transformation formula (\ref{trafo:intfield'}) of
$W^N_{g{\cal L}}$ ($W\in {\cal P}_0$ arbitrary) we have additionally
used:\\
- the master BRST-identity for
\begin{equation}
[Q_0,T_n(W,{\cal L}_0,...,{\cal L}_0)],\quad\quad\forall n\in\NN,\label{vs2}
\end{equation}
- the properties (\ref{G(W,N)=0}) and (\ref{G1(N,W)=0}) of $N_{0,0}$.

\section{Taking anomalies into account}

\subsection{General procedure and axial anomaly}

We recall that we understand by the expression 'anomaly' any term that
violates the master Ward identity  {\bf (N)}
and cannot be removed by an admissible finite renormalization of
the $T$-products. The aim of this section is to take anomalies
into account in the formulation of the MWI.
This has consequences for the normalization condition ${\bf (\tilde N)}$:
We want to normalize $T((\tilde\d^\nu V)\otimes W_1f_1\otimes...)$
such that (\ref{heur:tilde-d}) holds true: in the special case $W=1,\> V,
W_1,...,W_n\in {\cal P}_0$ the r.h.s. of ${\bf (\tilde N)}$ should agree
with the r.h.s. of {\bf (N)}. Hence, we will take the anomalies into account
also in ${\bf (\tilde N)}$.

We proceed inductively with respect to the order $n$. Since we are not
aware of a general proof that second order loop diagrams (i.e. $n=1$
in {\bf (N)}) are anomaly-free\footnote{In our proof of
charge conservation {\bf (N5) (charge)} (which is given in appendix
B of \cite{DF}) vacuum polarization plays an exceptional role, even to
second order: our general argumentation does not yield
$\d^\mu_x T_2(j_\mu(x)j_\nu(y))=0$, an explicit calculation was necessary.},
we start with that case. We set
\begin{eqnarray}
-a_{V,W}^{(2)\nu}(g,f)\=d T_2(V\d^\nu g\otimes Wf)+T_2((\d^\nu V)g
\otimes Wf)\nonumber\\
+i\sum_{\chi,\psi\in {\cal G}}(\pm) T_1\Bigl(\Delta^{\nu}_{\chi,\psi}
\bigl(\frac{\d V}{\d\chi}g,\frac{\d W}{\d\psi}f\bigr)\Bigr),
\quad V,W\in\tilde {\cal P}_0.\label{a:2}
\end{eqnarray}
For later purpose we let $V,W\in\tilde {\cal P}_0$ (not only $V,W\in
{\cal P}_0$ as in the MWI {\bf (N)}).
Due to causal factorization of the $T$-products we know that
$a_{V,W}^{(2)}(g,f)$ is local. Therefore, there exists a unique
$b_{V,W}^{(2)}(g,f)\in {\cal D}(\RR^4,{\cal P}_0)$ with\footnote{Remember
$T_1({\cal D}(\RR^4,{\cal P}_0))=T_1({\cal D}(\RR^4,\tilde {\cal P}_0))$ and that
$T_1$ is injective on ${\cal D}(\RR^4,{\cal P}_0)$.}
\begin{equation}
a_{V,W}^{(2)\nu}(g,f)=T_1(b_{V,W}^{(2)\nu}(g,f)).\label{b:2}
\end{equation}
Let us now assume that we have already defined $a_{V,W_1,...,W_m}^{(m+1)}$
and $b_{V,W_1,...,W_m}^{(m+1)}\in {\cal D}(\RR^4,{\cal P}_0)$
for all $m< n$. We then set
\begin{eqnarray}
-a_{V,W_1,...,W_n}^{(n+1)\nu}(g,f_1,...,f_n)\=d
T_{n+1}(V\partial^\nu g\otimes W_1f_1\otimes...\otimes W_nf_n)\nonumber\\
+T_{n+1}((\partial^\nu V)g\otimes W_1f_1\otimes...\otimes W_nf_n)\nonumber\\
+i\sum_{m=1}^n\sum_{\chi,\psi\in {\cal G}}(\pm) T_n\Bigl(
\Delta^{\nu}_{\chi,\psi}\bigl(\frac{\d V}{\d\chi}g,\frac{\d W_m}{\d
\psi}f_m\bigr)\otimes W_1f_1\otimes...\hat m...\otimes W_nf_n\Bigr)\nonumber\\
+\sum_{k=1}^{n-1}\sum_{1\leq m_1<...<m_k\leq n}(\pm)
T_{n+1-k}\Bigl(b_{V,W_{m_1},...,W_{m_k}}^{(k+1)\nu}(g,f_{m_1},...,f_{m_k})
\nonumber\\
\otimes W_1f_1\otimes...\hat m_1...\hat m_k...
\otimes W_nf_n\Bigr),\label{a:n+1}
\end{eqnarray}
where $V,W_1,...,W_n\in\tilde {\cal P}_0$ and
the possible signs $(\pm)$ come from the permutation of Fermi operators.
By causal factorization and the definition of the $(b^{(k+1)})_{k<n}$
we conclude that $a_{V,W_1,...,W_n}^{(n+1)}
(g,f_1,...,f_n)$ is local and, hence, that there exists a unique
$b_{V,W_1,...,W_n}^{(n+1)}(g,f_1,...,f_n)\in {\cal D}(\RR^4,{\cal P}_0)$ with
\begin{equation}
a_{V,W_1,...,W_n}^{(n+1)}(g,f_1,...,f_n)=
T_1(b_{V,W_1,...,W_n}^{(n+1)}(g,f_1,...,f_n)).\label{b:n+1}
\end{equation}
Obviously
\begin{gather}
  b^{(n+1)}:{\cal D}(\RR^4,\tilde {\cal P}_0)^{\otimes n+1}\rightarrow
{\cal D}(\RR^4,{\cal P}_0),\notag\\
Vg\otimes W_1f_1\otimes...\otimes W_nf_n
\rightarrow b_{V,W_1,...,W_n}^{(n+1)}(g,f_1,...,f_n)\label{b}
\end{gather}
is linear and symmetrical in all factors except the first one.
As a consequence of the normalization condition {\bf (N3)} the
$b_{V,W_1,...,W_n}^{(n+1)},\quad V,W_1,...,W_n\in
{\cal P}_0,\>n$ fixed, are not
independent. For example let $V$ be a sub-polynomial of $V'$ and $W_k$ a
sub-polynomial of $W'_k,\>\forall k=1,...,n$ ($V=V'$ and $W_k=W'_k$ is
admitted). Then $b_{V,W_1,...,W_n}^{(n+1)}\not= 0$ implies
$b_{V',W'_1,...,W'_n}^{(n+1)}\not= 0$\footnote{To see this we consider the
$(V',W'_1,...,W'_n)$-diagrams in which the additional factors of
$V',W'_1,...,W'_n$ are external legs. By amputating these external legs
we obtain all $(V,W_1,...,W_n)$-diagrams. {\bf (N3)} requires that the
non-amputated and amputated diagrams are equally normalized.}. However,
the $b_{V,W_1,...,W_n}^{(n+1)}$ are independent for {\it different} $n$,
because the violations of the MWI
coming from sub-diagrams are taken into account
in (\ref{a:n+1}) by the terms $\sum_{k=1}^{n-1}\sum (\pm)
T_{n+1-k}(b^{(k+1)}(...)\otimes ...)$.

Obviously the $b^{(k)}$ (\ref{a:n+1})-(\ref{b:n+1}) depend on the
normalization of the $T$-products. We assume that the latter fulfil
{\bf (N0)-(N3)} and ${\bf (\tilde N)}$ in the following modified form:
\begin{eqnarray}
{\bf (\tilde N)}&&T_{n+1}((\tilde\partial^\nu V)Wg\otimes W_1f_1
\otimes...\otimes W_nf_n)=\nonumber\\
&&T_{n+1}((\partial^\nu V)Wg\otimes
W_1f_1\otimes...\otimes W_nf_n)\nonumber\\
&&+i\sum_{m=1}^n\sum_{\chi,\psi\in\tilde {\cal G}}(\pm) T_n\Bigl(
\Delta^{\nu}_{\chi,\psi}\bigl(\frac{\partial V}
{\partial\chi}Wg,\frac{\partial W_m}{\partial\psi}f_m\bigr)
\otimes W_1f_1\otimes...\hat m...\otimes W_nf_n\Bigr)\nonumber\\
&&+\sum_{k=1}^{n}\sum_{1\leq m_1<...<m_k\leq n}(\pm)
T_{n+1-k}\Bigl(b_{V,W_{m_1},...,W_{m_k}}^{(k+1)\nu}(g,f_{m_1},...,f_{m_k})W
\nonumber\\
&&\otimes W_1f_1\otimes...\hat m_1...\hat m_k...\otimes W_nf_n\Bigr),
\label{tilde-N:neu}
\end{eqnarray}
where $V,W,W_1,...,W_n\in\tilde {\cal P}_0$.
Note that the sum over $k$ in the last term runs here up to $n$. Setting
$W=1$ and using the definition (\ref{a:n+1})-(\ref{b:n+1}), we in fact obtain
(\ref{heur:tilde-d}) (even for $V,W_1,...,W_n\in\tilde {\cal P}_0$), which
is the main reason for this modification of ${\bf (\tilde N)}$. In sect. 2
this implication relied on the validity of the MWI
{\bf (N)}. This assumption is not needed here to get
(\ref{heur:tilde-d}).

Also the modified ${\bf (\tilde N)}$ fixes the normalization of the
$T$-products of symbols with external derivatives in terms of
$T$-products without external derivatives, namely by the following
recursive procedure: For a monomial
$W=\prod_s\tilde \d^{a^{(s)}}\varphi_{r_s}\in\tilde {\cal P}_0,
\>\>\varphi_r\in {\cal P}_0,\>\> a^{(s)}\in(\NN_0)^4,$ we define $|W|=
\sum_s|a^{(s)}|$ where $|a^{(s)}|=a^{(s)}_0+...+a^{(s)}_3$. Let the
normalization of $T_n(W_1,...,W_n)$ with $|W_1|+...+|W_n|=0$ (i.e.
$W_1,...,W_n\in {\cal P}_0$) be given for all $n\in\NN$. Then the
determination of the $b_{W_1,...,W_n}^{(n)}$ and of the normalization
of $T_n(W_1,...,W_n)$ with $|W_1|+...+|W_n|>0$ goes in a double inductive
way: one makes a first induction with respect to the order $n$ and for
each fixed $n$ a second induction with respect to $|W_1|+...+|W_n|$.
More precisely let $T_l(W_1,...,W_l)$ and $b_{W_1,...,W_l}^{(l)}$ be
given for all $l\leq n$ and $W_1,...,W_l\in \tilde {\cal P}_0$,
and also for $l=n+1$ if $|W_1|+...+|W_{n+1}|<d
\quad (d\in\NN)$. Then we determine by
${\bf (\tilde N)}$ (\ref{tilde-N:neu}) the normalization of the
$T_{n+1}(W_1,...,W_{n+1})$ with $|W_1|+...+|W_{n+1}|=d$ (this step does
not take place for $d=0$, because $T_{n+1}$ is given in that case).
More precisely we use ${\bf (\tilde N)}$ for $|V|+|W|+|W_1|+...+|W_n|
=d-1$. Note that all $b^{(k+1)}$ and all $T$-products which appear in
this case on the r.h.s. of ${\bf (\tilde N)}$ are inductively given.
Finally, from (\ref{a:n+1})-(\ref{b:n+1}) we obtain the $b^{(n+1)}_
{V,W_1,...,W_n}$ with $|V|+|W_1|+...+|W_n|=d$. Again we point out that
thereby all terms which appear on the r.h.s. of (\ref{a:n+1}) are
inductively known. Starting this procedure with restricted
$T$-products $(T_n\vert_{\mathcal{D}(\RR^4,\mathcal{P}_0)^{\otimes
    n}})_{n\in\NN}$ which satisfy {\bf (N0)-(N3)} (such $T$-products
exist \cite{EG}) we end up with $T$-products which fulfil
{\bf (N0)-(N3)} and the modified ${\bf (\tilde N)}$.

To formulate the modified MWI we specialize to the case
$V,W_1,...,W_n\in {\cal P}_0$. Let us consider the set ${\cal T}$
of all sequences of $T$-products $(T_n)_{n\in\NN}$ which satisfy the
requirements of sect. 2.2 (in particular causality and the normalization
conditions {\bf (N0)-(N3)}) and the modified ${\bf (\tilde N)}$.
We now define
\begin{equation}
{\cal A}((T_n)_{n\in\NN})\=d (\bar b^{(n+1)})_{n\in\NN},\quad\quad
\forall (T_n)_{n\in\NN}\in {\cal T},
\end{equation}
where $\bar b^{(n+1)}$ is the restriction of $b^{(n+1)}$ (\ref{b}) to
${\cal D}(\RR^4,{\cal P}_0)^{\otimes n+1}$. The image ${\cal A}({\cal T})$
of this map is model dependent and it is usually hard work to get
information about ${\cal A}({\cal T})$.
If the zero-sequence (i.e. ${\bf 0}\=d (0,0,...)$) is an element of
${\cal A}({\cal T})$, which means that the model is anomaly-free,
we are in the situation of sect. 2.4: the MWI
is then the normalization condition which forbids all
$(T_n)_{n\in\NN}$ which are not an element of
${\cal A}^{-1}({\bf 0})$. If ${\bf 0}\not\in {\cal A}({\cal T})$
we choose a suitable (usually as simple as possible)
${\bf b}\in {\cal A}({\cal T})$ and {\it the
master Ward identity is then the normalization
condition that solely sequences $(T_n)_{n\in\NN}\in
{\cal A}^{-1}({\bf b})$ are allowed}.

We illustrate this by
the example of the axial anomaly. Let
${\cal P}_\psi$ be the linear space which is generated by
${\cal L}\=d A_\mu j^\mu,\,j_A,\,j_\pi$ (cf. (\ref{axan}))
and all sub-monomials thereof.
According to Bardeen \cite{Ba} the most simple ${\bf b}\in {\cal A}
({\cal T})$ reads
\begin{gather}
b_{V,W_1,...,W_n}^{(n+1)}=0,\quad\forall\> n+1\not= 3,\>\>\>V,W_1,...W_n
\in{\cal P}_\psi,\notag\\
b_{j_{A\nu},j^{\mu_1},j^{\mu_2}}^{(3)\nu}(g,f_1,f_2)=C
\epsilon^{\mu_1\mu_2\rho\tau}g(\d_\rho f_1)(\d_\tau f_2),\notag\\
b_{j_{A\nu},{\cal L},{\cal L}}^{(3)\nu}(g,f_1,f_2)=C
\epsilon^{\mu_1\mu_2\rho\tau}g\d_\rho (f_1A_{\mu_1})
\d_\tau (f_2A_{\mu 2}),\notag\\
b_{j_{A\nu},{\cal L},j^\mu}^{(3)\nu}(g,f_1,f_2)=
b_{j_{A\nu},j^\mu,{\cal L}}^{(3)\nu}(g,f_2,f_1)=C
\epsilon^{\mu_1\mu_2\rho\tau}g\d_\rho (f_1A_{\mu_1})\d_\tau f_2,\notag\\
b_{j_A,j_A,j_A}^{(3)}(g,f_1,f_2)=\frac{1}{3}
b_{j_A,j,j}^{(3)}(g,f_1,f_2)
\end{gather}
and $b_{V,W_1,W_2}^{(3)}=0$ for all other $(V,W_1,W_2)\in ({\cal P}_\psi)
^{\times 3}$, where $C$ is a well-known, fixed, complex number.
Then particular cases of the MWI read
\begin{gather}
T_{n+1}(j^\mu\d_\mu f\otimes {\cal L}g_1\otimes ...\otimes {\cal L}g_n)
=0,\notag\\
T_{n+1}(j_A^\mu\d_\mu f\otimes {\cal L}g_1\otimes ...\otimes {\cal L}g_n)
=2mT_{n+1}(j_\pi f\otimes {\cal L}g_1\otimes ...\otimes {\cal L}g_n)\notag\\
+\sum_{1\leq m_1<m_2\leq n}C\epsilon^{\mu_1\mu_2\rho\tau}
T_{n-1}(f\d_\rho (g_{m_1}A_{\mu_1})\d_\tau (g_{m_2}A_{\mu_2})\otimes
{\cal L}g_1\otimes ...\hat m_1...\hat m_2...\otimes {\cal L}g_n),
\end{gather}
which imply
\begin{gather}
j_{g{\cal L}}^\mu (\d_\mu f)=0,\notag\\
-j_{A\,g{\cal L}}^\mu (\d_\mu f)=2mj_{\pi\,g{\cal L}}(f)-
g_0^2\frac{C}{8}\epsilon^{\mu_1\mu_2\rho\tau}(F_{\rho\mu_1}F_{\tau\mu_2})
_{g{\cal L}}(f),
\end{gather}
where we assume $g(x)=g_0={\rm const.}, \forall x\in {\rm supp}\>f$.

Non-vanishing anomalies $b_{V,W_1,...,W_m}^{(m+1)}$ are not an obstacle
to fulfil the normalization condition {\bf (N4)} and hence the field
equation (\ref{fieldeq}) (see sect. 4.1), because (\ref{N4'}) still
solves {\bf (N4)} (\ref{N4}). But
the axial anomaly appears as an additional term in the charge conservation
{\bf (N5)} {\bf (charge)} (\ref{N5}) and in the generalized perturbative
gauge invariance {\bf (N6)} (\ref{N6}) and hence also in the master
BRST-identity, if axial fermions are present.
However, for the non-Abelian gauge models studied in
sect. 4.5, we expect that the master BRST-identity can be
satisfied in the relevant cases (\ref{vs1}) and (\ref{vs2}), and
that therefore our local construction of observables works.
But this remains to be proved.

\subsection{Energy momentum tensor: conservation and trace anomaly}

We follow the procedure in \cite{P:emt}.
Classically the canonical energy momentum tensor is the Noether current
belonging to translation invariance (in time and space).
Turning to QFT we consider
a real, free, scalar field $\phi$ of mass $m\geq 0$. (In the formalism
of appendix A we set $\varphi\=d\chi\=d\phi$ and choose $\epsilon =1$.)
The free canonical energy momentum tensor reads
\begin{equation}
\Theta^{\mu\nu}_{0\,{\rm can}}=\d^\mu\phi\d^\nu\phi-\frac{1}{2}g^{\mu\nu}
\d^\rho\phi\d_\rho\phi+\frac{1}{2}g^{\mu\nu}m^2\phi^2,\label{0,can}
\end{equation}
and this tensor is conserved due to the Klein-Gordon equation:
$\d_\mu \Theta^{\mu\nu}_{0\,{\rm can}}=0$.

Now we add an interaction of the form
\begin{equation}
{\cal L}=\lambda\phi^4.\label{interaction}
\end{equation}
The interacting canonical energy momentum tensor is not simply the
interacting field belonging to $\Theta^{\mu\nu}_{0\,{\rm can}}$, it
has an additional term
\begin{equation}
\Theta^{\mu\nu}_{{\rm can}\,g{\cal L}}(f)=
\Theta^{\mu\nu}_{0\,{\rm can}\,g{\cal L}}(f)
+g^{\mu\nu}{\cal L}_{g{\cal L}}(gf).\label{can}
\end{equation}
Let $W_1,...,W_n$ be polynomials in $\phi$ (without any
derivative). Applying twice the
definition (\ref{a:n+1})-(\ref{b:n+1}) we obtain the relation
\begin{gather}
T(W_1g_1\otimes...\otimes W_ng_n\otimes \Theta^{\mu\nu}_{0\,{\rm can}}
\d_\mu f)=\notag\\
-i\sum_{k=1}^nT(W_1g_1\otimes...\otimes (\d^\nu W_k)g_kf\otimes
...\otimes W_ng_n)-{\cal A}_1(f,g_1,...,g_n)=\notag\\
-{\cal A}_1(f,g_1,...,g_n)+i\sum_{k=1}^n\Bigl(
T(W_1g_1\otimes...\otimes W_k\d^\nu (g_kf)\otimes...\notag\\
...\otimes W_ng_n)+{\cal A}_{2,k}(f,g_1,...,g_n)\Bigr),\label{div(can)}
\end{gather}
where
\begin{gather}
{\cal A}_1(f,g_1,...,g_n)\=d\sum_{j=1}^n\sum_{m_1<...<m_j}T(b^{(j+1)}_{
\Theta^{\mu\nu}_{0\,{\rm can}},W_{m_1},...,W_{m_j}\,\mu}(f,g_{m_1},...,
g_{m_j})\notag\\
\otimes W_1g_1\otimes...\hat m_1...\hat m_j...\otimes W_ng_n),\notag\\
{\cal A}_{2,k}(f,g_1,...,g_n)\=d\sum_{l=1}^{n-1}\sum_{m_1<...<m_l\,
(m_j\not= k)}T(b^{(l+1)\nu}_{W_k,W_{m_1},...W_{m_l}}(g_kf,g_{m_1},...,g_{m_l})
\notag\\
\otimes W_1g_1 ...\hat m_1...\hat m_l...\otimes W_ng_n).
\end{gather}
In \cite{P:emt} it is shown that there exists a normalization (which is
compatible with {\bf (N0) - (N3)}\footnote{So far no external
derivatives are present. Hence, $(\tilde {\rm\bf N})$ plays no role.})
such that $-{\cal A}_1+i\sum_{k=1}^n
{\cal A}_{2,k}=0$. In the following we use this normalization. Then the
identity (\ref{div(can)}) and (\ref{can}) imply
\begin{equation}
\Theta^{\mu\nu}_{{\rm can}\,g{\cal L}}(\d_\mu f)=
-{\cal L}_{g{\cal L}}((\d^\nu g)f).\label{cons:can}
\end{equation}
The energy momentum tensor is only conserved in space-time regions
in which the coupling 'constant' $g$ is constant, in agreement with
the fact that translation invariance is broken by a non-constant $g$.

Unfortunately the trace of the canonical energy momentum tensor does
not vanish, even for free fields.
Following \cite{P:emt} and references cited therein, we assume
$m=0$ (and still $\mathcal{L}=\lambda \phi^4$)
and introduce the improved energy momentum tensor\footnote{In
massive theories it is already at the {\it classical} level impossible
to construct an energy momentum tensor $\Theta^{{\rm class}\,\mu\nu}$
which is conserved and traceless. Because the corresponding dilatation
current $D^{{\rm class}\,\mu}\equiv x_\nu\Theta^{{\rm class}\,\mu\nu}$
would be conserved, but a (non-vanishing) mass breaks dilatation
invariance.}. In (interacting) classical field theory it is defined by
\begin{equation}
\Theta^{{\rm class}\,\mu\nu}_{\rm imp}\=d
\Theta^{{\rm class}\,\mu\nu}_{\rm can}
-\frac{1}{3}[\d^\mu(\phi^{\rm class}\d^\nu \phi^{\rm class})-g^{\mu\nu}
\d^\rho(\phi^{\rm class}\d_\rho \phi^{\rm class})],\label{imp:class}
\end{equation}
where $\Theta^{{\rm class}\,\mu\nu}_{\rm can}$ is given by the same
formulas (\ref{0,can})-(\ref{can}) as in QFT. This improved tensor
is conserved and traceless. The latter relies on the field equation.

Now we are going to construct the corresponding tensor in QFT.
We apply the definition (\ref{a:n+1})-(\ref{b:n+1}) to
$T((\phi\d^\nu\phi)\d^\mu f\otimes ...)$ and $(\tilde {\rm\bf N})$
(\ref{tilde-N:neu}) to $T((\phi\tilde{\d}^\mu\d^\nu\phi) f\otimes
...)$. So we obtain
\begin{gather}
-T((\phi\d^\nu\phi)\d^\mu f\otimes W_1g_1\otimes...\otimes W_n g_n)=
T((\d^\mu\phi\d^\nu\phi)f\otimes W_1g_1\otimes...\otimes W_n g_n)\notag\\
+T((\phi\tilde{\d}^\mu\d^\nu\phi) f\otimes W_1g_1\otimes...\otimes W_n g_n)
+{\cal A}^{(n+1)\,\mu\nu}_{W_1,...,W_n}(f,g_1,...,g_n),
\label{imp:ward}
\end{gather}
where
\begin{gather}
{\cal A}^{(n+1)\,\mu\nu}_{W_1,...,W_n}(f,g_1,...,g_n)\=d
\sum_{k=1}^n\sum_{m_1<...<m_k}T(b^{(k+1)\,\mu}_{\phi\d^\nu\phi,W_{m_1},...,
W_{m_k}}(f,g_{m_1},...,g_{m_k})\notag\\
\otimes W_1g_1\otimes...\hat m_1...\hat m_k...\otimes W_ng_n).
\end{gather}
Here we have normalized $T(\d^\nu\phi,W_{m_1},...,W_{m_k})$ according to
(\ref{N4'}) {\bf (N4)}, which implies
\begin{equation}
b^{(k+1)}_{\d^\nu\phi,W_{m_1},...,W_{m_k}}=0\label{b=0}
\end{equation}
(there are no anomalies for tree-like diagrams, cf. sect 2.4).
Hence, there are no $T(b^{(k+1)}...)$-terms in the application
of $(\tilde {\rm\bf N})$ to $T((\phi\tilde{\d}^\mu\d^\nu\phi) f\otimes
W_1g_1\otimes...)$. In particular it follows
\begin{equation}
T((\phi\tilde{\d}^\mu\d^\nu\phi) f\otimes W_1g_1\otimes...\otimes W_n g_n)
=T((\phi\tilde{\d}^\nu\d^\mu\phi) f\otimes W_1g_1\otimes...\otimes W_n g_n).
\label{mu-nu}
\end{equation}
from  $(\tilde {\rm\bf N})$. For the interacting fields
the identity (\ref{imp:ward}) implies
\begin{equation}
-(\phi\d^\nu\phi)_{g{\cal L}}(\d^\mu f)
=(\d^\mu\phi\d^\nu\phi)_{g{\cal L}}(f)+
(\phi\tilde{\d}^\mu\d^\nu\phi)_{g{\cal L}}(f)+
{\cal A}^{\mu\nu}_g(f),\label{d(pdp)}
\end{equation}
where
\begin{equation}
{\cal A}^{\mu\nu}_g(f)\=d\sum_{n=1}^\infty i^n\sum_{r=1}^n
\frac{(-1)^{n-r}}{r!(n-r)!}
\bar{T}((g{\cal L})^{\otimes (n-r)}){\cal A}^{(r+1)\,\mu\nu}_
{{\cal L},...,{\cal L}}(f,g,...,g).
\end{equation}
Without further knowledge about $b^{(k+1)\,\mu}_{\phi\d^\nu\phi,{\cal L},...,
{\cal L}}$ we cannot interpret ${\cal A}^{\mu\nu}_g(f)$ as an interacting
field and ${\cal A}^{\mu\nu}_g(f)$ needs not to be symmetrical in
$\mu\leftrightarrow\nu$.\footnote{Due to {\bf (N0)} the anomaly
$a^{(k+1)\,\mu}_{\phi\d^\nu\phi,{\cal L},...,{\cal L}}=T_1(b^{(k+1)\,\mu}
_{\phi\d^\nu\phi,{\cal L},...,{\cal L}})$ has the form
\begin{gather}
a^{(k+1)\,\mu}_{\phi\d^\nu\phi,{\cal L},...,{\cal L}}(f,g,...,g)=
\int dx\,f(x)\int dy_1...dy_k\,g(y_1)...g(y_k)\notag\\
\{ P_4^{\mu\nu}(\d_1,...,\d_k)\delta (y_1-x,...,y_k-x)\notag\\
\sum_{j\leq l}:\phi (y_j)\phi (y_l):P_{2,jl}^{\mu\nu}(\d_1,...,\d_k)
\delta (y_1-x,...,y_k-x)\notag\\
+\sum_{j\leq l\leq r\leq s}:\phi (y_j)\phi (y_l)\phi (y_r)\phi (y_s):
P_{0,jlrs}^{\mu\nu}\delta (y_1-x,...,y_k-x)\},
\end{gather}
where $P^{\mu\nu}_{m,...}(\d_1,...,\d_k)$ is a polynomial
of degree $m$ in the partial derivatives $\d_{y_1},...,\d_{y_k}$, and
the expression in the $\{...\}$-bracket is symmetrical under permutations
of $y_1,...,y_k$. $P_{0,jlrs}^{\mu\nu}$ is \break
$\sim g^{\mu\nu}$ and, hence,
symmetrical in $\mu\leftrightarrow\nu$. But, e.g. for $k=2$ the
terms $(\w_1\d_1^\mu\d_2^\nu +\w_2\d_2^\mu\d_1^\nu)\delta(y_1-x,y_2-x)$
and $(:\phi^2(y_1):\d_1^\mu\d_2^\nu +:\phi^2(y_2):\d_2^\mu\d_1^\nu)
\delta(y_1-x,y_2-x)$ have not this $(\mu\leftrightarrow\nu)$-symmetry
and their
contributions to $a^{(k+1)\,\mu}_{\phi\d^\nu\phi,{\cal L},...,{\cal L}}
(\d_\mu f,g,...,g)-a^{(k+1)\,\nu}_{\phi\d^\mu\phi,{\cal L},...,{\cal L}}
(\d_\mu f,g,...,g)$ and hence to $I^{\mu\nu}_{g{\cal L}}(\d_\mu f)=
{\cal A}_g^{\mu\nu}(\d_\mu f)-{\cal A}_g^{\nu\mu}(\d_\mu f)$ (\ref{div(I)})
do not vanish.} By means of (\ref{mu-nu}) we find
\begin{equation}
(\phi\d^\nu\phi)_{g{\cal L}}(\d^\mu f)=(\phi\d^\mu\phi)_{g{\cal L}}
(\d^\nu f)-{\cal A}^{\mu\nu}_g(f)+{\cal A}^{\nu\mu}_g(f).\label{mu/nu}
\end{equation}
Now we define the improvement tensor
\begin{equation}
I^{\mu\nu}_{g{\cal L}}(f)\=d
-(\phi\d^\nu\phi)_{g{\cal L}}(\d^\mu f)+g^{\mu\nu}
(\phi\d_\rho\phi)_{g{\cal L}}(\d^\rho f).\label{I}
\end{equation}
By using (\ref{mu/nu}) we find that it is conserved up to
anomalous terms (i.e. terms which violate the MWI)
\begin{equation}
I^{\mu\nu}_{g{\cal L}}(\d_\mu f)={\cal A}_g^{\mu\nu}(\d_\mu f)-
{\cal A}_g^{\nu\mu}(\d_\mu f).\label{div(I)}
\end{equation}
To compute the trace, we first mention
\begin{gather}
T((\phi\tilde{\d}^\mu\d_\mu\phi) f\otimes W_1g_1\otimes...
\otimes W_n g_n)=\notag\\
i\sum_{m=1}^nT(W_1g_1\otimes...\otimes \phi\frac{\d W_m}{\d\phi}fg_m
\otimes...\otimes W_n g_n)\label{spur(T)}
\end{gather}
which is a consequence of $(\tilde {\rm\bf N})$ and (\ref{b=0}). Therefore,
\begin{equation}
(\phi\tilde{\d}^\mu\d_\mu\phi)_{g{\cal L}}(f)=-(\phi\frac{\d {\cal L}}
{\d\phi})_{g{\cal L}}(fg)=-4{\cal L}_{g{\cal L}}(fg).\label{spur1}
\end{equation}
From (\ref{I}), (\ref{d(pdp)}) and (\ref{spur1}) we obtain
\begin{equation}
\frac{1}{3}I^\mu_{\mu\,g{\cal L}}(f)=-(\d^\mu\phi\d_\mu\phi)_{g{\cal L}}(f)
+4{\cal L}_{g{\cal L}}(fg)-{\cal A}_{g\,\mu}^\mu (f)
=\Theta^\mu_{{\rm can}\,\mu\,g{\cal L}}(f)-{\cal A}_{g\,\mu}^\mu (f).
\label{spur}
\end{equation}
The improved energy momentum tensor is defined analogously
to (\ref{imp:class}), namely
\begin{equation}
\Theta^{\mu\nu}_{{\rm imp}\,g{\cal L}}(f)\=d
\Theta^{\mu\nu}_{{\rm can}\,g{\cal L}}(f)-
\frac{1}{3}I^{\mu\nu}_{g{\cal L}}(f).\label{imp}
\end{equation}
Our results (\ref{cons:can}), (\ref{div(I)}) and (\ref{spur})
yield that it is conserved and traceless up to anomalous terms:
\begin{eqnarray}
\Theta^{\mu\nu}_{{\rm imp}\,g{\cal L}}(\d_\mu f)&=&
\Theta^{\mu\nu}_{{\rm can}\,g{\cal L}}(\d_\mu f)-\frac{1}{3}
({\cal A}_g^{\mu\nu}(\d_\mu f)-{\cal A}_g^{\nu\mu}(\d_\mu f))\nonumber\\
&=&-{\cal L}_{g\,{\cal L}}((\d^\nu g)f)-\frac{1}{3}
({\cal A}_g^{\mu\nu}(\d_\mu f)-{\cal A}_g^{\nu\mu}(\d_\mu f)),\nonumber\\
\Theta^\mu_{{\rm imp}\,\mu\,g{\cal L}}(f)&=&{\cal A}_{g\,\mu}^\mu (f).
\label{sym:imp}
\end{eqnarray}
In the literature (\cite{P:emt} and references cited therein) it is shown
that the anomalous terms can be removed by suitable normalization
in one of the two equations in (\ref{sym:imp}), but not simultaneously
in both. Usually one puts the priority on the conservation and allows for
a trace anomaly. The latter breaks the dilatation invariance and gives
rise for anomalous dimensions of the interacting fields.

{\it Remark}: (1) We are going to show that the
trace anomaly is of order ${\cal O}(g^2)$
for the interaction (\ref{interaction}). We have to verify that
(\ref{div(can)}), (\ref{imp:ward}), (\ref{mu-nu}) and (\ref{spur(T)})
can be fulfilled without any anomalous terms $\mathcal{A}^{...}_{...}$
to first order in $g$. Due to {\bf (N3)} we have
\begin{gather}
T_2(\d^a\phi\tilde{\d}^b\d^c\phi ,\phi^4)(x,y)=
:\d^a\phi\d^{b+c}\phi (x)\phi^4(y):\label{zero}\\
+4\langle\Omega,T_2(\tilde{\d}^b\d^c\phi ,\phi)(x,y)\Omega\rangle:\d^a\phi (x)
\phi^3(y):\label{one-a}\\
+4\langle\Omega,T_2(\d^a\phi,\phi)(x,y)\Omega\rangle
:\d^{b+c}\phi (x)\phi^3(y):\label{one-b}\\
+6\langle\Omega,T_2(\d^a\phi\tilde{\d}^b\d^c\phi ,\phi^2)(x,y)\Omega\rangle
:\phi^2(y):.\label{two}
\end{gather}
For the tree diagrams (\ref{zero}), (\ref{one-a}) and (\ref{one-b})
the MWI holds true. An anomaly must come from
the loop diagram (\ref{two}), which is the two-legs sector.
We define the normalization of $\langle\Omega,T_2(\phi\tilde{\d}^\mu\d^\nu\phi
,\phi^2)\Omega\rangle$ by (\ref{imp:ward}) with $\mathcal{A}^{(2)\,\mu\nu}_
{\phi^2}\equiv 0$. The $(\mu\leftrightarrow\nu)$-symmetry
(\ref{mu-nu}) holds, because all tensors of rank two are $\sim g^{\mu\nu}$
or $\sim p^\mu p^\nu$, where $p$ is the momentum belonging to the relative
coordinate $(x-y)$. The $T$-products on the r.h. sides of
(\ref{div(can)}) and (\ref{spur(T)}) have four legs for $n=1$ and
$W_1={\cal L}=\lambda\phi^4$. Hence, it remains to show that there
exits a normalization such that
\begin{equation}
  \d^x_\mu \langle\Omega,T_2(\d^\mu\phi\d^\nu\phi,\phi^2)(x,y)\Omega\rangle=
\frac{1}{2}\d^\nu_x \langle\Omega,
T_2(\d^\rho\phi\d_\rho\phi,\phi^2)(x,y)\Omega\rangle
\end{equation}
(which is (\ref{div(can)})) and
\begin{equation}
  \d^x_\mu \langle\Omega,T_2(\phi\d^\mu\phi,\phi^2)(x,y)\Omega\rangle=
\langle\Omega,T_2(\d_\mu\phi\d^\mu\phi,\phi^2)(x,y)\Omega\rangle
\end{equation}
(which is (\ref{spur(T)})). An explicit
calculation shows that this can in fact be
done\footnote{The C-number distributions in (\ref{two}) for $b=0$ and
the relevant values of $a$ and $c$ have essentially been calculated in
the second paper of \cite{DHKS} (sect. 2 and appendix C).}.

\section{Conclusions}

The justifications to require the master Ward identity (as a
normalization condition for the time-ordered products) are the
following facts:

- In the classical limit $\hbar\rightarrow 0$ the MWI becomes an
identity which holds always true \cite{DF2}.

- The MWI has many, far-reaching and important
consequences (see sect. 4) which we would like to hold true in
QFT.\footnote{We discovered (or
invented) the MWI by searching for a local construction
of observables in non-Abelian quantum gauge theories. (In \cite{DF} this
construction is given for QED). We succeeded provided several normalization
conditions are fulfilled, see \cite{BDF}.
In order to prove that the latter have a common
solution we looked for a universal formulation of these normalization
conditions - and found the MWI.}

- It seems that the MWI can nearly always be satisfied:
it is compatible with the other normalization conditions (sect. 3),
and many consequences of the MWI (e.g. the field equation,
charge- and ghost-number conservation, conservation of the
energy momentum tensor
and perturbative gauge invariance ((\ref{N6}) with $j_1=...=j_n=0$)
for $SU(N)$-Yang-Mills theories) have already been proved in the
literature (sect. 4). The only counter-examples we know are the
usual anomalies of perturbative QFT.

\section{Appendix A: Feynman propagators}

Let $\varphi,\chi\in {\cal G}$ be the symbols corresponding to two massive
or massless free fields (without derivatives)
with the same mass and which satisfy the
Klein-Gordon or wave equation
\begin{equation}
(\w+m^2)\varphi =0,\quad\quad(\w+m^2)\chi=0,\quad\quad m\geq 0,
\end{equation}
and obey Bose or Fermi statistics. We assume that $T_1(\varphi g),\>g\in
{\cal D}(\RR^4)$ (anti-)commutes with all free
fields except $T_1(\chi h),\>h\in {\cal D}(\RR^4)$ and the same for
$\varphi$ and $\chi$ exchanged. The non-vanishing (anti-)commutator
is given by
\begin{equation}
\Delta_{\varphi,\chi}=\epsilon D_m,\label{D}
\end{equation}
where $D_m$ is the (massive or massless) Pauli-Jordan distribution
to the mass $m$, $\epsilon$ is a sign which
depends on $(\varphi,\chi)$ and we have extended
the notation (\ref{def:Delta}) to anti-commutators. For a bosonic
real scalar field it is $\chi=\varphi$ and for a bosonic complex
scalar field we have $\chi=\varphi^+$. In case of the fermionic
ghost fields of non-Abelian gauge theories
$\varphi$ and $\chi$ must be different: $\varphi=\tilde u_a,\>
\chi=u_a,\>\epsilon =1$ where $a$ is the colour index. Alternatively one
may also set $\varphi=u_a,\>\chi=\tilde u_a,\>\epsilon=-1$.
Spinor fields will be treated later.

According to our definition (\ref{Feyprop}) of the Feynman propagators
and the normalization condition {\bf (N0)},
$\Delta^F_{\d^a\varphi,\d^b\chi}$ contains undetermined local terms if
and only if
\begin{equation}
\omega\=d {\rm sd}(\Delta^F_{\d^a\varphi,\d^b\chi})-4\equiv
-2+|a|+|b|\geq 0,
\end{equation}
namely
\begin{equation}
\Delta^F_{\d^a\varphi,\d^b\chi}=\epsilon (-1)^{|b|}[\d^a\d^b D_m^F+
\sum_{|c|=0}^\omega C^{(a,b)}_c\d^c\delta],\label{D^F}
\end{equation}
where $D_m^F$ is the massive or massless Feynman propagator and the
$C^{(a,b)}_c\in\CC$ are constants.
We give an explicit list of the undetermined terms for the lowest values
of $|a|+|b|$:
\begin{gather}
\Delta^F_{\d^\mu\varphi,\d^\nu\chi}=-\epsilon [(\d^\mu\d^\nu D_m^F +
Cg^{\mu\nu}\delta)],\label{1-1}\\
\Delta^F_{\d^\mu\d^\nu\varphi,\chi}=\Delta^F_{\varphi,\d^\mu\d^\nu\chi}
=\epsilon [\d^\mu\d^\nu D_m^F -\frac{1}{4}g^{\mu\nu}\delta]\label{2-0}\\
-\Delta^F_{\d^\mu\d^\nu\varphi,\d^\lambda\chi}=\Delta^F_{\d^\lambda
\varphi,\d^\mu\d^\nu\chi}=\epsilon [\d^\mu\d^\nu\d^\lambda D_m^F\notag\\
+C_1g^{\mu\nu}\d^\lambda\delta-(\frac{1}{2} +2C_1)
(g^{\mu\lambda}\d^\nu\delta+g^{\nu\lambda}\d^\mu\delta)]\label{2-1}\\
\Delta^F_{\d^\mu\d^\nu\d^\lambda\varphi,\chi}=-\Delta^F_{\varphi,
\d^\mu\d^\nu\d^\lambda\chi}=\epsilon [\d^\mu\d^\nu\d^\lambda D_m^F\notag\\
-\frac{1}{6}(g^{\mu\nu}\d^\lambda\delta
+g^{\mu\lambda}\d^\nu\delta+g^{\nu\lambda}\d^\mu\delta)],\label{3-0}
\end{gather}
where we have taken account of Poincare covariance, symmetry with respect
to exchange of Lorentz indices and
\begin{equation}
\Delta^F_{\d^a\w\varphi,\d^b\chi}=-m^2 \Delta^F_{\d^a\varphi,\d^b\chi}=
\Delta^F_{\d^a\varphi,\d^b\w\chi}.
\end{equation}
With these formulas and $(\w +m^2) D_m^F=\delta$ we compute
$\delta^\mu_{\chi,\psi}\=d \partial^\mu\Delta^F_{\chi,\psi}
-\Delta^F_{\partial^\mu\chi,\psi}$ (\ref{delta2}):
\begin{gather}
\delta^\mu_{\varphi,\chi}=0,\label{delta:0-0}\\
\delta^\mu_{\varphi,\d^\nu\chi}=\epsilon Cg^{\mu\nu}\delta,
\label{delta:0-1}\\
\delta^\mu_{\d^\nu\varphi,\chi}=\epsilon \frac{1}{4}g^{\mu\nu}\delta,
\quad\quad\delta^\mu_{\d_\mu\varphi,\chi}=
\epsilon \delta,\label{delta:1-0}\\
\delta^\mu_{\d^\tau\varphi,\d^\nu\chi}=\epsilon [-(C+\frac{1}{2}+2C_1)
g^{\tau\nu}\d^\mu\delta +C_1g^{\mu\tau}\d^\nu\delta
-(\frac{1}{2}+2C_1)g^{\mu\nu}\d^\tau\delta],\label{delta:1'-1}\\
\delta^\mu_{\d_\mu\varphi,\d^\nu\chi}=-\epsilon
(1+C)\d^\nu\delta,\label{delta:1-1}\\
\delta^\mu_{\d_\mu\d_\tau\varphi,\chi}=\epsilon
\frac{3}{4}\d_\tau\delta,\label{delta:2-0}\\
\delta^\mu_{\d_\mu\d^\tau\varphi,\d^\nu\chi}=\epsilon [
(-\frac{1}{2}+C_1)\d^\nu\d^\tau\delta+(\frac{1}{2}+2C_1)
g^{\tau\nu}\w\delta -Cm^2g^{\tau\nu}\delta].\label{delta:2-1}
\end{gather}

For {\it spinor fields} with mass $m\geq 0$ obeying the
Dirac equation we have
\begin{equation}
\Delta_{\psi,\psq}=-(i\gamma^\mu\d_\mu +m)D_m
\end{equation}
and find
\begin{eqnarray}
\delta^\mu_{\gamma_\mu\psi,\psq}=-i\delta,\label{delta:psi-psq}\\
\delta^\mu_{\psq\gamma_\mu,\psi}=-i\delta.\label{delta:psq-psi}
\end{eqnarray}

\section{Appendix B: Explicit results for $\Delta^\mu$
used in the application of the MWI to the BRST-current}

Let $j_\mu$ be the free BRST-current (\ref{BRST-current}). We assume
that each symbol in
$W\in\mathcal{P}_0$ carries at most a first (internal) derivative
(no higher derivatives). Then the following
\begin{equation}
  \Delta^\mu_{\chi,\psi}
\bigl(\frac{\d j_\mu}{\d\chi}g,\frac{\d W}{\d\psi}f\bigr),\quad
\quad\chi,\psi\in\mathcal{G},
\end{equation}
do not vanish:
\begin{eqnarray}
\chi=\d_\tau A^\tau_a:&&\Delta^\mu_{\d_\tau A^\tau_a,A^\nu_b}
\bigl((\d^\mu u_a)g,\frac{\d W}{\d A^\nu_b}f\bigr)=
\frac{1}{4}(\d^\mu u_a)\frac{\d W}{\d A^\mu_a} gf,\label{j2}\\
&&\Delta^\mu_{\d_\tau A^\tau_a,\d^\sigma A^\nu_b}
\bigl((\d^\mu u_a)g,\frac{\d W}{\d (\d^\sigma A^\nu_b)}f\bigr)=\nonumber\\
&&(C_{A_a}+\frac{1}{2}+2C_{1A_a})[(\tilde\d^\mu\d_\mu u_a)
\frac{\d W}{\d (\d_\nu A^\nu_a)}gf\nonumber\\
&&+(\d_\mu u_a)\frac{\d W}{\d (\d_\nu A^\nu_a)}(\d^\mu g)f]\nonumber\\
&&-C_{1A_a}[(\tilde\d^\sigma\d^\nu u_a)
\frac{\d W}{\d (\d^\sigma A^\nu_a)}gf+(\d^\mu u_a)
\frac{\d W}{\d (\d^\nu A^\mu_a)}(\d^\nu g)f]\nonumber\\
&&+(\frac{1}{2}+2C_{1A_a})[(\tilde\d^\nu\d^\sigma u_a)
\frac{\d W}{\d (\d^\sigma A^\nu_a)}gf\nonumber\\
&&+(\d^\mu u_a)
\frac{\d W}{\d (\d^\mu A^\nu_a)}(\d^\nu g)f],\label{j3}
\end{eqnarray}
\begin{eqnarray}
\chi=\d_\mu u_a:&&\Delta^\mu_{\d_\mu u_a,\tilde u_b}
\bigl((\d_\tau A^\tau_a+m_a\phi_a) g,\frac{\d W}{\d \tilde
  u_b}f\bigr)=\nonumber\\
&&-(\d_\tau A^\tau_a+m_a\phi_a)\frac{\d W}{\d \tilde u_a}gf,\label{j4}\\
&&\Delta^\mu_{\d_\mu u_a,\d_\nu\tilde u_b}
\bigl((\d_\tau A^\tau_a+m_a\phi_a) g,
\frac{\d W}{\d (\d_\nu\tilde u_b)}f\bigr)=\nonumber\\
&&-(1+C_{u_a})
[(\tilde\d_\nu(\d_\tau A^\tau_a+m_a\phi_a))
\frac{\d W}{\d (\d_\nu\tilde u_a)}gf\nonumber\\
&&+(\d_\tau A^\tau_a+m_a\phi_a)\frac{\d W}{\d (\d_\nu\tilde u_a)}
(\d_\nu g)f,\label{j5}
\end{eqnarray}
\begin{eqnarray}
\chi=\d_\mu \d_\tau A^\tau_a:&&\Delta^\mu_{\d_\mu\d_\tau A^\tau_a,
A^\nu_b}\bigl(- u_a g,\frac{\d W}{\d A^\nu_b}f\bigr)=\nonumber\\
&&\frac{3}{4}[(\tilde\d^\nu u_a)\frac{\d W}{\d A^\nu_a}gf+
u_a\frac{\d W}{\d A^\nu_a}(\d^\nu g)f],\label{j6}\\
&&\Delta^\mu_{\d_\mu\d_\tau A^\tau_a,\d^\sigma
A^\nu_b}\bigl(- u_a g,\frac{\d W}{\d (\d^\sigma A^\nu_b)}f\bigr)=
(\frac{1}{2}-C_{1A_a})\cdot\nonumber\\
&&[(\tilde\d^\nu\tilde\d^\sigma u_a)
\frac{\d W}{\d (\d^\sigma A^\nu_a)}gf+(\tilde\d^\sigma u_a)
\Bigl(\frac{\d W}{\d (\d^\nu A^\sigma_a)}+\frac{\d W}{\d (\d^\sigma A^\nu_a)}
\Bigl)(\d^\nu g)f\nonumber\\
&&+u_a\frac{\d W}{\d (\d^\sigma A^\nu_a)}(\d^\nu\d^\sigma g)f]
-(\frac{1}{2}+2C_{1A_a})[(\tilde{\w}u_a)\frac{\d W}{\d (\d_\nu A^\nu_a)}
gf\nonumber\\
&&+2(\tilde\d_\nu u_a)\frac{\d W}{\d (\d_\tau A^\tau_a)}(\d^\nu g)f+
u_a\frac{\d W}{\d (\d_\tau A^\tau_a)}(\w g)f]\nonumber\\
&&+C_{A_a}m_a^2u_a\frac{\d W}{\d(\d_\tau A^\tau_a)}gf,\label{j7}
\end{eqnarray}
\begin{eqnarray}
\chi=u_a:&&\Delta^\mu_{u_a,\tilde u_b}
\bigl(-(\d_\mu(\d_\tau A^\tau_a+m_a\phi_a)) g,\frac{\d W}{\d \tilde u_b}f
\bigr)=0,\label{j8}\\
&&\Delta^\mu_{u_a,\d_\nu\tilde u_b}
\bigl(-(\d_\mu(\d_\tau A^\tau_a+m_a\phi_a)) g,
\frac{\d W}{\d (\d_\nu\tilde u_b)}f\bigr)=\nonumber\\
&&C_{u_a}(\d_\nu(\d_\tau A^\tau_a+m_a\phi_a))
\frac{\d W}{\d (\d_\nu\tilde u_a)}gf,\label{j9}
\end{eqnarray}
\begin{eqnarray}
\chi=\phi_a:&&\Delta^\mu_{\phi_a,\phi_b}
\bigl((\d_\mu u_a)g,\frac{\d W}{\d\phi_b}f\bigr)=0,\label{j9a}\\
&&\Delta^\mu_{\phi_a,\d_\nu\phi_b}
\bigl((\d_\mu u_a)g,\frac{\d W}{\d(\d_\nu\phi_b)}f\bigr)=
-C_{\phi_a}(\d_\mu u_a)\frac{\d W}{\d(\d_\mu\phi_a)}gf,\label{j9b}
\end{eqnarray}
\begin{eqnarray}
\chi=\d_\mu\phi_a:&&\Delta^\mu_{\d_\mu\phi_a,\phi_b}
\bigl(-u_ag,\frac{\d W}{\d\phi_b}f\bigr)=
u_a\frac{\d W}{\d\phi_a}gf,\label{j9c}\\
&&\Delta^\mu_{\d_\mu\phi_a,\d_\nu\phi_b}
\bigl(-u_ag,\frac{\d W}{\d(\d_\nu\phi_b)}f\bigr)=\nonumber\\
&&(1+C_{\phi_a})[(\tilde\d_\nu u_a)\frac{\d W}{\d(\d_\nu\phi_a)}gf
+u_a\frac{\d W}{\d(\d_\nu\phi_a)}(\d_\nu g)f],\label{j9d}
\end{eqnarray}
where we have used the explicit expressions (\ref{delta:0-0})-(\ref{delta:2-1})
for the $\delta^\mu$ and the definition (\ref{Delta2}) of
$\Delta^\mu$.

\vskip0.5cm
{\bf Acknowledgements:} We very much profitted from discussions with
Klaus Fredenhagen about all parts of this paper, including technical
details. He gave us important stimuli. Together with the first
author, he is working on a second paper about the master Ward identity
\cite{DF2}, which has influenced this paper. We thank Raymond Stora
for interesting comments, detailed questions and improvements of
various formulations. We are also grateful to Joachim Freund,
Dirk Prange and Karl-Henning Rehren for discussions.

This paper was mainly written at the 'II. Institut f\"ur Theoretische
Physik der Universit\"at Hamburg'.

\end{document}